\documentclass[aps,twocolumn,showpacs,preprintnumbers,nofootinbib,prd,superscriptaddress,10pt]{revtex4-1}

\usepackage{graphicx,amssymb,amsmath,amsthm,amsfonts,epsfig,epsf,fixmath}
\usepackage{bm}
\usepackage{dcolumn}
\usepackage{latexsym}
\usepackage{rotating}
\usepackage{longtable} 

\usepackage{enumerate}
\usepackage{tensor,multirow}
\usepackage{url}

\usepackage[linktocpage]{hyperref}
\usepackage[usenames]{color}
\usepackage{epstopdf}
\usepackage{bm}
\usepackage{dcolumn}
\usepackage[latin1]{inputenc}
\usepackage{latexsym}
\usepackage{rotating}
\usepackage{hyperref}
\usepackage{color}
\usepackage{longtable}

\usepackage{enumerate}
\usepackage{tensor,multirow}
\usepackage{url}
\usepackage{tikz,tikz-3dplot}
\usepackage{ulem}

\usepackage{slashed}

\usepackage[scr=boondox]{mathalfa}

\DeclareFontFamily{U}{min}{}
\DeclareFontShape{U}{min}{m}{n}{<-> udmj30}{}

\newcommand{\GSSI}{Gran Sasso Science Institute (GSSI), I-67100 L'Aquila, Italy}
\newcommand{\GranSasso}{INFN, Laboratori Nazionali del Gran Sasso, I-67100 Assergi, Italy}
\newcommand{\milan}{\affiliation{Dipartimento di Fisica ``G. Occhialini'', 
Universit\'a degli Studi di Milano-Bicocca, Piazza della Scienza 3, 20126 Milano, Italy}}
\newcommand{\infn}{\affiliation{INFN, Sezione di Milano-Bicocca, 
Piazza della Scienza 3, 20126 Milano, Italy}}

\newcommand{\nn}{\nonumber}

\begin{document}

\title{Analytical model of precessing binaries using post-Newtonian theory\\ in the extreme mass-ratio limit I: General Formalism
}

\author{Nicholas Loutrel}
 \email{nicholas.loutrel@unimib.it}
 \milan
 \infn
 \affiliation{Dipartimento di Fisica, ``Sapienza'' Universit\`a di Roma \& Sezione INFN Roma1, Piazzale Aldo Moro 5, 00185, Roma, Italy}

\author{Sajal Mukherjee}
 \email{sajal.mukherjee@pilani.bits-pilani.ac.in}
 \affiliation{Birla Institute of Technology and Science Pilani, Rajasthan, 333031, India}
\affiliation{Astronomical Institute of the Czech Academy of Sciences, Bo\v{c}n\'{i} II 1401/1a, CZ-141 00 Prague, Czech Republic}

\author{Andrea Maselli}
 \email{andrea.maselli@gssi.it}
 \affiliation{\GSSI}
 \affiliation{\GranSasso}
 
\author{Paolo Pani}
 \email{paolo.pani@uniroma1.it}
 \affiliation{Dipartimento di Fisica, ``Sapienza'' Universit\`a di Roma \& Sezione INFN Roma1, Piazzale Aldo Moro 5, 00185, Roma, Italy}


\begin{abstract}
    We develop a fully analytical waveform model for precessing binaries with 
    arbitrary spin vectors using post-Newtonian~(PN) theory in the extreme mass-ratio limit and a hierarchical multi-scale analysis.
    The analytical model incorporates leading PN order spin precession dynamics from spin-orbit, spin-spin, and quadrupole-monopole couplings, and 2PN order dissipative dynamics truncated to first order in the mass ratio $q \ll 1$. Due to the pure analytic nature of the model, the framework developed herein can readily be extended to both higher PN and higher-$q$ order.
    Although the PN series is asymptotic to this limit, our results can be used to estimate how precession affects the measurability of certain binary parameters, and to inform and compare with other waveform approximants, such as effective-one-body models, hybrid waveforms, and self-force calculations.
\end{abstract}

\maketitle

\section{Introduction}
Extreme mass-ratio inspirals~(EMRIs) are among the most interesting  sources for future gravitational-wave~(GW) detectors.
In a typical EMRI, a small compact object (the secondary) orbits around a supermassive 
black hole~(BH) (the primary), performing ${\cal O}(1/q)$ cycles before plunging, where 
$q\ll1$ is the binary mass ratio. Systems with a stellar mass secondary and $q< 10^{-4}$ 
emit GWs in the mHz regime, falling in the frequency bucket of the LISA 
space mission~\cite{LISA:2017pwj}. Less asymmetric binaries with lighter primaries 
would push the emission at higher frequencies, possibly entering the horizon of deciHz 
detectors~\cite{LGWA:2020mma}. Moreover, in case subsolar compact objects exist, 
EMRIs could assemble around stellar-mass or intermediate-mass BHs, providing a 
novel source for ground-based detectors~\cite{Miller:2020kmv,Barsanti:2021ydd}.

EMRI signals are chiefly emitted when the secondary probes the strong-field region near the 
primary. This, augmented by the large number of orbits performed before the plunge, 
allows for unprecedented precision in the parameter estimation, and makes EMRIs unique 
probes of strong gravity, to test both GW emission and the 
structure of spacetime near supermassive compact objects.

Projections based on future LISA detections show that tests of strong gravity with EMRIs 
will improve current constraints by several orders of magnitude~\cite{Babak:2017tow,Barausse:2020rsu,LISA:2022kgy,Baibhav:2019rsa}, 
including searches for extra fundamental charges and fields~\cite{Cardoso:2011xi,Yunes:2011aa,Pani:2011xj,Maselli:2020zgv,Maselli:2021men,Barsanti:2022ana,Barsanti:2022vvl,Liang:2022gdk,Zhang:2023vok,Zi:2022hcc,Lestingi:2023ovn}, anomalous multipole moments~\cite{Barack:2006pq,Babak:2017tow,Fransen:2022jtw,Raposo:2018xkf,Bena:2020see,Bianchi:2020bxa,Loutrel:2022ant,Kumar:2023bdf}, tests of the Kerr bound on the spin of the secondary~\cite{Piovano:2020ooe}, nonvanishing tidal Love 
numbers of the primary~\cite{Pani:2019cyc,Piovano:2022ojl,Datta:2021hvm}, horizon-scale structure~\cite{Datta:2019epe,Datta:2019euh,Maggio:2021uge}, tests of exotic compact objects~\cite{Pani:2010em,Macedo:2013qea,Destounis:2023khj} and of 
ultralight-boson condensates around BHs~\cite{Hannuksela:2018izj,Hannuksela:2019vip,Brito:2023pyl,Duque:2023cac}.

The large number of orbits in an EMRI is both a blessing and a curse. On the one hand, it provides 
a magnifying glass to measure and constrain the above effects to unprecedented levels, 
often well below percent level.
On the other hand, this requires an equally exceptional modelling of the complex and 
long EMRI signal in order to tame systematic errors (see~\cite{LISAConsortiumWaveformWorkingGroup:2023arg} 
for a recent review).
In particular, astrophysical EMRIs are expected to assemble mostly due multibody scattering events, and therefore to evolve on highly eccentric, non-equatorial 
orbits.
For the same reason, both the primary and secondary spin vectors are generically oriented and not aligned with each other nor with the orbital angular momentum. This implies that, at variance with stellar-mass binaries detected so far~\cite{GWTC-3}, precession in EMRIs is the rule rather than the exception and should therefore be accurately modelled.

State of the art perturbative self-force~(SF) models are currently able to provide 
waveforms accurate at first-post-adiabatic accuracy, i.e. yielding a $\mathcal{O}(q)$ 
phase error over the course of the inspiral~\cite{Pound:2019lzj,Warburton:2021kwk,Wardell:2021fyy}. Such models have been computed 
for quasi-circular inspirals of Schwarzschild BHs in General Relativity~\cite{Durkan:2022fvm,Miller:2020bft}.  Extending these 
results to generic orbits around a Kerr BH is a primary goal of current efforts~\cite{Green:2019nam,Dolan:2021ijg,Upton:2021oxf,Toomani:2021jlo,Osburn:2022bby,Spiers:2023cip,Spiers:2023mor,Miller:2023ers}, 
possibly including other effects, as 
geodesic resonances and the spin of the secondary~\cite{Nasipak:2021qfu,Piovano:2020zin,Mathews:2021rod,Drummond:2022xej,Upton:2023tcv,Drummond:2023loz,Upton:2023tcv}. Further, in the broader context of GW modeling and phenomenology, SF models overlap in validity with the effective-one-body (EOB) framework~\cite{Buonanno:1998gg,Buonanno:2005xu,Damour:2007xr,Ramos-Buades:2023ehm,Gamba:2021ydi}, which seeks to approximate the fully relativistic two-body problem through a deformed Schwarzschild/Kerr BH, with the deformation informed from the current limit of PN theory and calibrated to numerical relativity simulations. The two approaches, SF and EOB, have shown quantitative agreement for quasi-circular non-spinning EMRIs~\cite{Albertini:2023aol}.

With SF calculations beyond General Relativity having just started to develop consistent 
waveform models~\cite{Spiers:2023cva},  most of the studies about tests of gravity listed 
above are typically restricted to simplified settings, especially to circular and/or equatorial 
orbits around a Kerr BH. When the secondary spin is included, this is typically assumed 
to be aligned to the primary spin and to the binary angular momentum, forcing zero precession~\cite{Huerta:2011kt,Huerta:2011zi,Piovano:2022ojl,Piovano:2021iwv}.\footnote{See~\cite{Lynch:2023gpu,Drummond:2023wqc} for recent progress in modelling a misalignment between the primary spin and the orbital angular momentum at first post-adiabatic order.}
It is important to stress that, for EMRIs, there is no strong underlying motivation for this assumption 
other than simplicity.
Furthermore, although challenging to model, precession is crucial to disentangle certain effects due to the objects' multipolar structure~\cite{Loutrel:2023boq}.

The scope of this paper is to develop a new waveform model for precessing binaries with arbitrary spin vectors. By exploiting {\it both} the EMRI limit and the post-Newtonian~(PN) 
approximation, as well as using a multi-scale analysis, we can solve the equations of motion {\it fully analytically}. 

While the PN series is known to be asymptotic to the EMRI limit~\cite{Fujita:2012cm,Yunes:2008tw,Futamase:1983vsr},  
and accurate waveform modelling for EMRIs requires SF calculations, we believe 
our approach can be fruitful for a variety of purposes. For example, generation of 
SF templates is computationally expensive due to the long duration of the signal and 
high-harmonic content, although impressive progress has been achieved recently~\cite{Chua:2020stf,Hughes:2021exa,Katz:2021yft,Speri:2023jte}.
Our analytical waveform model can be very useful for fast, order-of-magnitude, measurability 
estimates on the relevance of precession effects, and for comparison/hybridization with 
other waveform models aiming to incorporate SF results, for example to describe less asymmetric 
binaries~\cite{Albertini:2022dmc,vandeMeent:2023ols,LISAConsortiumWaveformWorkingGroup:2023arg}.

In this paper we present the formalism and analytical computation. Parameter estimation will be discussed in a follow-up work~\cite{followup}.

The rest of the paper is organized as follows. 
In Sec.~\ref{sec:setup} we present the setup, framework, and main equations. The latter are then solved perturbatively in the $q\ll1$ and PN limit in Sec.~\ref{sec:prec}. Radiation-reaction effects and waveform approximants are computed in Sec.~\ref{sec:rr}, while in Sec.~\ref{sec:dephase} we explicitly present quantitative estimates for the EMRI GW phase introduced by precession.
We conclude in Sec.~\ref{sec:conclusion} with a discussion on possible extensions.
We use geometrized ($G=c=1$) units throughout.

\section{Setup \& Formalism} \label{sec:setup}

In this section, we provide the details of our formalism, specifically the reduction of the PN precession equations to the EMRI limit, and the details of our chosen spin-precessing waveform. 

\subsection{EMRI Limit of the PN Spin Precession Equations}

Consider the PN spin precession equations, which to second order in the spins of the individual bodies include the spin-orbit, spin-spin, and quadrupole-monopole couplings~\cite{Khan:2018fmp,Kesden:2014sla,Gerosa:2015tea,Chatziioannou:2017tdw}. To the leading PN order, the equations take the form
\begin{align}
    \label{eq:S1dot}
    \dot{\vec{S}}_{1} &= \vec{\Omega}_{1} \times \vec{S}_{1}\,,
    \\
    \label{eq:S2dot}
    \dot{\vec{S}}_{2} &= \vec{\Omega}_{2} \times \vec{S}_{2}\,,
    \\
    \label{eq:Lhat-dot}
    \dot{\hat{L}} &= - L^{-1} \left(\dot{\vec{S}}_{1} + \dot{\vec{S}}_{2}\right)\,,
\end{align}
where 
\begin{equation}
    \label{eq:Omega-1}
    \vec{\Omega}_{1} = \vec{\Omega}^{\rm SO}_{1} + \vec{\Omega}^{\rm SS}_{1} + \vec{\Omega}^{\rm QM}_{1}\,,
\end{equation}
with
\begin{align}
    \label{eq:spin-orbit}
    \vec{\Omega}^{\rm SO}_{1} &= \frac{1}{r^{3}} \left(2 + \frac{3}{2} \frac{m_{2}}{m_{1}}\right) L \hat{L}\,,
    \\
    \label{eq:spin-spin}
    \vec{\Omega}^{\rm SS}_{1} &= \frac{1}{2r^{3}} \left[ S_{2} - 3 \left(\vec{S}_{2} \cdot \hat{L}\right)\hat{L}\right] \,,
    \\
    \label{eq:quad-mon}
    \vec{\Omega}^{\rm QM}_{1} &= -\frac{3}{2r^{3}} \frac{m_{2}}{m_{1}} \left(\vec{S}_{1} \cdot \hat{L}\right) \hat{L}\,.
\end{align}
In the above expressions, $L = \mu \sqrt{M r}$, the masses of the two BHs are $(m_{1},m_{2})$, $\mu = m_{1}m_{2}/M$ is the reduced mass, and $M = m_{1}+m_{2}$ is the total mass. 
The expression for $\vec{\Omega}_{2}$ is obtained by taking $1\leftrightarrow2$ in Eqs.~\eqref{eq:Omega-1}-\eqref{eq:quad-mon}. The spins can be written in terms of dimensionless quantities as $\vec{S}_{A} = m_{A}^{2} \vec{\chi}_{A}$, with $A = (1,2)$. Eq.~\eqref{eq:Lhat-dot} is due to the fact that the total angular momentum, $\vec{J} = L \hat{L} + \vec{S}$ (where $\vec{S}=\vec{S}_{1} + \vec{S}_{2}$ is the total spin vector) is conserved on the precession timescale.

Now, consider taking the EMRI limit of these equations. Let the larger object have mass and spin $(m_{1},\vec{S}_{1})$, while the smaller object has mass and spin $(m_{2}, \vec{S}_{2})$. The limit is given by $m_{1} \gg m_{2}$, with 
$q = m_{2}/m_{1} \ll 1$. From this,
\begin{align}
    M &= (1+q) m_{1}\,, \qquad \mu = \frac{qm_{1}}{1+q} \,, 
    \\
    L &= \frac{q m_{1}^{2}}{v}\,, \qquad \vec{S}_{2} = q^{2} m_{1}^{2} \vec{\chi}_{2} = q^{2} \vec{\sigma}_{2}\,,
    \\
    \label{eq:J-emri}
    \vec{J} &= \vec{S}_{1} + \frac{q m_{1}^{2}}{v} \hat{L} + q^{2} \vec{\sigma}_{2}\,,
\end{align}
where $v=(M/r)^{1/2}$ is the orbital velocity, and $\vec{\sigma}_{2} = \vec{\chi}_{2} m_{1}^{2}$ is a ``renormalized" spin vector. The inclusion of quadrupole-monopole interactions introduces an additional constant of motion on the precession timescale, specifically
\begin{align}
    \label{eq:chi-eff}
    M^{2} \chi_{\rm eff} &= \left(1 + \frac{m_{2}}{m_{1}} \right) \left(\vec{S}_{1} \cdot \hat{L} \right) + \left(1 + \frac{m_{1}}{m_{2}} \right) \left(\vec{S}_{2} \cdot \hat{L} \right)\,,
    \\
    &= \left(\vec{S}_{1} \cdot \hat{L} \right) + q \left[\left( \vec{S}_{1} + \vec{\sigma}_{2}\right)\cdot \hat{L} \right] + q^{2} \left(\vec{\sigma}_{2} \cdot \hat{L}\right)\,,
\end{align}
where we have recast all expressions to highlight all $q$ factors. 
Inserting these into Eqs.~\eqref{eq:S1dot}-\eqref{eq:Lhat-dot}, and expanding in $q\ll1$ gives
\begin{align}
\label{eq:prec-1-emri}
    \frac{d\vec{\chi}_{1}}{d\tau} &= \left[q \vec{\Omega}_{1}^{(1)} + q^{2} \vec{\Omega}_{1}^{(2)} + {\cal{O}}(q^{3})\right] \times \vec{\chi}_{1}\,,
    \\
    \label{eq:prec-2-emri}
    \frac{d\vec{\chi}_{2}}{d\tau} &= \left[\vec{\Omega}_{2}^{(0)} + {\cal{O}}(q) \right] \times \vec{\chi}_{2}\,,
    \\
    \label{eq:prec-L-emri}
    \frac{d\hat{L}}{d\tau} &= \left[\vec{\Omega}_{L}^{(0)} + q \vec{\Omega}_{L}^{(1)} + {\cal{O}}(q^{2})\right] \times \hat{L}
\end{align}
with $\tau = t/m_{1}$, and
\begin{align}
    \label{eq:Omega11}
    \vec{\Omega}_{1}^{(1)} &= v^{5}\left(2 - \frac{3}{2} v \chi_{\rm eff}\right) \hat{L} \,,
    \\
    \label{eq:Omega12}
    \vec{\Omega}_{1}^{(2)} &= -v^{5} \left(\frac{9}{2} - 3 v \chi_{\rm eff} \right)\hat{L} + \frac{v^{6}}{2} \vec{\chi}_{2}
    \\
    \label{eq:Omega20}
    \vec{\Omega}_{2}^{(0)} &= \frac{3}{2} v^{5} \left(1 - v \chi_{\rm eff} \right) \hat{L} + \frac{v^{6}}{2} \vec{\chi}_{1}
    \\
    \label{eq:OmegaL0}
    \vec{\Omega}_{L}^{(0)} &= \frac{v^{6}}{2} \left(4 - 3 v \chi_{\rm eff}\right) \vec{\chi}_{1}\,,
    \\
    \label{eq:OmegaL1}
    \vec{\Omega}_{L}^{(1)} &= \frac{3}{2} v^{6} \left(-3 + 2 v \chi_{\rm eff} \right)
    \vec{\chi}_{1} + \frac{3}{2} v^{6} \left(1 - v \chi_{\rm eff}\right)\vec{\chi}_{2}\,,
\end{align}
We have used the definition of $\chi_{\rm eff}$ in Eq.~\eqref{eq:chi-eff} to replace all instances of $\vec{S}_{1} \cdot \hat{L}$. It is worth noting that Eqs.~\eqref{eq:prec-1-emri}-\eqref{eq:prec-L-emri} are expanded to different orders in $q$. The reason for this is two-fold: 1) it is a necessity to employ MSA when we study solutions to the precession equations in the EMRI limit in Sec.~\ref{sec:prec}, and 2) it is necessary to ensure that $\vec{J}$ is conserved on the precession timescale, due to the different scalings of the spin vectors with $q$ in Eq.~\eqref{eq:J-emri}. Indeed, it is straightforward to check that $d\vec{J}/dt = {\cal{O}}(q^{3})$, and thus $\vec{J}$ is conserved at this order in the mass ratio (neglecting radiation reaction). 

At this stage, there is a useful simplification that can be performed, by replacing 
$\vec{\chi}_{1}$ with $\vec{J}/m_1^2$.
The expansion of $\vec{J}$ in the EMRI limit is given by Eq.~\eqref{eq:J-emri}. 
Using this expansion to replace 
$\vec{\chi}_{1}$ in Eqs.~\eqref{eq:OmegaL0}-\eqref{eq:OmegaL1} will produce a remainder 
${\cal{O}}(q^{2})$, since the linear term in the mass ratio is proportional to $\hat{L}$ and 
$\hat{L} \times \hat{L} = 0$. Such ${\cal{O}}(q^{2})$ remainder can be neglected, being 
higher order than what we are presently considering. Likewise, performing the same procedure 
with Eq.~\eqref{eq:Omega20} yields a remainder ${\cal{O}}(q)$, which can also be 
neglected. 
Thus, to the order we are working in $q$, we may replace $\vec{\chi}_{1}$ with 
$\vec{J}/m_{1}^{2}$ without loss of accuracy. Doing so completely eliminates the 
precessing $\vec{\chi}_{1}$ from Eqs.~\eqref{eq:prec-2-emri}-\eqref{eq:prec-L-emri}, and 
replaces it with the fixed $\vec{J}$. As a result, the evolution of $\chi_{1}$ decouples from the other angular momenta. The latter, namely $\vec{\chi}_2$ and $\hat L$, are the only quantities we need to take into account 
when solving for the precession equations in the EMRI Limit. 

\subsection{Spin Precessing SPA Waveform}

Gravitational waveforms from spin precessing binaries 
were first studied in the PN approximation in~\cite{ApostolatosCutler}, with the first 
frequency domain templates being computed by 
Lang \& Hughes~\cite{Lang:2006bsg} using 
the stationary phase approximation (SPA)~\cite{Bender}. 
Standard SPA breaks down for spin precessing waveforms~\cite{Klein:2014bua}, and is not suitable for actual searches. However, the Lang \& Hughes waveform provides the necessary qualitative features to capture the effects of spin precession of the frequency domain waveform, 
and results in conservative estimates on parameter estimation~\cite{Lang:2011je,Yagi:2009zm}. As a matter of simplicity, we limit ourselves to this waveform model. 
We remark however that the formalism and the analysis 
of the precession equations, contained in Sec.~\ref{sec:prec} \&~\ref{sec:waveform_prec}, 
is general enough to be used in any 
precessing waveform. 

In the frequency domain, the Lang \& Hughes waveform for spin precessing quasi-circular binaries takes the form
\begin{equation}
    \label{eq:waveform}
    \tilde{h}_{I}(f) = \sqrt{\frac{5}{96}} \frac{{\cal{M}}^{5/6}}{\pi^{2/3} D_{L}} A_{I}(f) f^{-7/6} e^{i\Phi_{I}(f)}
\end{equation}
with frequency $f$, chirp mass ${\cal{M}}=\mu^{3/5}M^{2/5}$, luminosity distance $D_{L}$, and the subscript $I$ labelling 
the two possible data streams from the three arms of the LISA detector\footnote{While our analysis is independent of the specific GW detector, for concreteness we will focus on LISA, for which EMRIs are a primary target.}. The amplitude function $A_{I}(f)$ is a slowly varying function of frequency through the precession orbital angular momentum of the binary and the motion of the LISA constellation. For the present analysis, the phase is more important than the amplitude, but explicit expression for the latter can be found in Eq.~(2.29) of~\cite{Lang:2006bsg}. The waveform phase $\Phi(f)$ is a sum of multiple contributions, specifically
\begin{equation}
    \label{eq:waveform-phase}
    \Phi(f) = \Psi(f) - \varphi_{{\rm pol},I}(f) - \varphi_{D}(f) - \delta\Phi(f)\,,
\end{equation}
with $\Psi(f)$ the Fourier phase arising from the evolution of orbital quantities under radiation reaction, $\varphi_{\rm pol}(f)$ the polarization phase due to the time-(frequency-)varying polarization basis of the GWs, $\varphi_{D}$ the Doppler phase due to the motion of the LISA constellation, and $\delta\Phi$ the waveform precession phase due to the precessing orbital angular momentum. The first and last of these contributions are most important to the present analysis, and are the focus of the remainder of this paper. Expressions for the polarization and Doppler phases can be found in Eqs.~(2.30) \& (2.31) in~\cite{Lang:2006bsg}, respectively.

In the time domain, the Fourier phase $\Psi$ is given by
\begin{equation}
    \label{eq:phase-time-domain}
    \Psi(t) = 2 \pi f t - 2 \phi(t)
\end{equation}
with $\phi(t)$ the orbital phase, and for any given value of $f$. The factor of two multiplying the orbital phase comes from the leading PN approximation of GWs, specifically the quadrupole approximation. Coordinate time $t$ and orbital phase $\phi$ may be computed in TaylorF2 approximates in terms of $v=(2\pi M F)^{1/3}$, with $F$ the orbital frequency. Application of the stationary phase approximation to Eq.~\eqref{eq:phase-time-domain} enforces $f=2F$. Thus $v=(\pi M f)^{1/3}$, and the Fourier phase takes the form
\begin{equation}
    \Psi(f) = 2\pi f t_{c} - 2\phi_{c} - \frac{\pi}{4} + \frac{3}{128 \eta v^{5}} \left[1 + \sum_{n} \psi_{n} v^{n} \right]\,.
\end{equation}
The PN coefficients of the phase are well known for generic mass binaries, see for example~\cite{Damour:2000zb}. We will provide expressions in the EMRI limit in Sec.~\ref{sec:F2-orb}.

The evolution of the waveform precession phase is directly related to the precession equation of the orbital angular momentum through Eq.~(28) of~\cite{ApostolatosCutler}, specifically
\begin{equation}
    \label{eq:prec-phase0}
    \frac{d\delta \Phi}{dt} = \frac{(\hat{L}\cdot\vec{N})}{1 - (\hat{L}
    \cdot \vec{N})^{2}} \left(\hat{L} \times \vec{N}\right) \cdot \frac{d\hat{L}}{dt}\,,
\end{equation}
where $\vec{N}$ is the line of sight from the detector to the source. Since the LISA constellation is not fixed relative to the Solar System's barycenter, $\vec{N}$ is generally a function of time. Further, because of the dependence on $\hat{L}$, the solution for $\delta \Phi$ is intricately tied to the solutions to the precession equations. Solving for these equations is the main goal of this work, and the detailed procedure will be explained in the next sections. Once one obtains the solution within the time domain, the waveform precession phase can be computed in the frequency domain by application of the stationary phase condition.

\section{Precession Solutions in the EMRI Limit}
\label{sec:prec}

The full precession equations in Eqs.~\eqref{eq:S1dot}-\eqref{eq:quad-mon} were solved analytically in the compatable mass limit in~\cite{Kesden:2014sla,Gerosa:2015tea,Chatziioannou:2017tdw}. The procedure to obtain these solutions is to transform into a co-precessing frame, where the angular momenta undergo nutation parameterized by the magnitude of the total spin $S^{2}$, which can be solved for in terms of Jacobi elliptic functions. One then returns to the inertial frame and solves for the precession angle. Due to the multiple scales associated with the problem (orbital, precession, and radiation reaction), one must utilize multiple scale analysis (MSA) to obtain solutions with sufficient phase accuracy for GW modeling. In this context, the leading order secular evolution at each scale is obtained by averaging over the shorter scale(s), specifically the precession dynamics are averaged over the orbital timescale, and the radiation reaction effects are averaged over both the orbital and precession timescale.

Since we are seeking precessing solutions within the EMRI limit, we break from the previous work and focus on solutions to the reduced equations in Eqs.~\eqref{eq:prec-1-emri}-\eqref{eq:OmegaL1}. A tempting approach to solving Eqs.~\eqref{eq:prec-2-emri}-\eqref{eq:prec-L-emri} might be to consider a series solution of the form
\begin{align}
    \label{eq:power-series}
    \hat{L}(t) = \sum_{n=0} q^{n} \hat{L}^{(n)}(t)\,,
\end{align}
and likewise for $\vec{\chi}_{2}$. At leading order, the equations and their solutions are all bounded. However, for $\hat{L}^{(1)}$ the subsequent evolution equations contain a resonance at the frequency associated with $\vec{\Omega}_{L}^{(0)}$, causing the solution to diverge linearly in time. This resonance is an artifact of the expansion in Eq.~\eqref{eq:power-series}, and is thus, unphysical. Similar pathologies appear in the study of nonlinear and/or forced oscillators, and must be handled via MSA or renormalization methods.

This consideration implies that, unlike the comparable mass case in~\cite{Gerosa:2015tea,Chatziioannou:2017tdw} (see also very recent work \cite{MacUilliam:2024oif,Arredondo:2024nsl}) where there are only three scales, in the EMRI scenario, there are actually four: one for the orbital timescale $T_{\rm orb} \sim v^{-3}$, two for the precession timescale $T_{L} \sim v^{-6}$ and $T_{\chi_{2}} \sim q v^{-6}$, and one for the radiation reaction timescale $T_{\rm rr} \sim v^{-9}$. Of the two precession timescales, the former arises from the precession of the orbital angular momentum, while the latter arises from the precession of the secondary's spin and is suppressed by $q$. Note that the presence of so many scales is largely a result of using PN equations as an approximation to the systems we wish to study. In a realistic self-force calculation, the only new scale would result from the presence of the smaller mass and would solely be coupled to the mass ratio $q$. Indeed, this is what introduces the new scale within the precession dynamics, specifically through the second term in Eq.~\eqref{eq:prec-L-emri}. In addition, to handle this new scale, we do not average over the shorter scales as is typically done in PN problems. Instead, the MSA we perform on the precession dynamics is to eliminate the artificial resonances that lead to divergences in the precession solutions, thus renormalizing the amplitude. While we use MSA to achieve this, one can also use renormalization group techniques, which provide equivalent solutions (see Appendix~\ref{app:RG}). 

To set up the MSA, we define the long timescale $\tilde{\tau} = q \tau$, and seek solutions of the form
\begin{equation}
    \hat{L}(t) = \sum_{n=0}^{\infty} q^{n} \hat{L}^{(n)}(\tau, \tilde{\tau})\,.
\end{equation}
The total derivatives with respect to $\tau$ transform to
\begin{equation}
    \frac{d}{d\tau} = \frac{\partial}{\partial \tau} + q \frac{\partial}{\partial\tilde{\tau}}\,.
\end{equation}
Under these assumptions, after expanding in $q \ll 1$, Eqs.~\eqref{eq:prec-2-emri}-\eqref{eq:prec-L-emri} become
\begin{align}
    \label{eq:L0-msa}
    \frac{\partial \hat{L}^{(0)}}{\partial \tau} &= \omega_{L} \hat{e}_{z} \times \hat{L}^{(0)}(\tau,\tilde{\tau})\,,
    \\
    \label{eq:chi2-msa}
    \frac{\partial \vec{\chi}_{2}^{(0)}}{\partial \tau} &= \left[\omega_{SJ} \hat{e}_{z} + \omega_{SL} \hat{L}^{(0)}\right] \times \vec{\chi}_{2}^{(0)}\,,
    \\
    \label{eq:L1-msa}
    \frac{\partial \hat{L}^{(1)}}{\partial \tau} + \frac{\partial \hat{L}^{(0)}}{\partial \tilde{\tau}} &= \omega_{L} \hat{e}_{z} \times \hat{L}^{(1)}  - \omega_{L}^{(1)} \hat{e}_{z} \times \hat{L}^{(0)} 
    \nn \\
    &+ v \omega_{SL} \vec{\chi}_{2}^{(0)} \times \hat{L}^{(0)}
\end{align}
with the frequencies 
\begin{align}
    \label{eq:omega-defs}
    \omega_{L} &= \frac{Jv^{6}}{2m_{1}^{2}}(4-3v\chi_{\rm eff})\,, \qquad \omega_{L}^{(1)} = \frac{3Jv^{6}}{2m_{1}^{2}}(3 - 2 v \chi_{\rm eff})\,,
    \nonumber \\
    \omega_{SL} &= \frac{3}{2} v^{5} \left(1-v\chi_{\rm eff}\right)\,, \qquad \omega_{SJ} = \frac{Jv^{6}}{2m_{1}^{2}}\,.
\end{align}
These new equations, specifically Eqs.~\eqref{eq:L0-msa}-\eqref{eq:L1-msa}, can be solved iteratively up to linear order in $q$.

\subsection{$\hat{L}$ at ${\cal{O}}(q^{0})$}
\label{sec:L0}

To leading order in MSA, the evolution of the orbital angular momentum is given by Eq.~\eqref{eq:L0-msa}. The presence of the cross product therein couples the $x$- and $y$-components of $\hat{L}^{(0)}$ together. However, these can be decoupled by re-casting the equations in second order form, specifically
\begin{align}
    \label{eq:Lx-eqn}
    \frac{\partial^{2}\hat{L}_{x}^{(0)}}{\partial\tau^{2}} &= - \omega_{L}^{2} \hat{L}_{x}\,,
    \qquad
    \frac{\partial^{2}\hat{L}_{y}^{(0)}}{\partial\tau^{2}} = -\omega_{L}^{2} \hat{L}_{y}\,,
\end{align}
The general solutions are
\begin{align}
    \label{eq:Lx-0}
    \hat{L}_{x}^{(0)}(\tau,\tilde{\tau}) &= X(\tilde{\tau}) \cos[\alpha(\tau)] + Y(\tilde{\tau}) \sin[\alpha(\tau)]\,,
    \\
    \label{eq:Ly-0}
    \hat{L}_{y}^{(0)}(\tau,\tilde{\tau}) &= X(\tilde{\tau}) \sin[\alpha(\tau)] - Y(\tilde{\tau}) \cos[\alpha(\tau)]\,,
\end{align}
where $(X,Y)$ are purely functions of the long timescale $\tilde{\tau}$, and the \textit{precession angle} $\alpha$ is defined by
\begin{equation}
    \label{eq:alpha-eqn}
    \frac{d\alpha}{d\tau} = \omega_{L}\,,
\end{equation}
which reduces to $\alpha(\tau) = \omega_{L} \tau$ since $\omega_{L}$ is a constant in the absence of radiation reaction. However, once radiation reaction is included, one needs to directly integrate Eq.~\eqref{eq:alpha-eqn} to obtain the full evolution of $\alpha(\tau)$, i.e.
\begin{equation}
    \label{eq:alpha-rr}
    \alpha(v) = \alpha_{c} + \int dv \frac{\omega_{L}(v)}{dv/d\tau}\,.
\end{equation}
The $z$-component can trivially be solved to obtain
\begin{equation}
    \label{eq:Lz-0}
    \hat{L}_{z}^{(0)}(\tilde{\tau}) = Z(\tilde{\tau})
\end{equation}
which does not depend on the short timescale $\tau$. The functions $(X,Y,Z)$ cannot be determined at this order in the MSA, but due to the normalization of the orbital angular momentum obey
\begin{equation}
    \label{eq:XYZ-norm}
    [X(\tilde{\tau})]^{2} + [Y(\tilde{\tau})]^{2} + [Z(\tilde{\tau})]^{2} = 1\,.
\end{equation}
%

\subsection{$\vec{\chi}_{2}$ at ${\cal{O}}(q^{0})$}
\label{sec:chi20}

To leading order, the precession equation for $\vec{\chi}_{2}^{(0)}(\tau,\tilde{\tau})$ is given in Eq.~\eqref{eq:chi2-msa},
with the frequencies $\omega_{SL}$ and $\omega_{SJ}$ given in Eq.~\eqref{eq:omega-defs}. 
The original precession equations possess a sufficient number of constants of motion to define a co-precessing reference frame, and the same holds true at ${\cal{O}}(q^{0})$. We can exploit this fact to define a frame where the unit orbital angular momentum is fixed to be
\begin{align}
    \hat{L}^{(0)}_{\rm co-p} = \left[X(\tilde{\tau}), Y(\tilde{\tau}), Z(\tilde{\tau})\right]\,,
\end{align}
while $\hat{J}$ still defines the z-axis of the frame. To define a suitable basis in the co-precessing frame, allow two basis vectors to be $\hat{J}$ and
\begin{align}
    \hat{P} &= \frac{\hat{L}^{(0)} \times \hat{J}}{|\hat{L}^{(0)} \times \hat{J}|} 
    \nn \\
    &=\frac{\left[X \sin\alpha - Y \cos\alpha, - X \cos\alpha - Y \sin\alpha, 0 \right]}{\sqrt{1-Z^{2}}}\,,
\end{align}
the latter of these is orthogonal to the $LJ$-plane. A third can then be defined by
\begin{align}
    \hat{Q} &= \frac{\hat{J} \times \hat{P}}{|\hat{J} \times \hat{P}|} 
    \nn\\
    &= \frac{\left[X \cos \alpha + Y\sin\alpha, X\sin\alpha-Y\cos\alpha, 0\right]}{\sqrt{1-Z^{2}}}\,,
\end{align}
which, along with $\hat{L}$, spans the $LJ$-plane. The secondary's spin vector can then be written as
\begin{align}
    \label{eq:chi2-co-p}
    \vec{\chi}_{2}^{(0)} &= \chi_{2,P}(\tau) \hat{P} + \chi_{2,Q}(\tau) \hat{Q} + \chi_{2,J}(\tau) \hat{J}\,,
\end{align}
with the components satsifying
\begin{align}
    \frac{\partial\chi_{2,P}}{\partial\tau} &= \omega_{R}(\tilde{\tau})\chi_{2,J}  -\omega_{Q}(\tilde{\tau})\chi_{2,Q}\,,
    \\
     \frac{\partial\chi_{2,Q}}{\partial\tau} &= \omega_{Q}(\tilde{\tau}) \chi_{2,P}\,,
    \\
    \frac{\partial\chi_{2,J}}{\partial\tau} &= -\omega_{R}(\tilde{\tau})\chi_{2,P}\,.
\end{align}
where
\begin{align}
    \omega_{R}(\tilde{\tau}) &= \omega_{SL} \sqrt{[X(\tilde{\tau})]^{2} + [Y(\tilde{\tau})]^{2}}\,,
    \\
     \omega_{Q}(\tilde{\tau}) &= \omega_{SJ} - \omega_{L} + \omega_{SL} Z(\tilde{\tau})\,.
\end{align}
This system can be decoupled to obtain a single equation for $\chi_{2,P}$, specifically
\begin{equation}
    \frac{d^{2}\chi_{2,P}}{d\tau^{2}} + \omega_{P}^{2}(\tilde{\tau}) \chi_{2,P} = 0\,,
\end{equation}
where 
\begin{align}
    \omega_{P}(\tilde{\tau}) &= \sqrt{[\omega_{Q}(\tilde{\tau})]^{2} + [\omega_{R}(\tilde{\tau})]^{2}}\,,
\end{align}
The solution is then
\begin{equation}
    \label{eq:chiP-sol}
    \chi_{2,P} = \chi_{P,0} \cos[\gamma_{P}(\tau)] + \dot{\chi}_{P,0} \sin[\gamma_{P}(\tau)]\,,
\end{equation}
with $[\chi_{P,0}, \dot{\chi}_{P,0}]$ set by initial conditions, and the \textit{spin angle} $\gamma_{P}$ is defined through 
\begin{equation}
    \label{eq:gamma_P}
    \frac{\partial\gamma_{P}}{\partial\tau} = \omega_{P}(\tilde{\tau})\,.
\end{equation}
Much like the precession angle $\alpha$, $\gamma_{P}$ must be integrated under the effect of radiation reaction. In general, $[\chi_{P,0}, \dot{\chi}_{P,0}]$ are functions of the long time variable $\tilde{\tau}$. However, at the order in $q$ we are working, this dependence is irrelevant and they may be taken to be constants. 

The remaining two components satisfy equations that depend on $v$, even after they are transformed to $\gamma_{P}$ being the dependent variable, i.e.
\begin{align}
    \label{eq:chi2Q-eqn}
    \frac{d\chi_{2,Q}}{d\gamma_{P}} &= -\frac{\omega_{Q}(v)}{\omega_{P}(v)} \chi_{2,P}(\gamma_{P})\,,
    \\
    \label{eq:chi2J-eqn}
    \frac{d\chi_{2,J}}{d\gamma_{P}} &= - \frac{\omega_{R}(v)}{\omega_{P}(v)} \chi_{2,P}(\gamma_{P})\,.
\end{align}
To solve these exactly, one would have to solve integrals of the form
\begin{align}
    \chi_{2,J} \sim \int d\tau \; \omega_{R}[v(\tau)] e^{i\gamma_{P}(\tau)}
\end{align}
which are non-trivial. These can be evaluated by realizing that $\omega_{P}$ in Eq.~\eqref{eq:gamma_P} is always positive definite, and thus, there are no critical points of the phase $\gamma_{P}$. Then, the integral can be approximated by Laplace's method via repeated integration by parts, the first application of which produces a correction of order
\begin{equation}
    \frac{1}{\omega_{P}} \frac{dv}{d\tau} \frac{\partial \omega_{SL}}{\partial v} \sim {\cal{O}}(v^{8})
\end{equation}
while the leading order term is of order $\omega_{R}/\omega_{P} \sim 1 + {\cal{O}}(v)$. Thus, $\chi_{2,J}$ can be solved to high precision by assuming $\nu_{R} = \omega_{R}/\omega_{P}$ does not evolve appreciably on the radiation reaction timescale, with the result being
\begin{align}
    \label{eq:chiJ-sol}
    \chi_{2,J} &= \chi_{J,0} + \nu_{R} \left[-\chi_{P,0}\sin\gamma_{P} + \dot{\chi}_{P,0}\cos\gamma_{P} \right] + {\cal{O}}(v^{8})\,,
\end{align}
where $\chi_{J,0}$ is an integration constant. Likewise,
\begin{equation}
    \label{eq:chiQ-sol}
    \chi_{2,Q} = \chi_{Q,0} + \nu_{Q} \left[\chi_{P,0} \sin\gamma_{P} - \dot{\chi}_{P,0} \cos\gamma_{P}\right]+ {\cal{O}}(v^{7})\,,
\end{equation}
with $\nu_{Q} = \omega_{Q}/\omega_{P}$, and the components of the secondary's spin in the co-precessing frame are now solved. 

As a final point before proceeding, the solutions in Eqs.~\eqref{eq:chiP-sol} \&~\eqref{eq:chiJ-sol}-\eqref{eq:chiQ-sol} possess four integration constants, but $\vec{\chi}_{2}$ has only three components. One of these is redundant, which can be discovered by inserting these solutions and Eq.~\eqref{eq:chi2-co-p} into the original precession equations in Eq.~\eqref{eq:chi2-msa}. This is only satisfied if $\chi_{J,0} = \omega_{Q} \chi_{Q,0}/\omega_{R}$, thus eliminating the anomalous degree of freedom.

\subsection{$\hat{L}$ at ${\cal{O}}(q)$}
\label{sec:L1}

We are now left with solving Eq.~\eqref{eq:L1-msa}, which must simultaneously give us $\hat{L}^{(1)}(\tau)$ and $[X(\tilde{\tau}), Y(\tilde{\tau}), Z(\tilde{\tau})]$. Rearranging terms, we may write
\begin{align}
    \label{eq:L1-new}
    \frac{\partial \hat{L}^{(1)}}{\partial\tau} - \omega_{L} \hat{e}_{z} \times \hat{L}^{(1)} &= -\frac{\partial \hat{L}^{(0)}}{\partial \tilde{\tau}} - \omega_{L}^{(1)} \hat{e}_{z} \times \hat{L}^{(0)}, 
    \nn \\
    &+ v \omega_{SL} \vec{\chi}^{(0)}_{2} \times \hat{L}^{(0)}\,.
\end{align}
The left hand side above indicates that $\hat{L}^{(1)}$ will oscillate with phase variable $\alpha$ on the short timescale. The artificial resonance arises due to the right hand side of the above equation containing two types of terms, specifically those that oscillate solely with $\alpha$, and those that couple both $\alpha$ and $\gamma_{P}$. The former of these are what cause the artificial resonance, and can be seen if one re-writes the source term above, which only depends on $\hat{L}^{(0)}$ and $\vec{\chi}_{2}^{(0)}$, into a harmonic decomposition. More specifically
\begin{equation}
    \frac{\partial \hat{L}^{(0)}}{\partial \tilde{\tau}} = \hat{C}_{1}(\tilde{\tau}) \cos\alpha + \hat{S}_{1}(\tilde{\tau}) \sin\alpha + Z'(\tilde{\tau}) \hat{e}_{z}\,,
\end{equation}
with
\begin{align}
    \hat{C}_{1}(\tilde{\tau}) &= [X'(\tilde{\tau}), - Y'(\tilde{\tau}), 0]\,, \nn \\ \hat{S}_{1}(\tilde{\tau}) &= [Y'(\tilde{\tau}), X'(\tilde{\tau}), 0]
\end{align}
where $'$ corresponds to differentiation with respect to $\tilde{\tau}$. Further,
\begin{align}
    \hat{e}_{z} \times \hat{L}^{(0)} &= \hat{C}_{2}(\tilde{\tau}) \cos\alpha + \hat{S}_{2}(\tilde{\tau}) \sin\alpha\,,
    \\
    \hat{C}_{2}(\tilde{\tau}) &= [Y(\tilde{\tau}), X(\tilde{\tau}), 0]\,, \nn \\
    \hat{S}_{2}(\tilde{\tau}) &= [-X(\tilde{\tau}), Y(\tilde{\tau}), 0]\,.
\end{align}
Lastly, we may write
\begin{align}
    \label{eq:s2crossL0}
    \vec{\chi}_{2}^{(0)} \times \hat{L}^{(0)} &= \sum_{j,k=-1}^{1} \hat{E}_{jk}(\tilde{\tau}) e^{i(j\alpha + k \gamma_{P})}
    \nn \\
    &= \hat{C}_{3}(\tilde{\tau}) \cos\alpha + \hat{S}_{3}(\tilde{\tau}) \sin \alpha 
    \nn \\
    &+ \sum_{j=-1}^{1} \sum_{k\neq0} \hat{E}_{jk}(\tilde{\tau}) e^{i(j\alpha + k\gamma_{P})}
\end{align}
where
\begin{align}
    \hat{C}_{3}(\tilde{\tau}) &= \chi_{Q,0} {\cal{F}}(\tilde{\tau}) \left[Y(\tilde{\tau}), X(\tilde{\tau}), 0 \right]\,,
    \\
    \hat{S}_{3}(\tilde{\tau}) &= \chi_{Q,0} {\cal{F}}(\tilde{\tau}) \left[-X(\tilde{\tau}), Y(\tilde{\tau}), 0 \right]\,,
    \\
    {\cal{F}}(\tilde{\tau}) &= \frac{\omega_{Q}(\tilde{\tau})}{\omega_{R}(\tilde{\tau})} - \frac{Z(\tilde{\tau})}{\sqrt{[X(\tilde{\tau})]^{2} + [Y(\tilde{\tau})]^{2}}}\,.
\end{align}
The $[\hat{C}_{j}(\tilde{\tau}), \hat{S}_{j}(\tilde{\tau}), Z'(\tilde{\tau})]$ produce the artificial resonance and, thus, the sum total of them must vanish in order to remove this. Thus, the equations that $[X(\tilde{\tau}), Y(\tilde{\tau}), Z(\tilde{\tau})]$ must satisfy are
\begin{align}
    \label{eq:Z-eqn}
    Z'(\tilde{\tau}) &= 0\,,
    \\
    \label{eq:X-eqn}
    X'(\tilde{\tau}) &= \omega_{XY}(\tilde{\tau}) Y(\tilde{\tau})\,,
    \\
    \label{eq:Y-eqn}
    Y'(\tilde{\tau}) &= -\omega_{XY}(\tilde{\tau}) X(\tilde{\tau})\,,
\end{align}
with
\begin{equation}
    \omega_{XY}(\tilde{\tau}) =  -  \omega_{L}^{(1)} + v \chi_{Q,0} \frac{\omega_{SL}}{\omega_{R}(\tilde{\tau})} (\omega_{SJ} - \omega_{L})\,.
\end{equation}
Eq.~\eqref{eq:Z-eqn} implies that $Z$ is a constant, and thus, $[\omega_{Q},\omega_{R}, \omega_{XY}]$ are also constant as a result of the normalization condition in Eq.~\eqref{eq:XYZ-norm}. As such, we will drop the dependence on $\tilde{\tau}$ from the quantities. Choosing $Z = \hat{L}_{z,0}$, the solutions for $X(\tilde{\tau})$ and $Y(\tilde{\tau})$ are
\begin{align}
    \label{eq:X-sol}
    X(\tilde{\tau}) &= \hat{L}_{x,0} \cos[\lambda(\tilde{\tau})] - \hat{L}_{y,0} \sin[\lambda(\tilde{\tau})]\,,
    \\
    \label{eq:Y-sol}
    Y(\tilde{\tau}) &= -\hat{L}_{y,0} \cos[\lambda(\tilde{\tau})] - \hat{L}_{x,0} \sin[\lambda(\tilde{\tau})]\,.
\end{align}
where $\lambda(\tilde{\tau})$, which we call the \textit{renormalization angle}, is defined by
\begin{equation}
    \label{eq:lambda-eq}
    \frac{d\lambda}{d\tilde{\tau}} = \omega_{XY}\,, 
\end{equation}
and $[\hat{L}_{x,0},\hat{L}_{y,0},\hat{L}_{z,0}]$ are constants set by initial data. The function $\hat{L}^{(0)}(\tau,\tilde{\tau})$ is now completely determined and we have removed the artificial resonance in Eq.~\eqref{eq:L1-msa}.

We are now left with the task of determining $\hat{L}^{(1)}$, or more specifically its dependence on the short timescale $\tau$. The only remaining source on the right hand side of Eq.~\eqref{eq:L1-new} is the double sum in Eq.~\eqref{eq:s2crossL0}, i.e.
\begin{equation}
    \label{eq:L1-prec}
    \frac{\partial \hat{L}^{(1)}}{\partial \tau} - \omega_{L} \hat{e}_{z} \times \hat{L}^{(1)} = v \omega_{SL}\hat{\cal{S}}(\tau,\tilde{\tau})
\end{equation}
with
\begin{equation}
    \hat{\cal{S}}(\tau,\tilde{\tau}) = \sum_{j}\sum_{k\neq 0} \hat{E}_{jk}(\tilde{\tau}) e^{i[j\alpha(\tau) + k \gamma_{p}(\tau)]}\,.
\end{equation}
It is simplest to consider each component of this separately. The z-component of the above equation decouples, and after integrating, produces
\begin{equation}
    \hat{L}^{(1)}_{z} = \frac{v \omega_{SL}}{\omega_{P}} \sqrt{1-\hat{L}_{z,0}^{2}} \left[\chi_{P,0} \sin\gamma_{P} + \dot{\chi}_{P,0} \left(1 - \cos\gamma_{P}\right) \right]\,.
\end{equation}
The $x-$ and $y-$ components are coupled in the same way as in Sec.~\ref{sec:L0}. Decoupling them produces
\begin{align}
    \frac{\partial^{2} \hat{L}_{x,y}^{(1)}}{\partial \tau^{2}} + \omega_{L}^{2} \hat{L}_{x,y}^{(1)} &= v\omega_{SL} {\cal{T}}_{x,y}(\tau,\tilde{\tau}) \label{L1xy}
\end{align}
where
\begin{align}
    {\cal{T}}_{x}(\tau,\tilde{\tau}) &= \frac{\partial}{\partial\tau} \hat{\cal{S}}_{x}(\tau,\tilde{\tau}) - \omega_{L} \hat{\cal{S}}_{y}(\tau,\tilde{\tau})
    \\
    {\cal{T}}_{y}(\tau,\tilde{\tau}) &= \frac{\partial}{\partial \tau} \hat{\cal{S}}_{y}(\tau,\tilde{\tau}) + \omega_{L} \hat{\cal{S}}_{x}(\tau,\tilde{\tau})\,,
\end{align}
The sources ${\cal{T}}_{x,y}$ can be written in harmonics of $\alpha$ and $\gamma_{P}$, specifically
\begin{align}
    {\cal{T}}_{x,y} &= \Omega_{+} \left\{C_{x,y}^{(+)}(\tilde{\tau}) \cos[\delta_{+}(\tau)] + S_{x,y}^{(+)}(\tilde{\tau}) \sin[\delta_{+}(\tau)] \right\}
    \nn \\
    &+ \Omega_{-} \left\{C_{x,y}^{(-)}(\tilde{\tau}) \cos[\delta_{-}(\tau)] + S_{x,y}^{(-)}(\tilde{\tau}) \sin[\delta_{-}(\tau)]\right\}\,,
\end{align}
with $\Omega_{\pm} = 2\omega_{L} \pm \omega_{P}$, $\delta_{\pm} = \alpha \pm \gamma_{P}$, and the functions
\begin{align}
    \label{eq:CS-coeffs-1}
    C_{x}^{(\pm)}(\tilde{\tau}) = S_{y}^{(\pm)}(\tilde{\tau}) &= A_{\pm} \left[-\chi_{P,0} Y(\tilde{\tau}) \mp \dot{\chi}_{P,0} X(\tilde{\tau})\right]\,,
    \\
    \label{eq:CS-coeffs-2}
    S_{x}^{(\pm)}(\tilde{\tau}) = -C_{y}^{(\pm)}(\tilde{\tau}) &= A_{\pm} \left[\pm \dot{\chi}_{P,0} Y(\tilde{\tau}) - \chi_{P,0} X(\tilde{\tau}) \right]\,,
\end{align}
where
\begin{equation}
    A_{\pm} = \frac{L_{z,0} (\omega_{P} \pm \omega_{Q}) \pm \omega_{SL} \mp L_{z,0}^{2} \omega_{SL}}{2 \omega_{P} \sqrt{1 - L_{z,0}^{2}}}
\end{equation}
are constants on the precession timescale. It is straightforward to show that the solutions to Eq.~\eqref{L1xy} are
\begin{align}
    \label{eq:Lxy-1-sol}
    \hat{L}_{x,y}^{(1)}(\tau,\tilde{\tau}) &= \ell_{x,y} \cos[\alpha(\tau)] + \dot{\ell}_{x,y} \sin[\alpha(\tau)]
    \nn \\
    &+ \frac{v \omega_{SL}}{\omega_{P}} \left\{C_{x,y}^{(-)}(\tilde{\tau}) \cos[\delta_{-}(\tau)] + S_{x,y}^{(-)}(\tilde{\tau}) \sin[\delta_{-}(\tau)]\right\} 
    \nn \\
    &- \frac{v \omega_{SL}}{\omega_{P}} \left\{C_{x,y}^{(+)}(\tilde{\tau}) \cos[\delta_{+}(\tau)] + S_{x,y}^{(+)}(\tilde{\tau}) \sin[\delta_{+}(\tau)] \right\}
\end{align}
where $[\ell_{x,y},\dot{\ell}_{x,y}]$ are constants set by initial conditions. In general, these will be functions of the long timescale $\tilde{\tau}$, which would then be determined by proceeding to next order in our MSA. Since we are truncating our analysis at linear order $q$, we take these to be constants instead.

At linear order in $q$, the initial conditions for $\hat{L}$ are
\begin{equation}
    \lim_{t \rightarrow 0} \hat{L}^{(1)} = [0,0,0]\,, \qquad \lim_{t \rightarrow 0} \frac{d\hat{L}^{(1)}}{dt} = \lim_{t\rightarrow 0} \hat{\cal{S}}(\tau,\tilde{\tau}) \equiv \hat{\cal{S}}(0)\,,
\end{equation}
where the former comes from the standard perturbative description for specifying inital data, while the latter is enforced from the precession equations, more specifically Eq.~\eqref{eq:L1-prec}. Enforcing this, the constants $[\ell_{x,y},\dot{\ell}_{x,y}]$ are
\begin{align}
    \ell_{x,y} &=  \frac{v \omega_{SL}}{\omega_{P}} \left[C^{(+)}_{x,y}(0) - C^{(-)}_{x,y}(0) \right]\,,
    \\
    \dot{\ell}_{x,y} &= \frac{{\cal{S}}_{x,y}(0)}{\omega_{L}} + \frac{v \omega_{SL}}{\omega_{L} \omega_{P}} \left[(\omega_{L} + \omega_{P}) S^{(+)}_{x,y}(0) 
    \right.
    \nn \\
    &\left.
    - (\omega_{L} - \omega_{P}) S^{(-)}_{x,y}(0)  \right] \,.
\end{align}
This completes the solution to the precession equations to the desired order in $q$.

\subsection{Numerical Comparison}

Having obtained the solution to the PN precession equations in the EMRI limit through MSA, we test the accuracy of the solutions by comparing to numerical evolutions of Eqs.~\eqref{eq:prec-1-emri}-\eqref{eq:OmegaL1}. Since we only desire to understand the accuracy of the precession solutions, we do not consider the effect of radiation reaction at this point, which is instead discussed in Sec.~\ref{sec:rr}.

\begin{figure*}[hbt!]
    \centering
    \includegraphics[width=\textwidth]{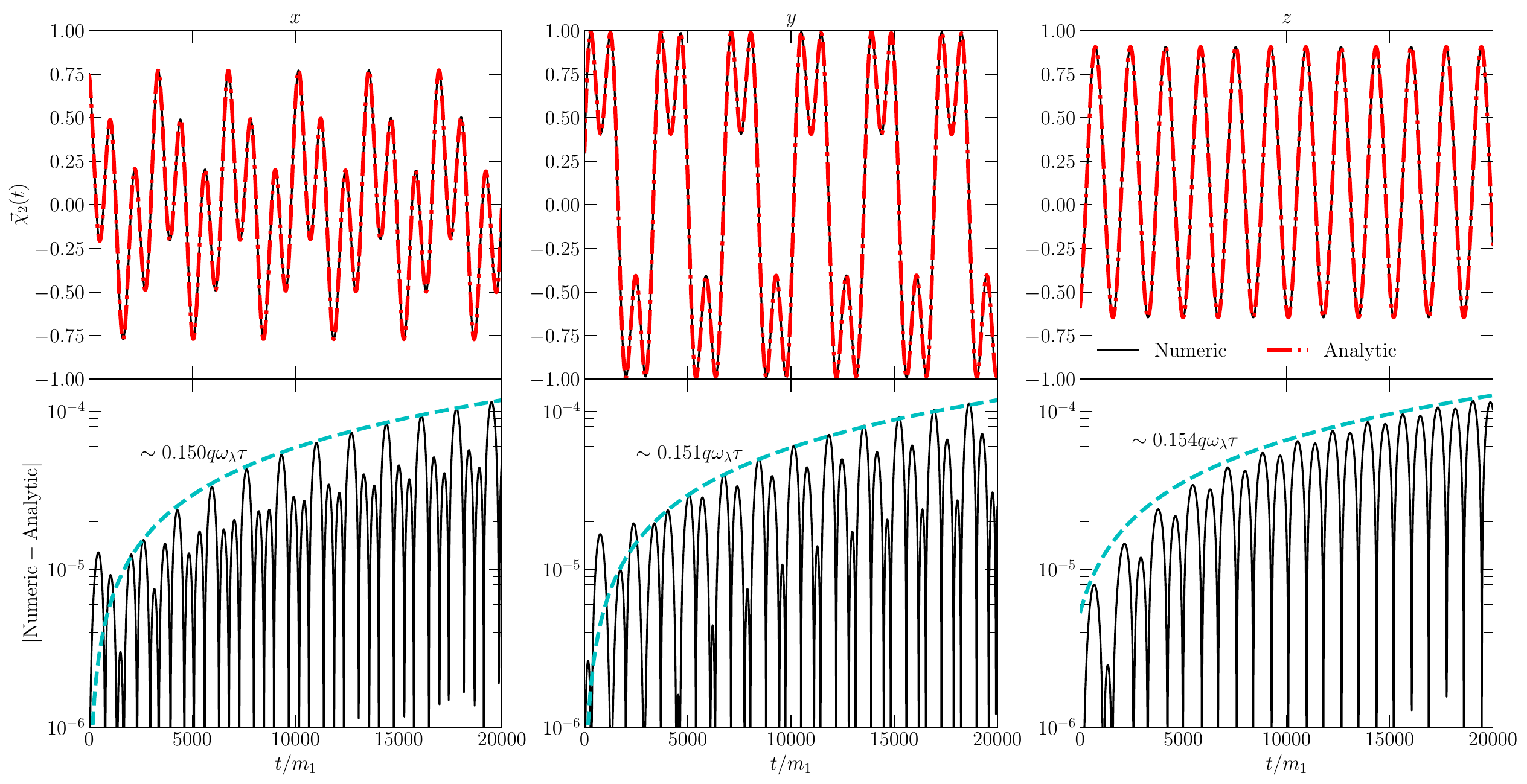}
    \caption{Top: Comparison between numerical evolution (black line) of $\chi_{2}$ and the analytic MSA solution (red dot-dashed) at order ${\cal{O}}(q^{0})$ for an EMRI with $q=10^{-5}$, $\beta_{0}=1.04$, $\chi_{\rm eff} = 0.5$, $\chi_{1}=0.99$, and $v=0.3193$. Bottom: Difference between the numerical and analytic solutions. The solutions generally de-phase due to undetermined remainders of ${\cal{O}}(q)$ in the MSA. The difference is bounded by the cyan dashed curves, which are directly proportional to the phase $q\omega_{\lambda}t \sim \lambda(t)$.}
    \label{fig:chi2-comp}
\end{figure*}

\begin{figure*}[hbt!]
    \centering
    \includegraphics[width=\textwidth]{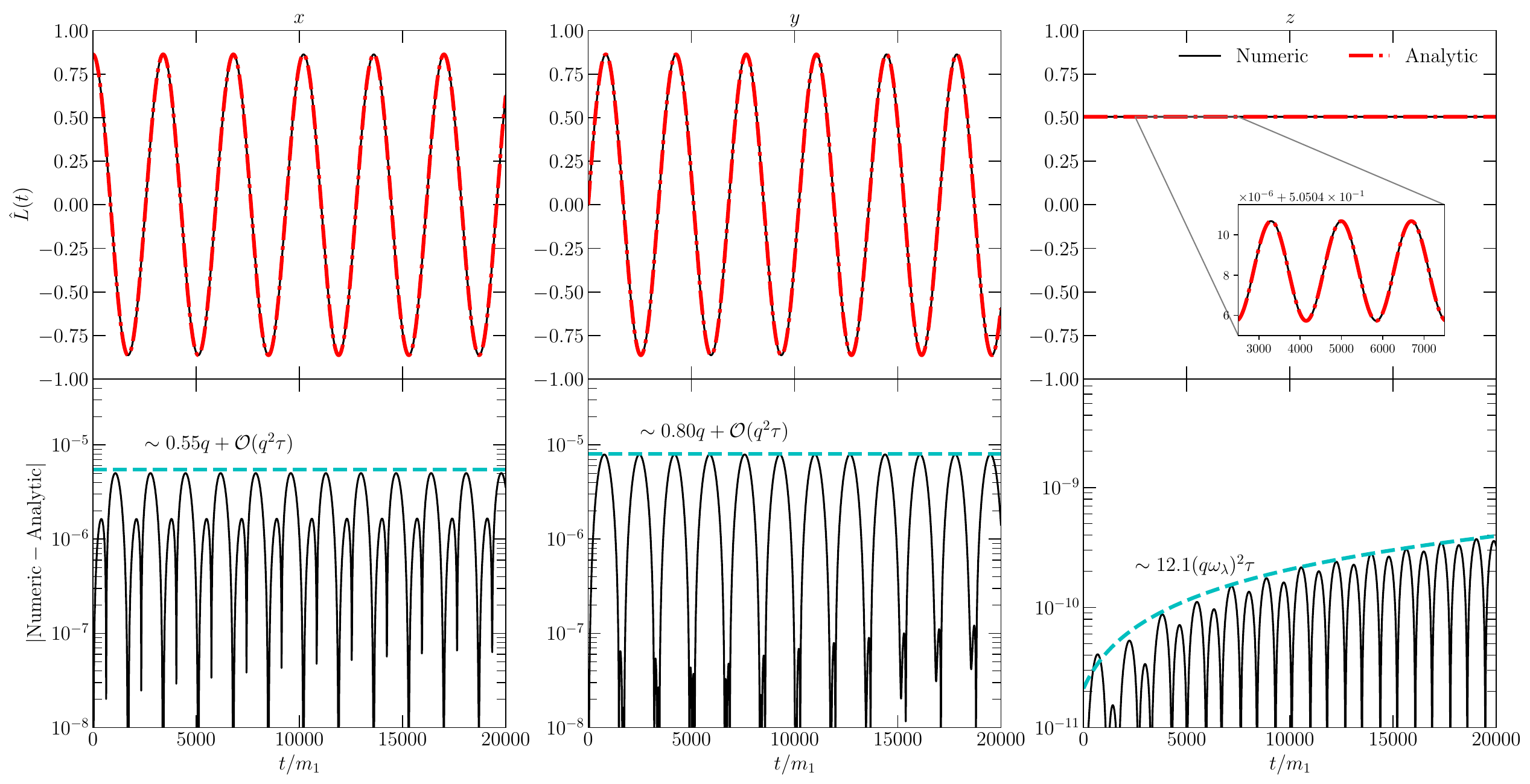}
    \caption{Top: Comparison between the numerical evolution (black solid line) of $\hat{L}$ and the analytic MSA solution (red dot-dashed line) at ${\cal{O}}(q)$ for an EMRI with the same parameters as Fig.~\ref{fig:chi2-comp}. The inset in the right plot shows a zoom of the behavior of $\hat{L}_{z}$, due to the fact that the nutational motion is suppressed by $q$. Bottom: Difference between the numerical and analytic solutions. The solutions generally dephase due to uncontrolled remainders in the MSA. For the $x-$ and $y-$components, the difference is bounded by the cyan dashed curves, which is directly proportional to the mass ratio $q$. The reason for this is that the constants of integration $\ell_{x,y}$ and $\dot{\ell}_{x,y}$ in Eq.~\eqref{eq:Lxy-1-sol} are technically functions of $\tilde{\tau}$ in the MSA. However, these can only obtained by going to higher order, whereas we have stopped at first order in $q$, leaving them as undetermined constants. For the $z-$component, the difference is primarily controlled by the undetermined remainder of ${\cal{O}}(q^{2})$.} 
    \label{fig:L-comp}
\end{figure*}

The numerical evolutions are performed in \texttt{Mathematica} with the \texttt{NDSolve} method. We use the \texttt{ImplicitRungeKutta} method with accuracy and precision tolerances set to $10^{-13}$. We consider an EMRI with $q=10^{-5}$, $\chi_{1} = 0.99$, $\chi_{2}=1$, and initial $\hat{L} = [\sin\beta_{0}, 0, \cos\beta_{0}]$ with $\beta_{0} = 1.04$. As a result of these choices, $\chi_{\rm eff} = 0.5$. We evolve the EMRI to $t_{f} = 10^{5} m_{1}$, which is long enough for us to estimate the error in the MSA. For this EMRI, the top panels in Figs.~\ref{fig:chi2-comp} \&~\ref{fig:L-comp} provide a comparison of the numeric and analytic solutions of the components of $\vec{\chi}_{2}$ and $\hat{L}$, respectively.

Generally, the analytic solutions dephase compared to the numerical evolutions, due to the latter only being accurate to finite order in the MSA, specifically ${\cal{O}}(q)$ for $\vec{\chi}_{2}$ and ${\cal{O}}(q^{2})$ for $\hat{L}$. To provide a complete estimate of how large the error can become throughout a complete EMRI coalescence, we empirically estimate the bounded curve as
\begin{equation}
    \label{eq:bound}
    \max | \hat{L}_{\rm num} - \hat{L}_{\rm an} | \le C q^{n} \tau\,,
\end{equation}
where $C$ is a number to be determined. Since we are neglecting radiation reaction for this comparison, the difference will generally grow linearly in $\tau=t/m_{1}$. However, this growth will change when radiation reaction is included, becoming more rapid closer to merger. To account for this, we recast Eq.~\eqref{eq:bound} in the form
\begin{equation}
    \max | \hat{L}_{\rm num} - \hat{L}_{\rm an} | \le C (q \omega_{\lambda})^{n} \tau \,,
\end{equation}
which is due to the fact that $\lambda$ is a proxy for the long timescale $\tilde{\tau} = q \tau$ through Eq.~\eqref{eq:lambda-eq}.

To find the value of $C$, we take each component of $\vec{\chi}_{2}$ and $\hat{L}$, and compute the difference between the MSA and numerical evolution, which are shown in the bottom panels of Figs.~\ref{fig:chi2-comp} \&~\ref{fig:L-comp}. We then find the points corresponding to the maxima of the oscillations, and use the \texttt{polyfit} function of \texttt{numpy} in \texttt{Python} to find $C$. The resulting curves are specificed by dot-dashed lines in the bottom panels of Fig.~\ref{fig:chi2-comp}-\ref{fig:L-comp}. The components of $\hat{L}$ act as outliers in this analysis. We find the bounding curves of the $x-$ and $y-$components to be approximately constants proportional to the mass ratio $q$ to high accuracy. The reason for this behavior arises from the constants $\ell_{x,y}$ and $\dot{\ell}_{x,y}$, which should generically be functions of the long time-scale $\tilde{\tau}$. However, to completely obtain these, one would have to proceed to higher order than we currrently desire in the MSA to completely fix these functions. The near-constant error in these two components is then a result of truncating the MSA at the desired order. On the other hand, the error in the $z-$components scales with $\omega_{\lambda}^{2}$ due to a remainder of ${\cal{O}}(q^{2})$. As a result, the error on $\hat{L}_{z}$ is better controlled than the other components of $\hat{L}$. Note that the analytic solutions can be improved by proceeding to higher order in the MSA, but this goes outside the scope of this study.

%

\section{Radiation Reaction \& Precession Phase}
\label{sec:rr}

The precession solutions of the previous section provide a solution to the binary dynamics on the precession timescale. To include radiation reaction within these solutions, we invoke the use of MSA again, now relying on the fact that the radiation reaction timescale is longer than the precession timescale. This can be seen readily from the definitions of the precession timescale $T_{\rm pr} = 1/\omega_{L} \sim v^{-6}$ and radiation reaction timescale $T_{\rm rr} = 1/\dot{v} \sim v^{-9}$, and thus $T_{\rm pr}/T_{\rm rr} \sim v^{3}$. As a result, the evolution of the binary on the precession timescale found in the previous section still holds, but all precessional constants (such as $J$, $\omega_{L}$, etc.) become time dependent. Meanwhile, the evolution of the binary on the radiation reaction timescale will be given to leading order by the precession average of relevant quantities containing $[\hat{L}, \vec{\chi}_{2}]$.

At this stage, it is worth pointing out that the quantities $[\chi_{Q,0}, \chi_{P,0}, \dot{\chi}_{P,0}]$ are not well behaved in the aligned limit. This can be seen by taking our analytic solutions for the components of $\vec{\chi}_{2}$ and taking the limit $t\rightarrow0$ in a PN expansion. In this limit, we may solve for the values of these parameters in terms of the more reasonable initial components of $\vec{\chi}_{2}$ in a (non-precessing) Cartesian reference frame. For each component, to leading PN order we have
\begin{align}
    \chi_{Q,0} &= \sqrt{1-L_{z,0}^{2}} \left(\hat{L}_{x,0} \chi_{2,0}^{x} + \hat{L}_{y,0} \chi_{2,0}^{y} + \hat{L}_{z,0} \chi_{2,0}^{z} \right) + {\cal{O}}(v)\,,
    \\
    \chi_{P,0} &= \frac{\hat{L}_{y,0} \chi_{2,0}^{x} - \hat{L}_{x,0} \chi_{2,0}^{y}}{\sqrt{1 - L_{z,0}^{2}}} + {\cal{O}}(v)\,,
    \\
    \dot{\chi}_{P,0} &= \frac{\chi_{2,0}^{z}(1 - \hat{L}_{z,0}^{2}) - \hat{L}_{z,0} (\hat{L}_{x,0} \chi_{2,0}^{x} + \hat{L}_{y,0} \chi_{2,0}^{y})}{\sqrt{1 - \hat{L}_{z,0^{2}}}} + {\cal{O}}(v)\,,
\end{align}
where $\chi_{2,0}^{x,y,z}$ are the initial values of the secondary spin in a non-precessing reference frame, and are regular for aligned EMRIs. One can see from the above expressions that one of these components will vanish when $\hat{L}_{z,0}\rightarrow1$, while the other two will diverge. In order to ensure regularity of all quantities that enter the waveform, we define parameters
\begin{align}
    \bar{\chi}_{Q,0} &= \frac{\chi_{Q,0}}{\sqrt{1-\hat{L}_{z,0}^{2}}}\,, 
    \\
    [\bar{\chi}_{P,0}, \dot{\bar{\chi}}_{P,0}] &= \sqrt{1 - \hat{L}_{z,0}^{2}} [\chi_{P,0}, \dot{\chi}_{P,0}]\,,
\end{align}
which are also regular in the aligned limit.

\subsection{Precession Averages}
\label{sec:avg}

To begin, we provide explicit precession averages of necessary quantities. More explicitly, within the radiation reaction equation for $\dot{v}$, a term $(\hat{L}\cdot \vec{S}_{1,2})$ will appear at 1.5PN order, whereas a term $(\vec{S}_{1}\cdot \vec{S}_{2})$ at 2PN order. While we did not explicitly solve for the evolution of $\vec{S}_{1}$ from the precession equations, the conservation of $\vec{J}$ on the precession timescale ensures that $\vec{S}_{1} = \vec{J} - \vec{L} - \vec{S}_{2}$. For our computations, the two averages of relevance are $(\hat{L}\cdot \vec{S}_{2})$ and $(\vec{S}_{1} \cdot \vec{S}_{2})$, due to the fact that any instances of $(\hat{L}\cdot \vec{S}_{1})$ can be substituted with $\chi_{\rm eff}$ in Eq.~\eqref{eq:chi-eff}.

The two averages of relevance will both be suppressed by $q^{2}$, due to their dependence on the secondary spin. At ${\cal{O}}(q^{2})$, both are independent of the evolution of $\hat{L}$, only being time dependent through the phase $\gamma_{P}(t)$. Hence, at leading order, the averages only need to be performed with respect to $\gamma_{P}$, specifically
\begin{equation}
    \langle f(\gamma_{P}) \rangle_{\gamma} = \frac{1}{2\pi} \int_{-\pi}^{\pi} d\gamma_{P} f(\gamma_{P})\ .
\end{equation}
Inserting the precession solutions and taking the average reveals
\begin{align}
    \label{eq:LdotS2-avg}
    \langle \hat{L} \cdot \vec{S}_{2} \rangle_{\gamma} &= \frac{q^{2} m_{1}^{2} \chi_{Q,0}}{\omega_{R}} \left(\omega_{Q} L_{z,0} + \omega_{R} \sqrt{1-L_{z,0}^{2}} \right) + {\cal{O}}(q^{3})\,,
    \\
    \label{eq:S1dotS2-avg}
    \langle \vec{S}_{1} \cdot \vec{S}_{2} \rangle_{\gamma} &= q^{2} m_{1}^{2} J \chi_{Q,0} \frac{\omega_{Q}}{\omega_{R}} + {\cal{O}}(q^{3})\,.
\end{align}
It is worth noting that the above averages only depend on one component of the secondary spin, specifically $\chi_{J,0}$. 
The remaining two components are contained in the oscillatory corrections to these, and will be necessary to ensure that the various phases appearing in the waveform will contain all of the components of $\vec{\chi}_{2}$. Thus, we write
\begin{align}
    \left(\hat{L} \cdot \vec{S}_{2}\right) &= \langle \hat{L} \cdot \vec{S}_{2} \rangle_{\gamma} + q^{2} m_{1}^{2} {\cal{D}}_{L,2} \Delta \chi(\tau) + {\cal{O}}(q^{3})\,,
    \\
    \left(\vec{S}_{1} \cdot \vec{S}_{2} \right) &= \langle \vec{S}_{1} \cdot \vec{S}_{2} \rangle_{\gamma} + q^{2} m_{1}^{4} {\cal{D}}_{1,2} \Delta\chi(\tau) + {\cal{O}}(q^{3})
\end{align}
with
\begin{align}
    \Delta \chi(\tau) &= \dot{\bar{\chi}}_{P,0} \cos[\gamma_{P}(\tau)] - \bar{\chi}_{P,0} \sin[\gamma_{P}(\tau)]\,,
    \\
    \label{eq:DL2-def}
    {\cal{D}}_{L,2} &=  L_{z,0} \frac{\omega_{SL}}{\omega_{P}} - \frac{\omega_{Q}}{\omega_{P}} \,,
    \\
    \label{eq:D12-def}
    {\cal{D}}_{1,2} &= \frac{J \omega_{R}}{m_{1}^{2}\omega_{P} \sqrt{1 - L_{z,0}^{2}}} \,.
\end{align}
%

\subsection{Evolution of $J$}
\label{sec:j-evo}

Within the PN two-body problem, the evolution of the total angular momentum obeys at leading PN order
\begin{equation}
    \label{eq:dJdL}
    \frac{dJ}{dL} = \frac{J^{2} + L^{2} - S^{2}}{2JL}\,,
\end{equation}
where we recall that $S^{2}$ is the magnitude of the total spin vector $\vec{S} = \vec{S}_{1} + \vec{S}_{2}$. From the results of Sec.~\ref{sec:avg}, the precession average can be readily taken, obtaining
\begin{equation}
    \label{eq:djdh}
    \bigg\langle\frac{dj}{dh}\bigg\rangle_{\gamma} = \frac{j^{2} + h^{2} - \chi_{1}^{2}}{2jh} - q^{2} \frac{\chi_{Q,0} \omega_{Q}}{\omega_{R} h} + {\cal{O}}(q^{3})
\end{equation}
where we have normalized all angular momenta by $m_{1}^{2}$, specifically $J = j m_{1}^{2}$ and $L = h m_{1}^{2}$, with $h=q/v$. In the comparable mass case, the evolution for $J(L)$ can be directly solved in closed form after taking the precession average (see~\cite{Chatziioannou:2017tdw}). However, because of the presence of $J$ in Eq.~\eqref{eq:S1dotS2-avg}, Eq.~\eqref{eq:djdh} is more complicated and an exact closed-form expression form has not been found. Physically, this arises due to the fact that, in the comparable mass case, the orbital angular momentum constitutes the largest contribution to the total angular momentum budget of the binary. In the EMRI limit, it's the primary's spin that takes up this role, unless the primary is nearly non-spinning. 

Instead of a closed-form solution, one can seek a perturbative solution in $q$, with ansatz $j(h) = j_{0}(h) + q^{2} j_{2}(h) + {\cal{O}}(q^{3})$. The analytic solutions for $j_{0}$ and $j_{2}$ are
\begin{align}
    \label{eq:j0-exact}
    j_{0}(h) &= \sqrt{\chi_{1}^{2} + h^{2} + c_{1} h}\,,
    \\
    \label{eq:j2-exact}
    j_{2}(h) &= \frac{L_{z,0} \chi_{Q,0}}{\sqrt{1 - L_{z,0}^{2}}} + \frac{h [c_{2} + \chi_{J,0} \ell_{1}(h)]}{j_{0}(h)} 
    \nn \\
    &+ \frac{c_{1} h \chi_{J,0}[\ell_{+}(h) - \ell_{-}(h) - \ln\chi_{1}]}{2\chi_{1} j_{0}(h)}
\end{align}
where $[c_{1}, c_{2}]$ are integration constants that must be set by initial conditions, and 
\begin{align}
    \ell_{1} &= \ln\left[2j_{0}(h) - c_{1} - 2h \right]\,,
    \\
    \ell_{\pm} &= \ln\left[j_{0}(h) - h \pm \chi_{1} \right]\,.
\end{align}

The above solution is obtained without taking either a complete EMRI or PN expansion of the right hand side of Eq.~\eqref{eq:dJdL}, but is obtained by taking $h=q/v$ as an independent variable. In~\cite{Chatziioannou:2017tdw}, it was found that for near equal mass binaries, a naive PN expansion of expression similar to Eq.~\eqref{eq:j0-exact} resulted in a significant loss of accuracy compared to numerical integration of the PN precession and radiation reaction equations. The reason for this poor convergence results from the dependence of $h$ (or $L$) on the PN expansion variable $v$, which always enters such expressions coupled to $q$. Taking an expansion in $v$ then assumes particular limits of the ratio $q/v$, an assumption that can be broken as the binary inspirals. For the EMRIs under consideration, $q\sim10^{-6}-10^{-4}$, and thus the ratio $q/v$ only becomes of order unity when $q\sim v$, which will only occur for EMRIs so widely separated that their GW emission is typically outside of the detection band. As a result, we expand Eqs.~\eqref{eq:j0-exact}-\eqref{eq:j2-exact} in $q\ll1$ to obtain our final expression for $j(v)$,
\begin{align}
    \label{eq:j-emri}
    j(v) = \chi_{1} + \frac{q c_{1}}{2 v \chi_{1}} + q^{2} \left[L_{z,0}\bar{\chi}_{Q,0} - \frac{c_{1}^{2} - 4 \chi_{1}^{2}}{8 v^{2} \chi_{1}^{3}} \right] + {\cal{O}}(q^{3})\,.
\end{align}
Before continuing, it is worth noting that the above expression for $j(v)$ has a pole when $\chi_{1} = 0$. This is a result of the non-uniform nature of the $q\ll1$ expansion of Eq.~\eqref{eq:j0-exact}. If $\chi_{1} \sim q/v$, then the expansion used to obtain Eq.~\eqref{eq:j-emri} is no longer valid. This is not unexpected, since a similar problem arises in the comparable mass case~\cite{Chatziioannou:2017tdw}. For the EMRI system used in Figs.~\ref{fig:chi2-comp} \&~\ref{fig:L-comp}, $q/v \sim 3\times 10^{-5}$ and $\chi_{1} = 0.99$. Thus, in order for the expansion in Eq.~\eqref{eq:j-emri} to not be valid, the primary would effectively have to be non-spinning to reasonably high precision. Since we are concerned with EMRIs with a spinning primary, we use Eq.~\eqref{eq:j-emri} for the remainder of our analysis. If one desires to extend the analyses herein to slowly spinning primaries, one must instead use Eqs.~\eqref{eq:j0-exact}-\eqref{eq:j2-exact}, and take caution when expanding in either $q\ll1$ or $v \ll 1$.

\begin{figure}[hbt!]
    \centering
    \includegraphics[width=\columnwidth]{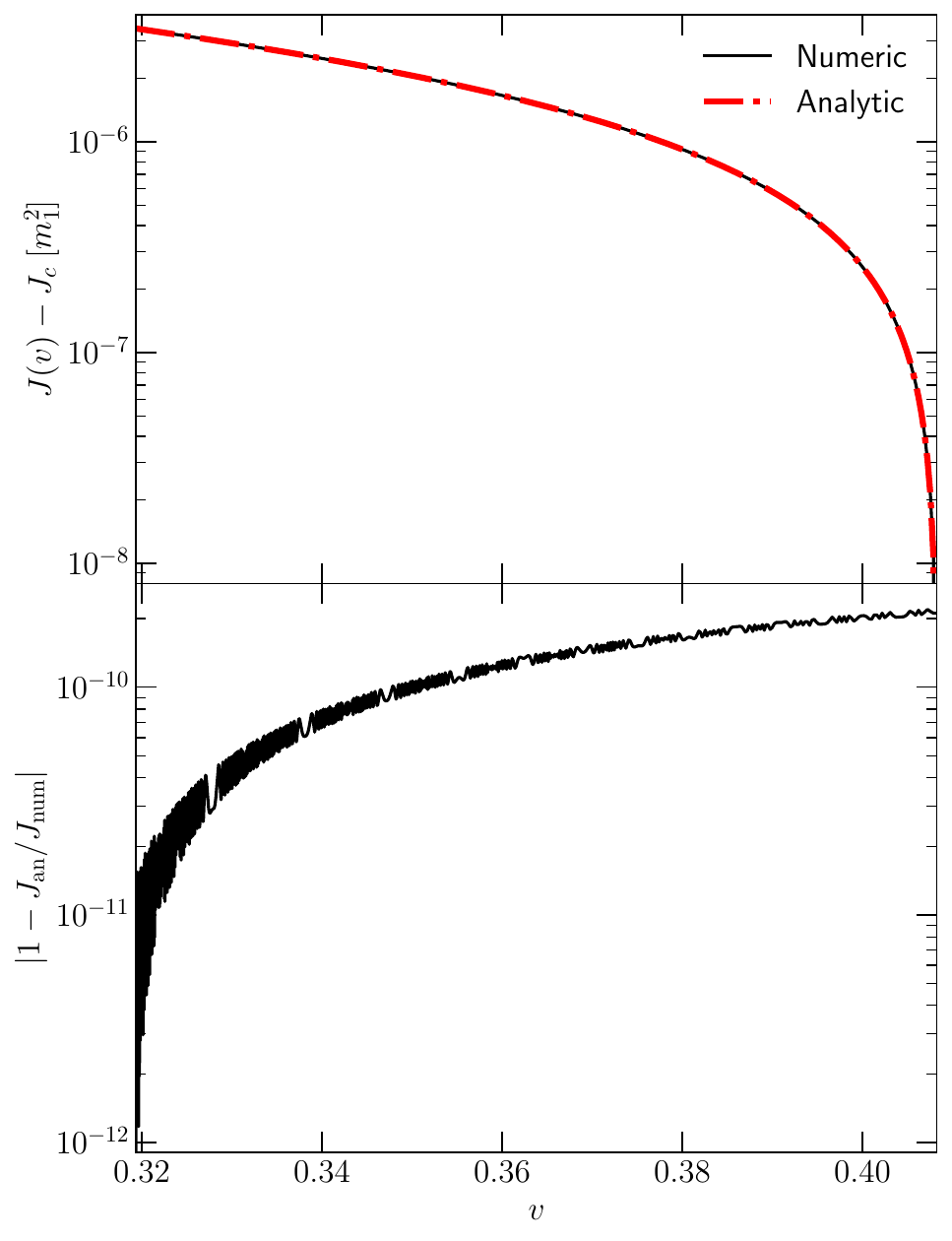}
    \caption{Top: Comparison of the analytic expression for the evolution of the total angular momentum $J$ under radiation reaction in Eq.~\eqref{eq:j-emri} (red dot-dashed line) to an exact solution obtained by numerically integrating the coupled PN precession and radiation reaction equations (black solid line). For an EMRI, $J$ does not vary significantly over the inspiral,  since the largest contribution to the total angular momentum is the primary's spin for this EMRI, and the effect of radiation reaction constitutes a first post-adiabatic (1PA) correction. As a result, the comparison is displayed relative to the value at the end of the numerical integration $J_{c}$. Bottom: Relative error between the numerical and analytic solutions for $J(v)$.}
    \label{fig:Jcomp}
\end{figure}

In Fig.~\ref{fig:Jcomp}, we provide an explicit comparison between the analytic solution $j(v)$ in Eq.~\eqref{eq:j-emri} 
against the numerical integration of the coupled PN precession and radiation reaction equations. Unlike the comparisons in Figs.~\ref{fig:chi2-comp}-\ref{fig:L-comp}, we don't restrict the comparison to a short time window, but allow the EMRI to full evolve up to the Schwarzschild innermost stable circular orbit (ISCO). We simply choose this as a convenient stopping point of the numerical integration due to the long computation time of EMRI inspirals. For the EMRI with the same physical parameters as Figs.~\ref{fig:chi2-comp}-\ref{fig:L-comp}, the numerical integration is performed up to time $\sim 1.45\times10^{7} m_{1}$, which corresponds to $\sim 2.31$ years for a $10^{6} M_{\odot}$ primary. The numerical integration is performed using \texttt{Mathematica}'s \texttt{NDSolve} function with the \texttt{ImplicitRungeKutta} method, and accuracy and precision goals set to $10^{-13}$. For this EMRI system, the total integration takes approximately seven minutes on current laptop processors without parallelization. On the other hand, the analytic approximations developed herein can be evaluated much more rapidly, highlighting one of the main strength's of our analytical model.

The top panel of Fig.~\ref{fig:Jcomp} provides a direct comparison of the numerical and analytic solutions as a function of the orbital velocity $v$. For the EMRI described above, the total angular momentum does not vary significantly over the inspiral, since the angular momentum budget is largely controlled by the primary. Specifically, $J$ only changes by approximately a few parts per million of its initial value, and thus the comparison in the top panel of Fig.~\ref{fig:Jcomp} is performed relative to the final value at the end of the coalescence, $J_{c}$. The bottom panel displays the relative error between the two solutions, highlight that the EMRI expansion used to obtain Eq.~\eqref{eq:j-emri} does not induce significant error.

\subsection{Expansion of $dv/dt$ \& TaylorF2 Approximants for Orbital Quantities}
\label{sec:F2-orb}

The relevant quantity where these averages will occur is the evolution of the orbital velocity, which in PN theory satisfies
\begin{equation}
    \frac{dv}{dt} = \frac{a_{0}(q)}{3M} v^{9} \left[1 + \sum_{n} a_{n}(q,S_{A}) v^{n} \right]\ ,
\end{equation}
where the $a_{n}$ coefficients up to 2PN order are
\begin{align}
    a_{0}(q) &= \frac{96\eta}{5}\,, \qquad a_{1}(q) = 0\,,
    \\
    a_{2}(q) &= -\frac{743}{336} - \frac{11}{4} \eta\,,
    \\
    a_{3}(q) &= 4\pi - \beta_{3}\,,
    \\
    a_{4}(q) &= \frac{34103}{18144} + \frac{13661}{2016}\eta + \frac{59}{18} \eta^{2} - \sigma_{4}\,.
\end{align}
The coefficients $(\beta_{3},\sigma_{4})$ are the 1.5PN spin-orbit and 2PN spin-spin corrections, specifically
\begin{align}
    \beta_{3} &= \frac{1}{M^{2}} \sum_{A} \left(\frac{113}{12} + \frac{25}{4} \frac{m_{B}}{m_{A}} \right) \left(\vec{S}_{A} \cdot \hat{L}\right)\,,
    \\
    \sigma_{4} &= \frac{1}{\mu M^{3}} \left[\frac{247}{48} \vec{S}_{1}\cdot\vec{S}_{2} - \frac{728}{48} \left(\vec{S}_{1}\cdot\hat{L}\right)\left(\vec{S}_{2}\cdot\hat{L}\right) \right]
    \nn \\
    &+ \sum_{A} \frac{1}{M^{2} m_{A}^{2}} \left[\frac{233}{96} S_{A}^{2} - \frac{719}{96} \left(\vec{S}_{A} \cdot \hat{L} \right)\right]\,.
\end{align}
To expand these quantities in the EMRI limit, we substitute all instances of $\vec{S}_{1}\cdot\hat{L}$ with $\chi_{\rm eff}$ in Eq.~\eqref{eq:chi-eff}, replace all remaining instances of dot products between angular momenta with their precession averages in Eqs.~\eqref{eq:LdotS2-avg}-\eqref{eq:S1dotS2-avg}, and finally series expand about $q\ll1$, followed by $v\ll1$. The end result is
\begin{align}
    \label{eq:dvdt-emri}
    \frac{dv}{dt} &= \frac{32}{5} \frac{q}{m_{1}} v^{9} \left[1 + \sum_{n=2} \bar{a}_{n} v^{n} + q \sum_{n=0} \bar{a}_{n}^{(1)} v^{n} 
    \right.
    \nn \\
    &\left.
    + q \Delta\chi(\gamma_{P}) \sum_{n=3} \bar{d}_{n} v^{n} + {\cal{O}}(q^{2})\right]\,,
\end{align}
where the coefficients $[\bar{a}_{n},\bar{a}_{n}^{(1)}, \bar{d}_{n}]$ are given, to 2PN order, in Appendix~\ref{app:pn-coeffs}.

Having obtained Eq.~\eqref{eq:dvdt-emri}, the two relevant orbital quantities we must solve for under radiation reaction are the orbital phase $\phi$ and coordinate time $t$, the former of which is related to $v$ through
\begin{equation}
    \frac{d\phi}{dt} = \Omega = \frac{v^{3}}{m_{1}(1+q)}\,,
\end{equation}
while the latter can be computed from the reciprocal of Eq.~\eqref{eq:dvdt-emri}. TaylorF2 approximants are expression of the form $\phi(v)$ and $t(v)$, which map to Fourier frequency through the stationary phase approximation (SPA). By using the reciprocal of Eq.~\eqref{eq:dvdt-emri}, and PN expanding, the evolution of $\phi(v)$ can be written as
\begin{equation}
    \phi(v) = \phi_{\rm sec}(v) + \phi_{\rm osc}(v)\,,
\end{equation}
and likewise for $t(v)$, where $\phi_{\rm sec}$ is a purely monotonic (secular) function of $v$, and $\phi_{\rm osc}$ is an oscillatory correction due to precession. The TaylorF2 approximates for the secular part of the orbital phase and coordinate time take the form
\begin{align}
    \label{eq:phi-TF2}
    \phi_{\rm sec}(v) &= \phi_{c} - \frac{1}{32 q v^{5}} \left[1 + \sum_{n=0} \sum_{k=0} q^{k} \phi_{n}^{(k-1)} v^{n} + {\cal{O}}(q^{2}) \right]\,,
    \\
    \label{eq:t-TF2}
    t_{\rm sec}(v) &= t_{c} - \frac{5 m_{1}}{256 q v^{8}} \left[1 + \sum_{n=0} \sum_{k=0} q^{k} t_{n}^{(k-1)} v^{n}  + {\cal{O}}(q^{2}) \right]\,,
\end{align}
with $[\phi_{c}, t_{c}]$ integration constants. The coefficients $[\phi_{n}^{(k)}, t_{n}^{(k)}]$ are given explicitly, up to 2PN order, in Appendix~\ref{app:pn-coeffs}.

The oscillatory corrections require a more careful procedure. The historical method of doing has been to perform MSA, similar to Sec.~\ref{sec:prec}, where, in this case, the two timescales are the precession timescale of the secondary spin $t_{\rm pr,2} \sim 1/\omega_{P}$ and the radiation reaction timescale $t_{\rm rr} \sim 1/(dv/dt)$. However, for the current problem, this isn't actually necessary, and one can simply use the same method, namely Laplace's method, used to solve Eqs.~\eqref{eq:chi2Q-eqn}-\eqref{eq:chi2J-eqn} to obtain the oscillatory corrections. After one application of Laplace's method, we find for the oscillatory corrections
\begin{align}
    \label{eq:phi-osc}
    \phi_{\rm osc}(v) &= - \frac{1}{32q v^{5}} \left[q^{2} \chi_{T}(\gamma_{P}^{\rm sec}) \sum_{n=7} \phi_{n}^{(\rm osc)} v^{n} + {\cal{O}}(q^{3}) \right] \,,
    \\
    \label{eq:t-osc}
    t_{\rm osc}(v) &= -\frac{5 m_{1}}{256 q v^{8}} \left[q^{2} \chi_{T}(\gamma_{P}^{\rm sec}) \sum_{n=7} t_{n}^{(\rm osc)} v^{n} + {\cal{O}}(q^{3}) \right]\,,
\end{align}
where
\begin{align}
    \chi_{T}(\gamma_{P}) &= \int d\gamma_{P} \; \Delta \chi(\gamma_{P}) 
    \nn \\
    &=  \bar{\chi}_{P,0} \cos \gamma_{P} + \dot{\bar{\chi}}_{P,0} \sin \gamma_{P}\,,    
\end{align}
and the $t_{n}^{\rm osc}$ coefficients are given in Appendix~\ref{app:pn-coeffs}. The higher order corrections are suppressed by the factor $dv/d\gamma_{P} \sim t_{\rm pr,2}/t_{\rm rr} \sim q v^{4}$. This is the same factor that appears at each order in MSA of the radiation reaction effects, and thus the two methods are equivalent. Note that, because of the nature of our perturbative expansion, only the secular part of $\gamma_{P}$ appears in the argument of $\chi_{T}$ for the oscillatory corrections. An explicit expression for this is given in the next section.

At the level of approximation we are currently working, the coefficients of the secular contributions only contain one component of $\vec{\chi}_{2}$, namely $\bar{\chi}_{Q,0}$, which enters at 1.5PN order and first order in the mass ratio $q$. The remaining two components $[\bar{\chi}_{P,0}, \dot{\bar{\chi}}_{P,0}]$ are contained in the coefficients of the oscillatory correction in Eqs.~\eqref{eq:phi-osc}-\eqref{eq:t-osc}, which enter at 3.5PN order and second order in the mass ratio. The effect of the latter two components is, thus, highly suppressed compared to $\bar{\chi}_{Q,0}$. These will enter into the secular part of the phases if one goes to higher order in the small mass ratio expansion. However, we do not do so here, because the averages in Eqs.~\eqref{eq:LdotS2-avg}-\eqref{eq:S1dotS2-avg} will depend on the first order correction $\vec{\chi}_{2}^{(1)}$, which has not been explicitly computed here. There is nothing stopping one from proceeding to higher order in the MSA of Sec.~\ref{sec:prec}, besides the increasing analytic complexity that comes with high order MSA computations. For our purposes, this goes outside the scope of this paper, and we truncate the expansions of the secular and oscillatory contributions at the orders indicated in Eqs.~\eqref{eq:phi-TF2}-\eqref{eq:t-TF2} \& Eqs.~\eqref{eq:phi-osc}-\eqref{eq:t-osc}, respectively. We leave higher order computations to future work.

With the orbital phase and time in hand, we can readily compute the stationary phase of the waveform from Eqs.~\eqref{eq:phi-TF2} \& \eqref{eq:t-TF2}. The stationary phase condition is still $f = 2F$, and the PN expansion parameter is $v=(2\pi M F)^{1/3} = (\pi M f)^{1/3}$, where the second equality holds only after applying the stationary phase condition. The secular part of the waveform's stationary phase, up to 2PN order, is now
\begin{align}
    \label{eq:Fourier-phase}
    \Psi_{\rm sec}(f) &= 2\pi f t(f) - 2 \phi(f) - \frac{\pi}{4}\,,
    \nn \\
    &= 2\pi f t_{c} - 2\phi_{c} - \frac{\pi}{4} 
    \nn \\
    &+ \frac{3}{128 q v^{5}} \left[1 + \sum_{n,k} q^{k} \psi_{n}^{(k-1)} v^{n} \right]\,,
\end{align}
where
\begin{align}
    \psi_{n}^{(-1)} &= - \frac{5}{3} t_{n}^{(-1)} + \frac{8}{3} \phi_{n}^{(-1)}\,,
    \\
    \psi_{n}^{(0)} &= \begin{cases}
        \frac{5}{3} - \frac{5}{3} t_{0}^{(0)} + \frac{8}{3} \phi_{0}^{(0)} & n=0\\
        -\frac{5}{3} t_{n}^{(0)} + \frac{5}{3} t_{n}^{(-1)} + \frac{8}{3} \phi_{n}^{(0)} & n>0
    \end{cases}\,,
\end{align}
with $[t_{n}^{(-1)}, t_{n}^{(0)}, \phi_{n}^{(-1)}, \phi_{n}^{(0)}]$ given in Appendix~\ref{app:pn-coeffs}. On the other hand, the oscillatory correction is
\begin{equation}
    \Psi_{\rm osc}(f) = -\frac{3}{128q v^{5}} \left[q^{2} \chi_{T}(\gamma_{P}^{\rm sec}) \sum_{n=7} \psi_{n}^{(\rm osc)} v^{n} \right]\,,
\end{equation}
with
\begin{equation}
    \psi_{n}^{(\rm osc)} = \frac{5}{6} t_{n}^{(\rm osc)} - \frac{8}{3} \phi_{n}^{(\rm osc)}\,,
\end{equation}
such that $\Psi(f) = \Psi_{\rm sec}(f) + \Psi_{\rm osc}(f)$. This completes the computation of TaylorF2 approximants for all orbital quantities.

\subsection{TaylorF2 Approximants for Precessional Quantities}
\label{sec:F2-prec}

Having solved for the orbital quantities under radiation reaction, we now turn to the effect of radiation reaction upon precessional quantities, and seek TaylorF2-style approximants for them. The quantities of relavance are the three phases that parameterize the precession solutions in Sec.~\ref{sec:prec}, namely the precession angle $\alpha$, the spin angle $\gamma_{P}$, and the renormalization angle $\lambda$, which are defined in Eqs.~\eqref{eq:alpha-eqn},\eqref{eq:gamma_P}, \&~\eqref{eq:lambda-eq}, respectively.

The method for obtaining the TaylorF2 approximants for these is nearly identical to that of Sec.~\ref{sec:F2-orb}, with two small caveats. For the spin angle $\gamma_{P}$, the frequency $\omega_{P}$ defined in Eq.~\eqref{eq:gamma_P} is a complicated function of $(q,v)$ and must be PN expanded to obtain the approximant. For the renormalization angle $\lambda$, this quantity is already suppressed by a factor of $q$ and, thus, we only solve for its evolution at leading order in the mass ratio. Following the procedure of Sec.~\ref{sec:F2-orb}, the total (secular plus oscillatory) solutions are
\begin{widetext}
\begin{align}
    \label{eq:alpha-F2}
    \alpha(v) &= \alpha_{c} - \frac{5 \chi_{1}}{32 q v^{2}} \left[1 + \sum_{n} \sum_{k=0} q^{k} \left( \alpha_{n}^{(k-1)} + \alpha_{n}^{l,(k-1)}\ln v\right)v^{n}  + q^{2} \chi_{T}(\gamma_{P}^{\rm sec}) \sum_{n} \alpha_{n}^{(\rm osc)} v^{n}\right]\,,
    \\
    \label{eq:gamma-F2}
    \gamma(v) &= \gamma_{c} - \frac{5}{64 q v^{3}} \left[1 + \sum_{n} \sum_{k=0} q^{k} \left(\gamma_{n}^{(k-1)} + \gamma_{n}^{l,(k-1)} \ln v \right)v^{2} + q^{2} \chi_{T}(\gamma_{P}^{\rm sec}) \sum_{n} \gamma_{n}^{(\rm osc)} v^{n} \right]\,,
    \\
    \label{eq:lambda-F2}
    \lambda(v) &= \lambda_{c}  +  \frac{45 \chi_{1}}{128 v^{2}} \left[1 + \sum_{n} \left(\lambda_{n}^{(k-1)} + \lambda_{n}^{l,(k-1)} \ln v\right)v^{n} + q \chi_{T}(\gamma_{P}^{\rm sec}) \sum_{n} \lambda_{n}^{(\rm osc)} v^{n} \right]\,,
\end{align}
\end{widetext}
where $[\alpha_{c},\gamma_{c},\lambda_{c}]$ are integration constants, and the secular coefficients $[\alpha_{n}^{(k)},\alpha_{n}^{l,(k)},\gamma_{n}^{(k)},\gamma_{n}^{l,(k)},\lambda_{n}^{(0)},\lambda_{n}^{l,(0)}]$ and oscillatory coefficients $[\alpha_{n}^{(\rm osc)}, \gamma_{n}^{(\rm osc)}, \lambda_{n}^{(\rm osc)}]$ are given explicitly in Appendix~\ref{app:pn-coeffs}. Note that the secular parts of these quantities can be obtained by setting the oscillatory coefficients to zero in the above expressions. Further, unlike the orbital quantities, the secular parts of these expression contain terms dependent on the logarithm of $v$. These natuarlly appear at sufficiently high order in the PN expansion, and appear here due to the starting PN order of these precessional quantities. These will also appear for the orbital quantities in Eqs.~\eqref{eq:phi-TF2}-\eqref{eq:t-TF2} if one proceeds to higher PN order. 

\subsection{Waveform Precession Phase $\delta \Phi$}
\label{sec:waveform_prec}

The evolution of the precession phase of the waveform is given in Eq.~(28) of~\cite{ApostolatosCutler}, specifically
\begin{equation}
    \label{eq:prec-phase}
    \frac{d\delta \Phi}{dt} = \frac{(\hat{L}\cdot\vec{N})}{1 - (\hat{L}
    \cdot \vec{N})^{2}} \left(\hat{L} \times \vec{N}\right) \cdot \frac{d\hat{L}}{dt}\,,
\end{equation}
where $\vec{N}$ is the line of sight from the detector to the source. Since the precession phase $\delta \Phi$ is a scalar quantity, it does not matter which frame one chooses to compute it in, i.e. the inertial frame of the binary versus the inertial frame of the detector. For simplicity, we choose to do the computation in the former where $\vec{J}$ is fixed and defines the z-axis of the coordinate system. Then,
\begin{equation}
    \label{eq:Nvec}
    \vec{N} = \left[\sin \theta_{N} \cos\phi_{N}, \sin\theta_{N} \sin\phi_{N}, \cos\theta_{N} \right]\,,
\end{equation}
and applying Eq.~\eqref{eq:prec-L-emri}, Eq.~\eqref{eq:prec-phase} becomes
\begin{align}
    \label{eq:prec-phase-new}
    \frac{d\delta \Phi}{d\tau} &= \frac{(\hat{L}\cdot \vec{N})}{1 - (\hat{L}\cdot \vec{N})^{2}} \left\{\omega_{L} {\cal{P}}_{LN}\left(\vec{J}\right) 
    \right.
    \nn \\
    &\left.
    + q \left[v \omega_{SL} {\cal{P}}_{LN}\left(\vec{\chi}_{2}\right)- \omega_{L}^{(1)} {\cal{P}}_{LN}\left(\vec{J}\right) \right]\right\}\ ,
\end{align}
where
\begin{equation}
    \label{eq:PLN-eqn}
    {\cal{P}}_{LN}\left(\vec{A}\right) = \left( \hat{L} \cdot \vec{A} \right) \left( \vec{L} \cdot \vec{N} \right) - \left( \vec{A} \cdot {\vec{N}} \right)
\end{equation}
for an arbitrary vector $\vec{A}$. Generically, $(\theta_{N}, \phi_{N})$ will be functions of time due to the fact that the LISA detector is not fixed relative to the inertial frame of the binary. However, we argue that this effect can be neglected in the computation of the precession phase in Appendix~\ref{app:no-LISA}.

The right hand side of Eq.~\eqref{eq:prec-phase-new} must be expanded in $q\ll1$, and we seek a solution of the form
\begin{equation}
    \delta \Phi = \delta \Phi^{(0)}(\tau) + q \delta \Phi^{(1)}(\tau) + {\cal{O}}(q^{2})\,.
\end{equation}
Formally, one should perform MSA to solve for the precession phase. However, we employ a shortcut that reproduces the results of such an MSA, and does not result in a loss of accuracy. Because of the algebraic complexity of solving for the precession phase, we split the computation into two separate parts below.

\subsubsection{$\delta\Phi$ at ${\cal{O}}(q^{0})$}
At leading order in $q\ll1$, one simply has to take $q\rightarrow 0$ and $\hat{L} \rightarrow \hat{L}^{(0)}$ in Eq.~\eqref{eq:prec-phase-new}. The definition of $\vec{N}$ is given in Eq.~\eqref{eq:Nvec}, while $\hat{L}^{(0)}$ is given by Eqs.~\eqref{eq:Lx-0}-\eqref{eq:Ly-0},~\eqref{eq:Lz-0}, \&~\eqref{eq:X-sol}-\eqref{eq:Y-sol}. Thus, we have
\begin{align}
    \hat{L}\cdot \vec{N} &= \hat{L}_{z,0} \cos\theta_{N} + \sqrt{1-\hat{L}_{z,0}^{2}} \sin\theta_{N} \cos\bar{\alpha}\,,
\end{align}
and, at ${\cal{O}}(q^{0})$, Eq.~\eqref{eq:prec-phase} becomes
\begin{equation}
    \label{eq:dPhi-red}
    \frac{d \delta \Phi^{(0)}}{d\alpha} = -\frac{b_{0} + b_{1} \cos\bar{\alpha} + b_{2} \cos^{2}\bar{\alpha}}{d_{0} - d_{1} \cos\bar{\alpha} - d_{2} \cos^{2}\bar{\alpha}} \equiv {\cal{F}}_{0}(\bar{\alpha})\,,
\end{equation}
where $\bar{\alpha} = \alpha + \lambda - \phi_{N} + \phi_{L}$ with $\phi_{L} = \arctan(\hat{L}_{y,0}/\hat{L}_{x,0})$, $\lambda$ is given by Eq.~\eqref{eq:lambda-eq}, and the $(b_{i}, d_{i})$ coefficients are only functions of the constants $(\hat{L}_{z,0}, \theta_{N})$ and are given in Appendix~\ref{app:prec-coeffs}. The presence of $\lambda$ in this expression would normally require the application of MSA to solve for $\delta \Phi^{(0)}$. However, the shortcut in Sec.~\ref{sec:F2-orb} to obtain the oscillatory corrections to the TaylorF2 approximants may also be employed here. More explicitly, the solution is
\begin{align}
    \label{eq:dPhi-integral}
    \delta \Phi^{(0)}(\tau) &= \int d\alpha \; {\cal{F}}_{0}[\bar{\alpha}(\alpha)]\,,
    \nn \\
    &= \int d\bar{\alpha} \left(1 + q \nu_{XY}\right)^{-1} {\cal{F}}(\bar{\alpha})\,.
\end{align}
where to obtain the second equality we have performed a change of variables, and the factor in the parentheses comes from
\begin{equation}
    \frac{d\bar{\alpha}}{d\alpha} = 1 + q \nu_{XY}(\tau)\ ,
\end{equation}
with $\nu_{XY}(\tau) = \omega_{XY}/\omega_{L}$, which only varies on the radiation reaction timescale. Further, ${\cal{F}}_{0}(\bar{\alpha})$ only varies on the more rapid precession timescale, and $\bar{\alpha}$ does not possess any stationary points. Thus, we may  again employ Laplace's method to obtain
\begin{align}
    \label{eq:dPhi-expression}
    \delta\Phi^{(0)}(\tau) &= \delta_{c} +  {\cal{N}}_{\Phi} \bar{\alpha}(\tau) + \frac{\sqrt{2}}{d_{2}} \left[h_{-}{\cal{E}}_{-}(\tau) -h_{+} {\cal{E}}_{+}(\tau) \right]
    \nn \\
    &+ {\cal{O}}(q^{2} v^{3})
\end{align}
where the remainder is determined by $d\nu_{XY}/d\bar{\alpha}$ expanded in $q\ll1$ and $v\ll1$, and $\delta_{c}$ is an integration constant. The purely oscillatory (non-secular) functions ${\cal{E}}_{\pm}$ are explicitly
\begin{equation}
    {\cal{E}}_{\pm}(\tau) = \tan^{-1}\left[\frac{\beta_{\pm} \sin\bar{\alpha}(\tau)}{1 - \beta_{\pm}\cos\bar{\alpha}(\tau)}\right]
\end{equation}
quantities $({\cal{N}}_{\Phi}, h_{\pm}, \beta_{\pm})$ are functions of the $(b_{i}, d_{i})$ coefficients and are given explicitly in Appendix~\ref{app:prec-coeffs}. 

\subsubsection{$\delta\Phi$ at ${\cal{O}}(q)$}

At first order in the mass ratio,
\begin{align}
    \label{eq:dPhi-1}
    \frac{d\delta\Phi^{(1)}}{d\gamma_{P}} &= \frac{\omega_{L}}{\omega_{P}} \frac{{\cal{B}}_{1}(\alpha,\lambda,\gamma_{P},v)}{d_{0} - d_{1} \cos\bar{\alpha} - d_{2} \cos^{2}\bar{\alpha}} 
    \nn \\
    &+ \frac{\omega_{L}}{\omega_{P}}\frac{{\cal{B}}_{2}(\alpha,\lambda,\gamma_{P},v)}{(d_{0} - d_{1} \cos\bar{\alpha} - d_{2} \cos^{2}\bar{\alpha})^{2}}\,,
\end{align}
where ${\cal{B}}_{1,2}$ are complicated functions of time through $[\alpha,\lambda,\gamma_{P}]$ on the precession timescale and $v$ on the radiation reaction timescale. We do not provide explicit expression for these, but they can be readily derived from Eq.~\eqref{eq:prec-phase-new} by inserting $\hat{L} \rightarrow \hat{L}^{(0)} + q \hat{L}^{(1)}$ and $\vec{\chi}_{2} \rightarrow \vec{\chi}_{2}^{(0)}$, and expanding in $q\ll1$. In terms of these vector quantities
\begin{align}
    {\cal{B}}_{1} &= \left(\hat{L}^{(0)} \cdot \vec{N} \right)^{2} \left(\hat{L}^{(1)} \cdot \hat{J} \right) + \nu_{SL} \left(\hat{L}^{(0)} \cdot \vec{N} \right) {\cal{P}}_{LN}^{(0)}\left(\vec{\chi}_{2}^{(0)}\right)
    \nn \\
    &- \nu_{1} \left(\hat{L}^{(0)} \cdot \vec{N} \right) {\cal{P}}_{LN}^{(0)}\left( \hat{J} \right)\,,
    \\
    {\cal{B}}_{2} &= \left(\hat{L}^{(1)} \cdot \vec{N} \right) \Bigg\{2 \left(\hat{L}^{(0)} \cdot \hat{J}\right) \left(\hat{L}^{(0)} \cdot \vec{N}\right) 
    \nn\\
    &- \left(\hat{J} \cdot \vec{N} \right) \left[1 + \left(\hat{L}^{(0)} \cdot \vec{N} \right)^{2} \right] \Bigg\}\ ,
\end{align}
where $\nu_{SL} = v \omega_{SL}/\omega_{L}$, $\nu_{1} = \omega_{L}^{(1)}/\omega_{L}$, and ${\cal{P}}_{LN}^{(0)}$ is given by Eq.~\eqref{eq:PLN-eqn} with the replacement $\hat{L} \rightarrow \hat{L}^{(0)}$.

There is no exact closed-form solution to Eq.~\eqref{eq:dPhi-1}, but an approximate solution can be obtained from the following considerations. The ratios $[\nu_{SL}, \nu_{1}]$ only vary on the radiation reaction timescale, and not significantly. For the example EMRI system studied in Fig.~\ref{fig:Jcomp}, $\nu_{SL}$ varies by only $\sim 2\%$ over the course of the inspiral, while $\nu_{1}$ varies by $\sim 0.6\%$. Thus, we treat these factors as effectively constant when solving Eq.~\eqref{eq:dPhi-1}. Further, $\gamma_{P}$ varies more rapidly than $\alpha$ and $\lambda$, which can be seen from the PN scaling of these quantities in Eqs.~\eqref{eq:alpha-F2}-\eqref{eq:lambda-F2}. More specifically, $d\alpha/d\gamma_{P} = \omega_{L}/\omega_{P} \sim {\cal{O}}(v)$. Since we are working in a PN expansion, we define $\xi = \omega_{L}/\omega_{P}$ to be an order keeping parameter and perform MSA using $\xi \ll 1$. Eq.~\eqref{eq:dPhi-1} can now be solved by using
\begin{equation}
    \frac{d}{d\gamma_{P}} = \frac{\partial}{\partial\gamma_{P}} + \xi \frac{\partial}{\partial \alpha}
\end{equation}
and seeking a solution of the form
\begin{align}
    \delta \Phi^{(1)} = \delta \Phi^{(1)}_{0,{\rm sec}}(\alpha,\lambda) + \sum_{k=1} \xi^{k} &\left[\delta \Phi^{(1)}_{k,{\rm osc}}(\alpha, \lambda, \gamma_{P})\right .\nonumber\\
    &\left . + \delta \Phi^{(1)}_{k,{\rm sec}}(\alpha, \lambda)\right] \,.
\end{align}
Note that $\delta \Phi^{(1)}_{0,{\rm sec}}$ does not have an oscillatory contribution so from now on we define $\delta \Phi^{(1)}_{0,{\rm sec}}=\delta \Phi^{(1)}_{0}$ for the ease of notation. We write the functions ${\cal{B}}_{1,2}$ as
\begin{align}
    {\cal{B}}_{1,2} &= \langle {\cal{B}}_{1,2} \rangle_{\gamma}^{(0)}(\alpha, \lambda) + \delta {\cal{B}}_{1,2}^{(0)}(\alpha, \lambda, \gamma_{P})
    \nn \\
    &+ \xi \left[\langle {\cal{B}}_{1,2}\rangle_{\gamma}^{(1)}(\alpha,\lambda) + \delta {\cal{B}}_{1,2}^{(1)}(\alpha,\lambda,\gamma_{P}) \right]\,,
\end{align}
where $\delta {\cal{B}}_{1,2}^{(0,1)}$ are purely oscillatory functions in $\gamma_{P}$, i.e. $\langle \delta {\cal{B}}_{1,2}^{(0,1)} \rangle_{\gamma} = 0$. 

At first order in $\xi$, Eq.~\eqref{eq:dPhi-1} becomes
\begin{align}
    \frac{d\delta\Phi_{0}^{(1)}}{d\alpha} + \frac{\partial \delta\Phi_{1,{\rm osc}}^{(1)}}{\partial \gamma_{P}} &= \frac{\langle {\cal{B}}_{1}\rangle_{\gamma}^{(0)} + \delta {\cal{B}}_{1}^{(0)}}{d_{0} - d_{1} \cos\bar{\alpha} - d_{2} \cos^{2}\bar{\alpha}}
    \nn \\
    &+ \frac{\langle {\cal{B}}_{2} \rangle_{\gamma}^{(0)} + \delta {\cal{B}}_{2}^{(0)}}{(d_{0} - d_{1} \cos\bar{\alpha} - d_{2} \cos^{2}\bar{\alpha})^{2}}
\end{align}
Averaging the above equation with respect to $\gamma_{P}$ eliminates the second term on the left hand side, as well as the $\delta{\cal{B}}_{1,2}^{(0)}$ terms on the right hand side.
This decouples the dynamics of the secular $\delta \Phi_{0}^{(1)}$ from the oscillatory $\delta \Phi_{1,{\rm osc}}^{(1)}$. The averages $\langle{\cal{B}}_{1,2}\rangle_{\gamma}^{(0)}$ take the simple forms
\begin{align}
    \langle {\cal{B}}_{1} \rangle_{\gamma}^{(0)} &= b_{0}^{(1)} + b_{1}^{(1)} \cos\bar{\alpha} + b_{2}^{(1)} \cos^{2}\bar{\alpha}\,,
    \\
    \langle {\cal{B}}_{2} \rangle_{\gamma}^{(0)} &= \left(\sum_{k=1}^{3} c_{k}^{(1)} \cos^{k}\bar{\alpha}\right) + \sin\bar{\alpha} \left(\sum_{k=0}^{2} s_{k}^{(1)} \cos^{k}\bar{\alpha}\right)
\end{align}
with the coefficients $[b_{i}^{(1)}, c_{i}^{(1)}, s_{i}^{(1)}]$ given in Appendix~\ref{app:prec-coeffs}. Due to the superposition of the source term in Eq.~\eqref{eq:dPhi-1}, we can split the solution for $\delta\Phi^{(1)}$ into the linear combination of solutions sourced from ${\cal{B}}_{1}$ and ${\cal{B}}_{2}$ separately, specifically $\delta\Phi_{0}^{(1)} = \delta\Phi_{0}^{({\cal{B}}_{1})} + \delta\Phi_{0}^{({\cal{B}}_{2})}$. The necessary integral with respect to $\alpha$ can be performed using the same method in Eq.~\eqref{eq:dPhi-integral}, with the end result being
\begin{align}
    \label{eq:dPhi-sec-B1}
    \delta \Phi_{0}^{({\cal{B}}_{1})} &= -{\cal{N}}_{\Phi}^{({\cal{B}}_{1})} \bar{\alpha} + \frac{\sqrt{2}}{d_{2}} \left[h_{+}^{({\cal{B}}_{1})} {\cal{E}}_{+}(\tau) - h_{-}^{({\cal{B}}_{1})} {\cal{E}}_{-}(\tau)\right]\,,
    \\
    \label{eq:dPhi-sec-B2}
    \delta \Phi_{0}^{({\cal{B}}_{2})} &= {\cal{N}}_{\Phi}^{({\cal{B}}_{2})} \bar{\alpha} + \frac{\kappa}{\Delta_{0}^{3/2}} {\cal{L}}(\tau) 
    \nn \\
    &+ \frac{\sqrt{2}}{\Delta_{0}^{3/2} \Delta_{1}} \left[\frac{g_{+}}{\sqrt{\Delta_{+}}} {\cal{E}}_{+}(\tau) + \frac{g_{-}}{\sqrt{\Delta_{-}}} {\cal{E}}_{-}(\tau)\right]
    \nn \\
    &+ \frac{\sigma_{0} + \sigma_{C} \cos\bar{\alpha} + \sigma_{S}^{(1)} \sin\bar{\alpha} + \sigma_{S}^{(2)} \sin(2\bar{\alpha})}{2d_{2} \Delta_{0} \Delta_{1} (d_{0} - d_{1} \cos\bar{\alpha} - d_{2} \cos^{2}\bar{\alpha})}
\end{align}
with
\begin{equation}
    {\cal{L}}(\tau) = \log\left[\frac{\sqrt{\Delta_{0}} - d_{1} - 2d_{2} \cos\bar{\alpha}(\tau)}{\sqrt{\Delta_{0}} + d_{1} + 2d_{2} \cos\bar{\alpha}(\tau)}\right]\,,
\end{equation}
and $[{\cal{N}}_{\Phi}^{({\cal{B}}_{1,2})}, h_{\pm}^{({\cal{B}}_{1})}, \Delta_{1},\kappa, g_{\pm},\sigma_{0},\sigma_{C},\sigma_{S}^{(1,2)}]$ are given in Appendix~\ref{app:prec-coeffs}. Generically, one could also have integration constants in Eqs.~\eqref{eq:dPhi-sec-B1}-\eqref{eq:dPhi-sec-B2}, but these can be eliminated by re-defintion of $\delta_{c}$ in Eq.~\eqref{eq:dPhi-expression}. For the oscillatory contribution, the solution is trivially given by
\begin{align}
    \label{eq:dPhi-1-osc}
    \delta \Phi_{1,{\rm osc}}^{(1)} &= \frac{1}{d(\bar{\alpha})}\int d\gamma_{P} \; \delta {\cal{B}}_{1}^{(0)}(\alpha,\lambda,\gamma_{P})
    \nn \\
    & + \frac{1}{[d(\bar{\alpha})]^{2}} \int d\gamma_{P} \; \delta {\cal{B}}_{2}^{(0)}(\alpha,\lambda,\gamma_{P})\,,
    \nn \\
    &= \left\{\frac{{\cal{B}}_{1}^{C}(\bar{\alpha})}{d(\bar{\alpha})} + \frac{{\cal{B}}_{2}^{C}(\bar{\alpha})}{[d(\bar{\alpha})]^{2}} \right\} \cos\gamma_{P} 
    \nn \\
    &+ \left\{\frac{{\cal{B}}_{1}^{S}(\bar{\alpha})}{d(\bar{\alpha})} + \frac{{\cal{B}}_{2}^{S}(\bar{\alpha})}{[d(\bar{\alpha})]^{2}} \right\} \sin\gamma_{P}
\end{align}
where $[{\cal{B}}_{1,2}^{C},{\cal{B}}_{1,2}^{S}]$ are functions of $\bar{\alpha}$ given explicitly in Appendix~\ref{app:prec-coeffs}. We stop our computation here, since it suffices for capturing the leading order behavior of the secondary spin. The methodology can be extended to higher order is one desires more phase accuracy.

Finally, we have numerically checked that the phase contribution of $\delta\Phi^{(1)}_{1,{\rm sec}}$ is negligible, so we do not provide analytical results for this term.

\begin{figure}[hbt!]
    \centering
    \includegraphics[width=\columnwidth]{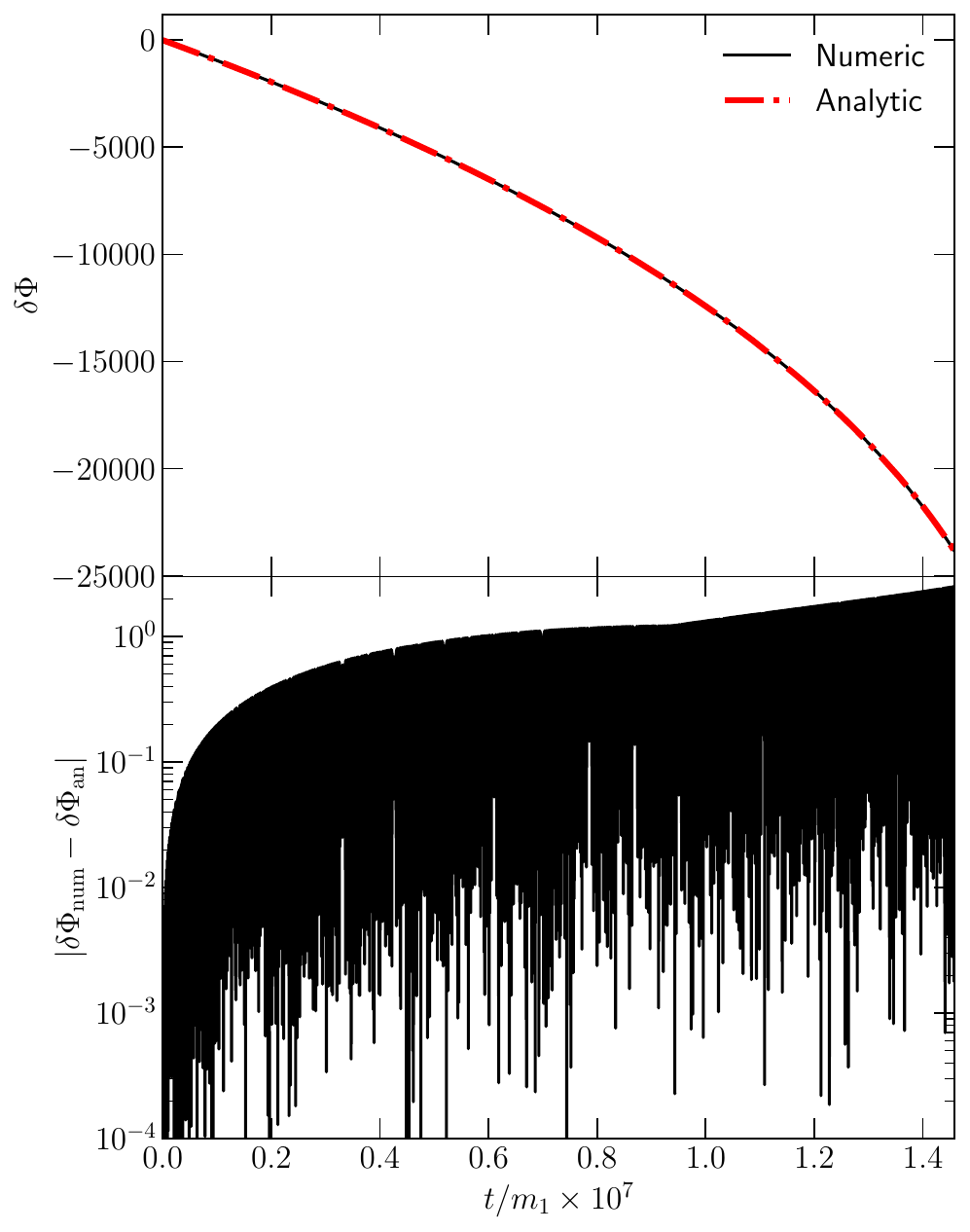}
    \caption{Top: Comparison of the analytical result for the waveform precession phase $\delta\Phi$ found by combining Eqs.~\eqref{eq:dPhi-expression},\eqref{eq:dPhi-sec-B1}-\eqref{eq:dPhi-sec-B2}, \&~\eqref{eq:dPhi-1-osc} (red dashed line) to an exact solution obtained by numerically integrating Eq.~\eqref{eq:prec-phase} (black solid line). Bottom: Absolute error between the analytic and numerical solutions.}
    \label{fig:prec_comp}
\end{figure}

In Fig.~\ref{fig:prec_comp}, we provide a comparison of our analytic solution for the total $\delta\Phi$ to numerical integration of Eq.~\eqref{eq:prec-phase} for the same EMRI system in Fig.~\ref{fig:Jcomp}. The dephasing (bottom panel) between the two solutions is $\sim2$ radians over the full coalescence, indicating the accuracy of the approximations used to obtain the analytic expression in Eqs.~\eqref{eq:dPhi-expression},\eqref{eq:dPhi-sec-B1}-\eqref{eq:dPhi-sec-B2}, \&~\eqref{eq:dPhi-1-osc}. It is worth remembering that this total de-phasing is over $>2$ years of inspiral, while over the last year of the inspiral, the dephasing is $\sim 1$ radian for this EMRI. While typically this could impact parameter estimation, the goal of this study is not to develop the most accurate model possible for EMRIs, but a model that has the necessary qualitative features to forecast uncertainties on the secondary's spin. If one desired more phase accuracy, one could carry out the analysis here to higher order. This completes the computation of the waveform precession phase.

%

\section{Waveform Phasing Due To Spin Precession And Secondary Spin}
\label{sec:dephase}

Finally, with the analytical results of the previous sections at hand, we can now quantify the GW phase introduced by spin precession and by the presence of a (non)precessing secondary spin. The total waveform phase is given in Eq.~\eqref{eq:waveform-phase}, with the Fourier phase $\Psi(f)$ and precession phase $\delta\Phi(f)$ being the most relevant for our analysis. We define the quantity
\begin{equation}
    \Psi_{T}(f;\theta^{a}, \chi_{2}^{a}) = \Psi(f;\theta^{a},\chi_{2}^{a}) - \delta\Phi(f;\theta^{a},\chi_{2}^{a})\,, 
\end{equation}
with $\chi_{2}^{a} = [\bar{\chi}_{Q,0}, \bar{\chi}_{P,0}, \dot{\bar{\chi}}_{P,0}]$, and $\theta^{a} = [m_{1}, q, L_{z,0}, \phi_{L}, c_{1}, \chi_{1}, \chi_{\rm eff}, \theta_{N}, \phi_{N}, t_{c}, \phi_{c}, \delta_{c}]$ are the waveform parameters not associated with the secondary's spin. Further, we define the total phase accumulated by $\Psi_{T}$ over one year of inspiral to be
\begin{equation}
    \label{eq:Delta-Psi-T}
    \Delta \Psi_{T}(\theta^{a}, \chi_{2}^{a}) = \Psi_{T}(f_{\rm ISCO}; \theta^{a}, \chi_{2}^{a}) - \Psi_{T}(f_{\rm 1yr}; \theta^{a}, \chi_{2}^{a})\,.
\end{equation}
where $f_{\rm ISCO}$ is the Fourier frequency associated with the Kerr ISCO, and $f_{\rm 1yr}$ is the Fourier frequency one year before the EMRI reaches the ISCO. 

\begin{figure*}
    \centering
    \includegraphics[width=\textwidth]{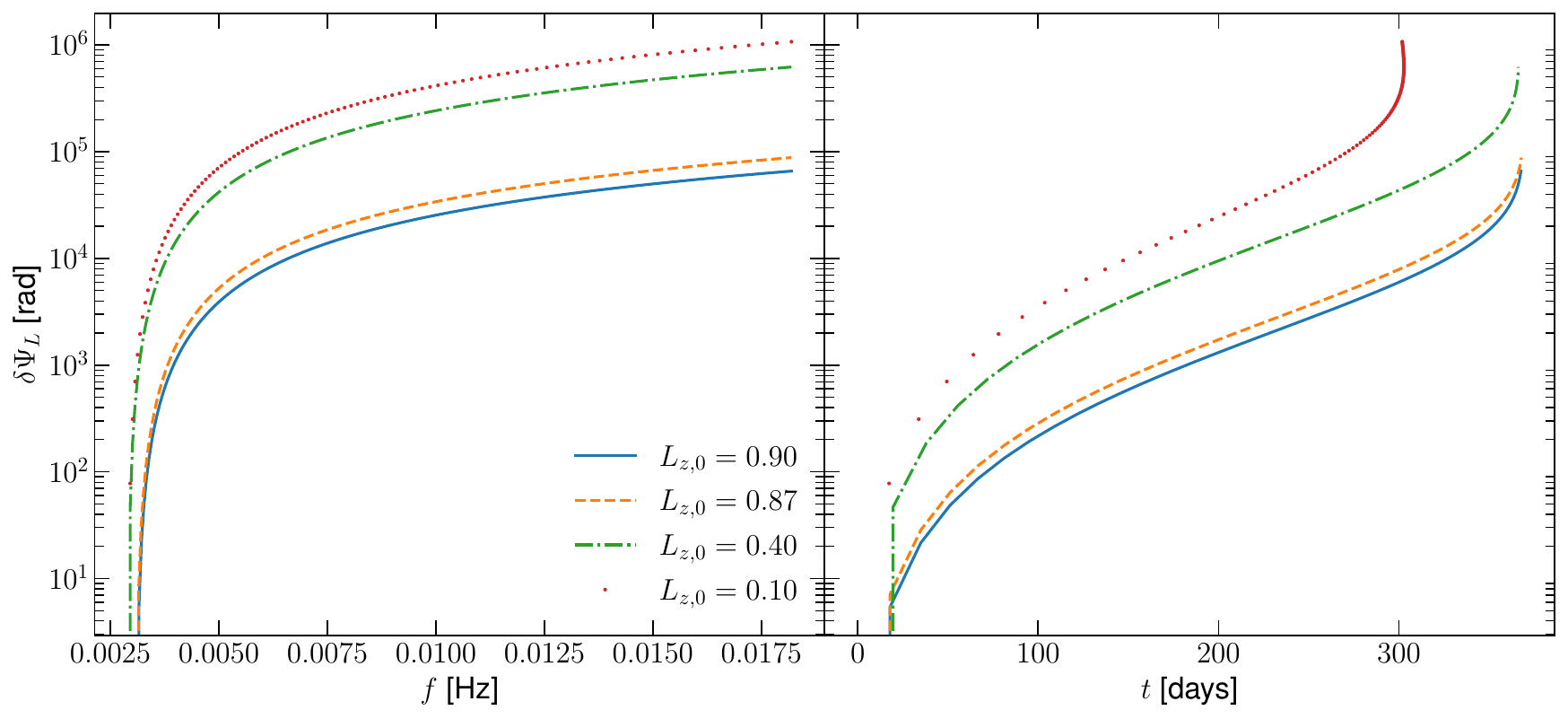}
    \caption{GW de-phasing 
    \eqref{eq:dPsiL-eq} between an aligned EMRI $(L_{z,0} = 1)$ and a precessing configuration with varying misalignment (colored curves, smaller values of $L_{z,0}$ correspond to large misalignment). The EMRIs in this figure has a non-spinning secondary, and the precession is induced by the spin-orbit coupling between the primary spin and orbital angular momentum. Further, we fix $\chi_{1} = 0.90$ and $q=10^{-5}$.}
    \label{fig:dephase-no-spin}
\end{figure*}

From the total accumulated phase in Eq.~\eqref{eq:Delta-Psi-T}, we consider two quantities that act as measures of the impact of spin precession on the waveform. The first is the de-phasing between an aligned (non-precessing) EMRI and an EMRI undergoing spin-orbit precession (without a spinning secondary), specifically
\begin{equation}
    \label{eq:dPsiL-eq}
    \delta \Psi_{L}(\theta^{a}) = \Delta \Psi_{T}(\theta^{a}, \chi_{2}^{a} = 0) - \Delta \Psi_{T}(\theta^{a}_{\rm align}, \chi_{2}^{a}=0)\,,
\end{equation}
where $\theta^{a}_{\rm align}$ are the parameters in the aligned limit. The quantity $\delta \Psi_{L}$ provides us with a measure of how many radians of phase the leading-order spin-orbit interaction introduces in the waveform. For this comparison, the spin of the secondary is neglected, while we take $\chi_{\rm eff} = \chi_{1} L_{z,0}$, which follows from the expansion of Eq.~\eqref{eq:chi-eff} about $q\ll1$. The aligned limit, specified by $\theta^{a}_{\rm align}$, is given by $L_{z,0} = 1$. We use the TaylorF2 approximation for $\Psi(f)$ given in Eq.~\eqref{eq:Fourier-phase}, while for $\delta\Phi(f)$ we use the analytic approximation given in Sec.~\ref{sec:waveform_prec} with $\alpha$ and $\gamma_{P}$ given by Eqs.~\eqref{eq:alpha-F2} \& \eqref{eq:gamma-F2}, respectively. For the analysis carried out here, we fix the orientation of the line of sight vector $\vec{N}$, which enters the precession phase $\delta\Phi$, to $\theta_{N}=\pi/6$ and $\phi_{N}=\pi/4$. We fix $\chi_{1}=0.9$ and $q=10^{-5}$, and study how the de-phasing $\delta\Psi_{L}$ varies with increasing misalignment, i.e. with decreasing $L_{z,0}$. Fig.~\ref{fig:dephase-no-spin} shows the results of this comparison, revealing that increasing misalignment produces greater de-phasing compared to the aligned limit, owing to the increasing precession effects on the binary. The total dephasing is generally large, typically $\gtrsim10^{4}$ radians or larger, even for EMRIs with small misalignment. The reason for this is that the misalignment enters the GW phase at leading order in the mass ratio ${\cal{O}}(q^{-1})$ and at 1.5PN order, ${\cal{O}}(v^{3})$, through $\chi_{\rm eff}$. Thus, the contribution for generic spin-orbit precession is large compared to a non-precessing EMRI.

The second quantity we can consider is the de-phasing between an EMRI with a spinning secondary and one without it,
\begin{equation}
    \delta \Psi_{\chi_{2}}(\theta^{a}, \chi_{2}^{a}) = \Delta\Psi_{T}(\theta^{a}, \chi_{2}^{a}) - \Delta\Psi_{T}(\theta^{a}, \chi_{2}^{a}=0)\,,\label{eq:dephasing}
\end{equation}
where $\theta^{a}$ is held fixed between the two inspirals. Studying the above quantity is useful since, assuming the absence of parameter degeneracies, 
a de-phasing greater than one radian would substantially impact a matched-filter search, leading to a significant loss of detected events, and being potentially detectable~\cite{Lindblom:2008cm}. The quantity $\delta\Psi_{\chi_{2}}$ is dependent on the parameters $\theta^{a}$, but many of these have negligible impact on the de-phasing. The three most important parameters we have identified are $q$, $L_{z,0}$, and $\chi_{1}$. The first should be obvious since the secondary's spin is coupled to the mass ratio. The z-component of the orbital angular momentum's initial orientation $L_{z,0}$ controls the amplitude of precession effects, specifically, more misalignment (smaller $L_{z,0}$) leads to stronger precession effects. Lastly, the primary's spin controls the location of the ISCO, and the total angular momentum through Eq.~\eqref{eq:j-emri}, modulating how much the primary influences the precession of the secondary. 

\begin{figure*}
    \centering
    \includegraphics[width=\textwidth]{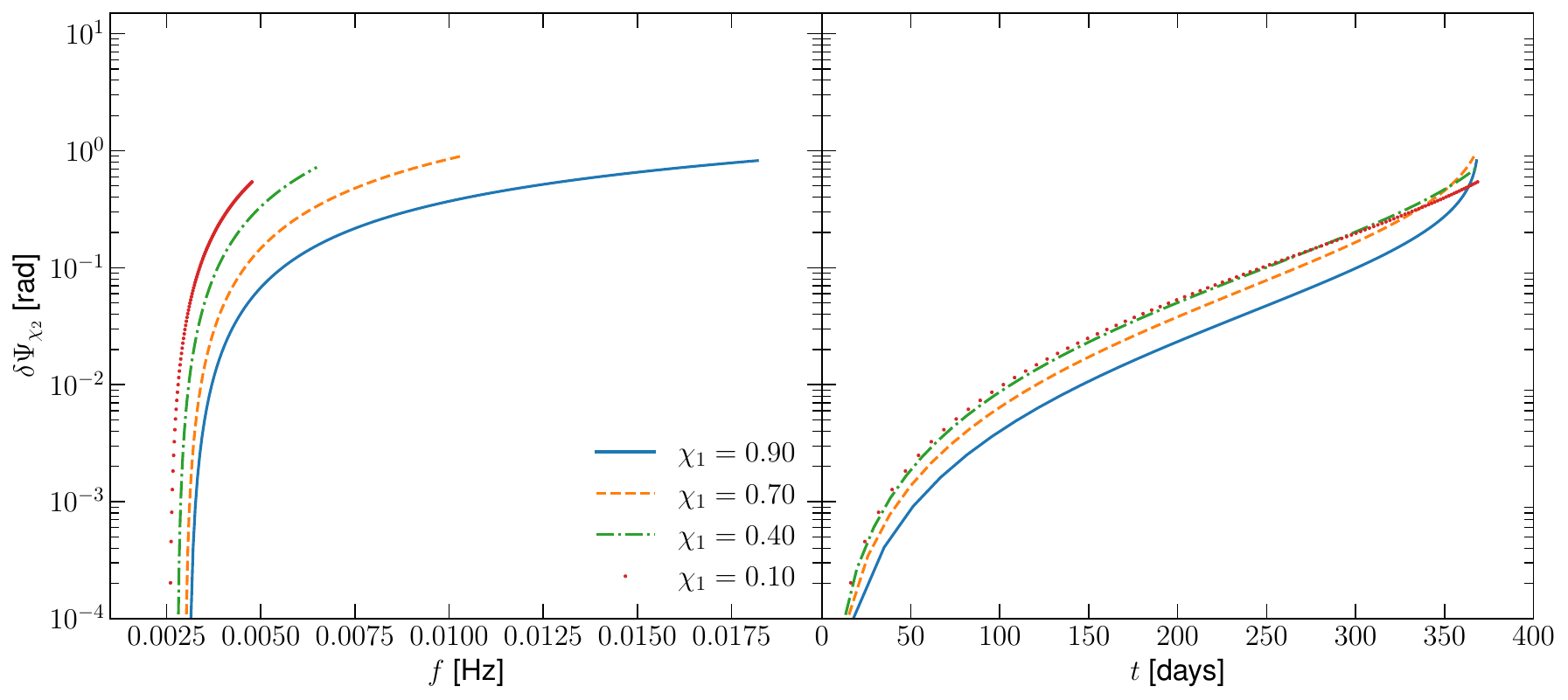}
    \caption{GW de-phasing 
    \eqref{eq:dephasing} between an EMRI with a spinning secondary and one without, 
    as a function of the frequency 
    (left panel) and of the time (right 
    panel). We consider binaries with 
    mass ratio $q=10^{-5}$, evolving 
    for one year until the ISCO. Colored 
    curves refer to different values 
    of the primary spin $\chi_1$, while 
    we fix $L_{z,0}=0.87, \bar{\chi}_{Q,0}=1, \bar{\chi}_{P,0}=0 =\dot{\bar{\chi}}_{P,0}$.}
    \label{fig:dephase-chi1-xQ_1p0}
\end{figure*}

\begin{figure*}
    \centering
    \includegraphics[width=\textwidth]{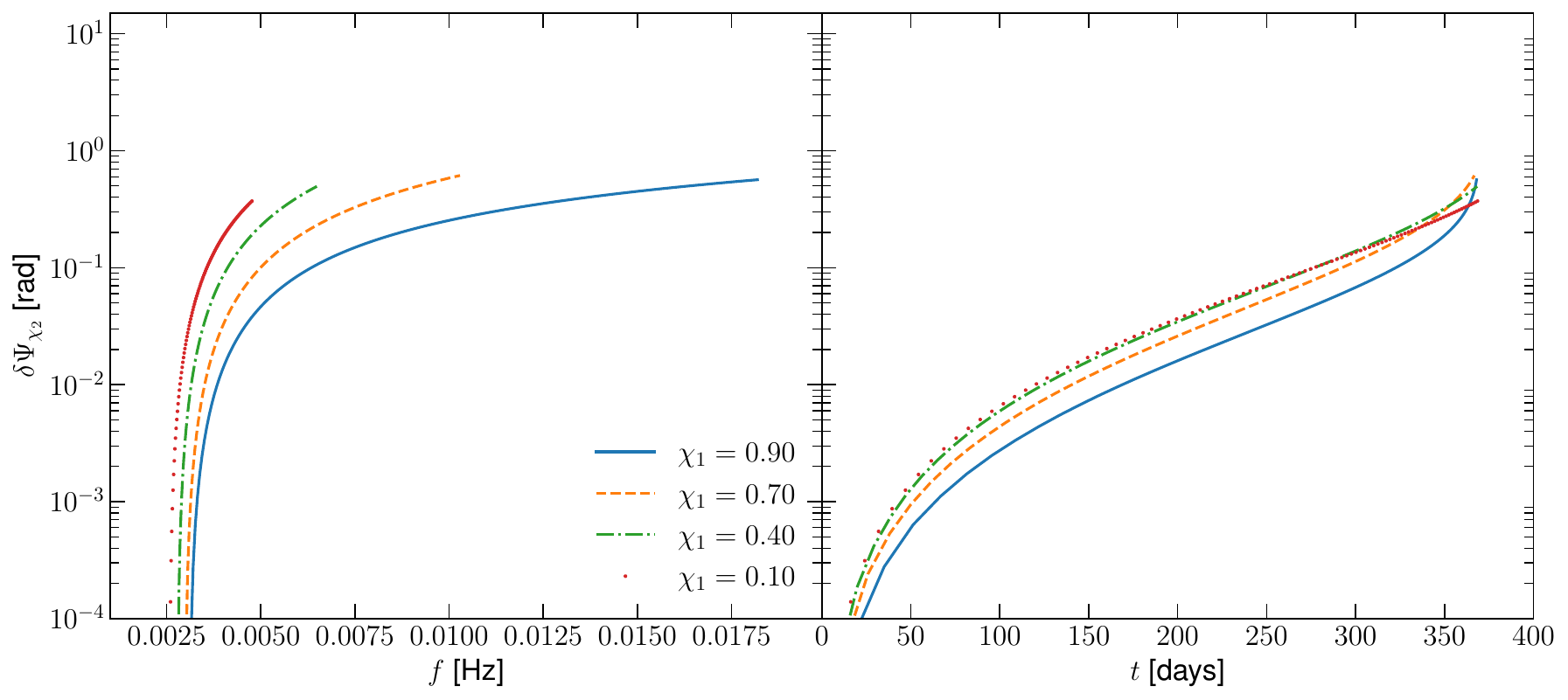}
    \caption{Same as Fig.~\eqref{fig:dephase-chi1-xQ_1p0} but for $L_{z,0}=0.87, \bar{\chi}_{Q,0}=0.687, \bar{\chi}_{P,0}=-0.259, \dot{\bar{\chi}}_{P,0}=-0.635$. The latter two component of the secondary spin contribute to the waveform phase in oscillatory corrections, which are suppressed by $q^{2}$, while $\bar{\chi}_{Q,0}$ is only suppressed by $q$. As a result, there is less de-phasing compared to the case in Fig.~\ref{fig:dephase-chi1-xQ_1p0}.}
    \label{fig:dephase-chi1-xQ_0p69}
\end{figure*}

Fig.~\ref{fig:dephase-chi1-xQ_1p0} shows the de-phasing $\delta\Psi_{\chi_{2}}$ as a function of frequency (left) and time (right) for various values of the primary's spin, and with $\bar{\chi}_{Q,0} = 1$ and $\bar{\chi}_{P,0} = 0 = \dot{\bar{\chi}}_{P,0}$. In this case, the secondary's spin is a constant in the co-precessing reference frame of Sec.~\ref{sec:chi20}, and thus only undergoes simple precession (no nutation). The total de-phasing over one year of inspiral is typically $0.7-1.0$ radian, depending on the value of the primary's spin $\chi_{1}$. Larger values of of $\chi_{1}$ produce slightly more de-phasing than lower values, owing to the ISCO being smaller for prograde orbits and allowing for a larger number of total GW cycles. Fig.~\ref{fig:dephase-chi1-xQ_0p69} shows the same comparison, but for the secondary spin with orientation $[\bar{\chi}_{Q,0}, \bar{\chi}_{P,0}, \dot{\bar{\chi}}_{P,0}] = [0.687, -0.259, -0.635]$. The de-phasings in this case are slightly smaller than those in Fig.~\ref{fig:dephase-chi1-xQ_1p0}, due to the fact that the components $[\bar{\chi}_{P,0}, \dot{\bar{\chi}}_{P,0}]$ mainly contribute to the phase through oscillatory corrections, which are suppressed by $q^{2}$. Thus, the component $\bar{\chi}_{Q,0}$ produces the largest contribution to the phase, and will likely be the best recovered component of the secondary's spin when performing parameter estimation. Note that this is not unexpected, since it is known that the component of the secondary spin parallel to the total angular momentum, which is related to $\bar{\chi}_{Q,0}$, produces the dominate contribution to the GW phase~\cite{Witzany:2019nml,Witzany:2023bmq,Skoupy:2023lih,Burke:2023lno}.

\begin{figure*}
    \centering
    \includegraphics[width=\textwidth]{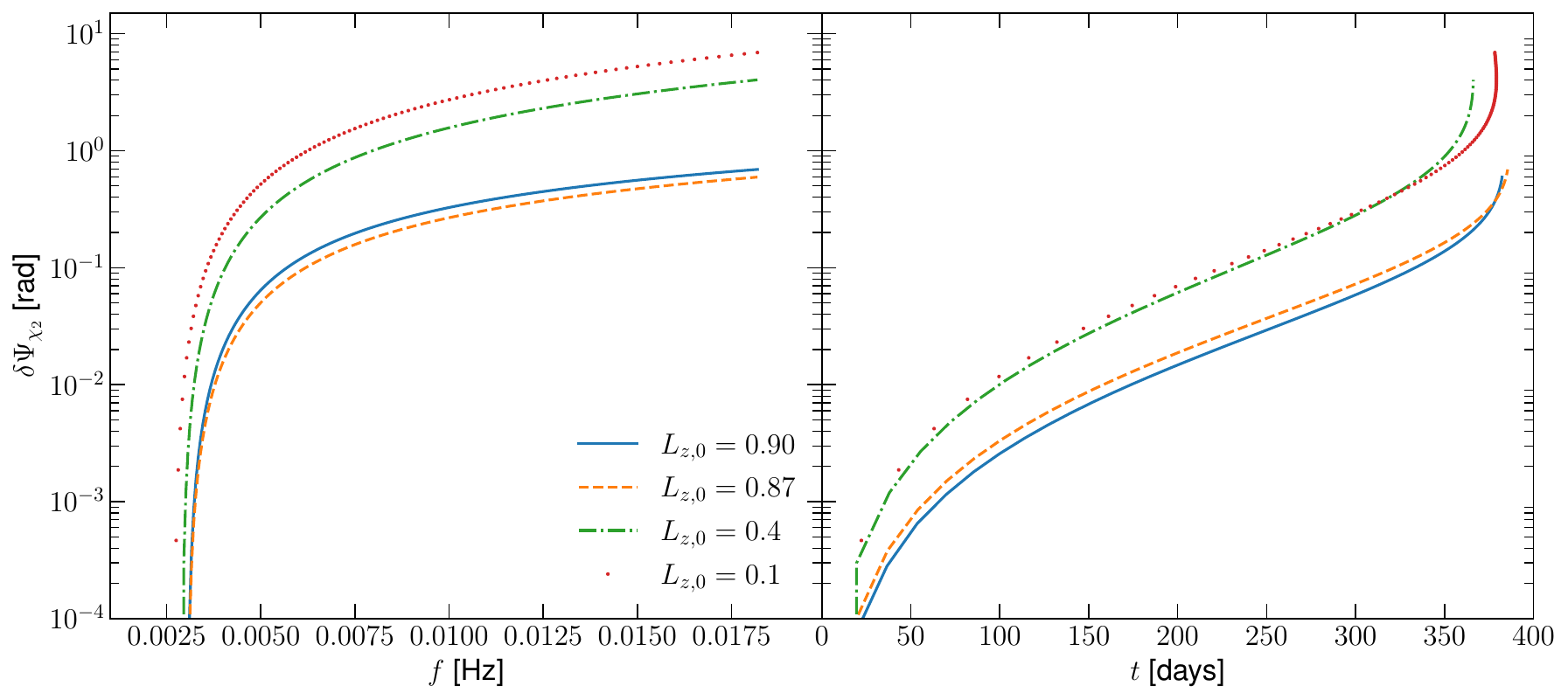}
    \caption{Same as Fig.~\eqref{fig:dephase-chi1-xQ_1p0} but for $\chi_{1}=0.90$ and varying the value of $L_{z,0}$. Smaller values of $L_{z,0}$ correspond to systems with greater initial misalignment between the orbital angular momentum and primary spin, leading to more precession cycles in the waveform, and thus, larger $\delta\Psi_{\chi_{2}}$.}
    \label{fig:dephase-Lz0}
\end{figure*}

\begin{figure*}
    \centering
    \includegraphics[width=\textwidth]{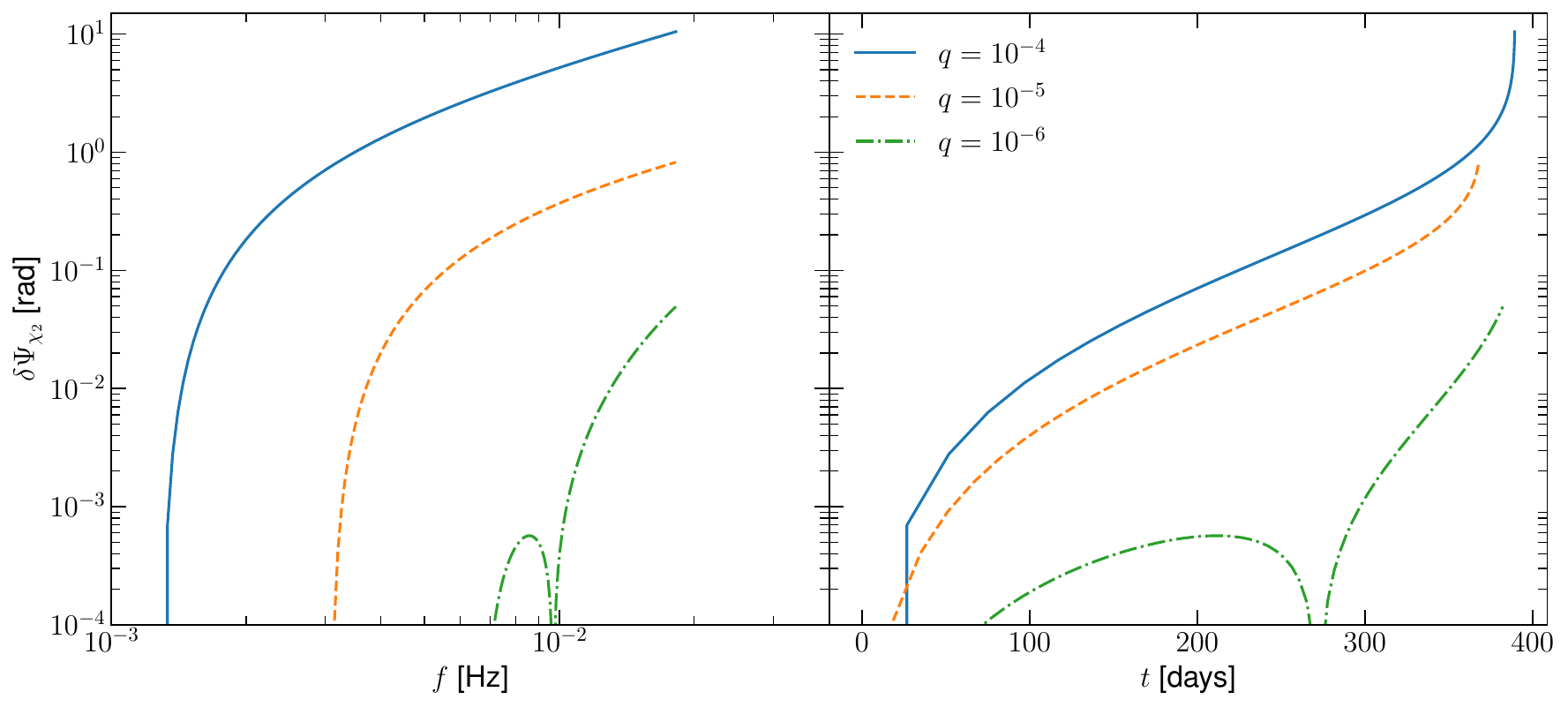}
    \caption{Same as Fig.~\eqref{fig:dephase-chi1-xQ_1p0} but for $\chi_{1}=0.90, L_{z,0}=0.87$, and varying the value of the mass ratio $q$. Since the secondary's spin is suppressed by the mass ratio, larger (more comparable mass systems) produce more de-phasing over a fixed one year duration of inspiral.}
    \label{fig:dephase-q}
\end{figure*}

In Fig.~\ref{fig:dephase-Lz0}, we investigate how the initial misalignment of the orbital angular momentum, encoded in $L_{z,0}$, impacts the phase accumulated due to the secondary's spin. The limit $L_{z,0}\rightarrow 1$ corresponds to alignment between the orbital angular momentum and primary spin, and decreasing values provide greater misalignment. Generally, more misalignment produces more precession cycles, thus increasing the impact of the secondary spin on the waveform's phase. For the largest misalignment studied here, specifically $L_{z,0}=0.1$, $\delta\Psi_{\chi_{2}}\sim {\cal{O}}(10)$ radians after one year of inspiral. The case $L_{z,0}=0.87$ provides the smallest de-phasing, which is due to the fact that $\hat{L}$ is nearly aligned with the line of sight vector $\vec{N}$. Lastly, in Fig.~\ref{fig:dephase-q}, we study the impact of the mass ratio on the dephasing. Due to the fact that spin effects are suppressed by the mass ratio, more comparable mass systems (larger $q$) produce more dephasing over one year of inspiral than more disparate mass systems.

The PN analysis carried out here suggests that we may expect ${\cal{O}}(1)-{\cal{O}}(10)$ radians of phase from EMRIs with a precessing secondary. To understand if this is reasonable, we may compare to aligned results obtains using Teukolsky fluxes in~\cite{Piovano:2020ooe}. In Fig.~2 therein, it was found that one may reasonably expect $\sim{\cal{O}}(10)$ radians of dephasing due to an aligned spinning secondary. In our case, Fig.~\ref{fig:dephase-Lz0} shows that close to alignment (solid line), only $\sim{\cal{O}}(1)$ radians are accumulated, an order of magnitude less than tha found in~\cite{Piovano:2020ooe}. We owe the difference to the fact that we are using the PN approximation to model an EMRI system, which is known to not be a reasonably accurate approximation of the dynamics of EMRIs. Nevertheless, this comparison shows that the dephasing we obtain from our naive PN analysis is a conservative estimate of the effects of the secondary's spin. As a result, we expect that results on the uncertainty of the secondary's spin components obtained in a parameter estimation study that makes use of our waveform model will also be conservative estimates of those obtained from a proper self-force waveform that includes a precessing secondary. Such an analysis goes outside the scope of this paper, and we plan to address this in future work \cite{followup}.

\section{Conclusion and future work} \label{sec:conclusion}

In this work we have developed a consistent and fully analytical framework to describe precessing binaries with generic spin vectors and a large mass ratio asymmetry, exploiting both the PN theory in the EMRI limit, as well as a hierarchical multi-scale analysis. We have solved the equations of motion for precessing quantities at the leading order in the mass ratio.

As a key result of our formalism, we have 
computed the PN phase corrections due to 
precession within the TaylorF2 GW approximant 
in the frequency domain, including contributions 
from both the primary and secondary spin vectors with generic orientation. 
We have therefore developed a fully analytical 
waveform model for precessing EMRIs with 
arbitrary spin vectors.

The agreement between our analytical results and numerical integrations of the equations of motion is well controlled. While we find excellent agreement to ${\cal O}(q^0)$, the maximum dephasing to ${\cal O}(q)$ is less controlled, up to two radians for a full evolution lasting a few years up to the ISCO of the primary.
Note that our perturbative procedure can be extended to higher orders in $q$, and there is nothing preventing one from including high order MSA computations, which will presumably make the final result more accurate. 

In any case, the intrinsic error of the analytical waveform are typically smaller than those of the precession and secondary-spin effects that we have focused on. Furthermore, despite this intrinsic error, our analytical template can be useful to forecast the order-of-magnitude of the 
parameter errors inferred by future 
observations (using either Fisher-matrix or 
more sophisticated Monte Carlo Markov Chain approaches), or 
for comparison with other waveforms in 
certain regimes. The model developed here 
can also be used for producing hybrid 
waveforms aiming to describe less 
asymmetric binaries by matching SF models that include PN contributions and the EOB formalism.

In a follow-up work~\cite{followup}, we will use 
our TaylorF2 model to perform parameter estimation 
using a Fisher-matrix analysis. Preliminary results 
show that, although the PN series is known to be only asymptotic to
 the EMRI regime (and is therefore not a faithful representation of the signal), it nevertheless provides reliable 
results for what concerns the errors of the 
waveform parameters. Thus, it can be exploited 
to estimate how precession affects the 
measurability of certain binary parameters, 
especially those that feature degeneracies in the nonprecessing case. A detailed study on this problem 
will appear in~\cite{followup}.

\begin{acknowledgments}
We thank Gabriel Piovano for interesting comments on the draft.
This work is partially supported by the MIUR PRIN Grant 2020KR4KN2 ``String
Theory as a bridge between Gauge Theories and Quantum Gravity'', by the FARE programme (GW-NEXT, CUP:~B84I20000100001),
and by the EU Horizon 2020 Research and Innovation Programme under the Marie Sklodowska-Curie Grant Agreement No. 101007855. N.L. is supported by ERC Starting Grant No.~945155--GWmining, 
Cariplo Foundation Grant No.~2021-0555, MUR PRIN Grant No.~2022-Z9X4XS, 
and the ICSC National Research Centre funded by NextGenerationEU. S.M. is thankful to the GravNet grant for providing financial support to carry out the present research, and Sapienza University of Rome for a warm hospitality where a part of this work was done. S.M. is also thankful to the Inspire Faculty Grant DST/INSPIRE/04/2022/001332, DST, Government of India, and Lumina Quaeruntur No. LQ100032102 of the Czech Academy of Sciences, for 
support.
\end{acknowledgments}

\appendix

\section{Arguments for neglecting the LISA constellation motion in the calculation of the precession phase}
\label{app:no-LISA}

Consider, in addition to the line of sight from binary to detector $\vec{N}$, the line of sight from binary to the barycenter of the ecliptic $\vec{N}_{E}$, which is parameterized by a new set of angles $(\theta_{NE}, \phi_{NE})$. We assume that the inertial frame of the ecliptic is fixed relative to the inertial frame of the binary, and thus $(\theta_{NE}, \phi_{NE})$ are fixed. Then, one has the distance from the binary to the ecliptic's barycenter $\vec{R}_{E} = R_{E} \vec{N}_{E}$, the distance from the binary to the LISA detector $\vec{R}(t) = R(t) \vec{N}(t)$, and the distance from the ecliptic's barycenter to the LISA detector $r_{D} = r_{D} \vec{n}_{D}(t)$. Here $r_{D} = 1 {\rm AU}$, and these vectors satisfy $\vec{R}(t) = \vec{R}_{E} + \vec{r}(t)$. Re-arranging, we have
\begin{equation}
    \label{eq:N-relation}
    \vec{N} = \frac{R_{E}}{R(t)} \vec{N}_{E} + \frac{r(t)}{R(t)} \vec{n}_{D}(t)\,.
\end{equation}
One can also show using the law of cosines that $R(t) \sim R_{E} + {\cal{O}}[r(t)/R(t)]$. Combining this with Eq.~\eqref{eq:N-relation}, we realize that $\vec{N} \sim \vec{N}_{E} + {\cal{O}}(r/R_{E})$, where $R_{E}$ can be taken as the luminosity distance to the source. Since $r_{D} = 1{\rm AU}$, and $R_{E} \sim 100 {\rm Mpc} - 10 {\rm Gpc}$, the ratio $r/R_{E} \sim 10^{-14} - 10^{-16}$, which is approximately the limit of double precision accuracy. Hence, we can consider $(\theta_{N},\phi_{N})$ as fixed when considering the precession phase of the waveform $\delta\Phi$. The LISA constellation's motion does, however, enter the definitions of the beam pattern functions.

\section{Coefficients of TaylorF2 Approximates}
\label{app:pn-coeffs}

In this Section we provide the explicit expression for the coefficients appearing in various 
PN quantities in Sec.~\ref{sec:rr}. For $dv/dt$ expanded in small mass ratio in 
Eq.~\eqref{eq:dvdt-emri}, the PN coefficients are
\begin{align}
    \bar{a}_{2} &= -\frac{743}{336}\,, \qquad \bar{a}_{3} = 4\pi - \frac{113}{12} \chi_{\rm eff}\,,
    \\
    \bar{a}_{4} &= \frac{34103}{18144} - \frac{233}{96} \chi_{1}^{2} + \frac{719}{96} \chi_{\rm eff}^{2}\,,
    \\
    \bar{a}_{0}^{(1)} &= -3\,, \qquad \bar{a}_{2}^{(1)} = \frac{435}{112}\,,
    \\
    \bar{a}_{3}^{(1)} &= \frac{1}{12} \left(-144 \pi + 377 \chi_{\rm eff} + 38 \bar{\chi}_{Q,0} \right)\,,
    \\
    \bar{a}_{4}^{(1)} &= \frac{215}{189} + \frac{1165}{96} \chi_{1}^{2} - \frac{719}{32} \chi_{\rm eff}^{2} - \frac{133}{16} L_{z,0} \chi_{1} \bar{\chi}_{Q,0} 
    \nn \\
    &+ \frac{1}{24} \chi_{\rm eff} \bar{\chi}_{Q,0}\,.
    \\
    \bar{d}_{3} &= \frac{19}{6} {\cal{D}}_{L,2}\,, \qquad \bar{d}_{4} = \frac{1}{24} {\cal{D}}_{L,2} \chi_{\rm eff} - \frac{247}{48} {\cal{D}}_{1,2}\,,
\end{align}
with ${\cal{D}}_{L,2}$ and ${\cal{D}}_{1,2}$ velocity dependent factors given in Eqs.~\eqref{eq:DL2-def}-\eqref{eq:D12-def}. To leading order in a PN and small mass ratio expansion
\begin{align}
    {\cal{D}}_{L,2} &\sim -\chi_{1} v + {\cal{O}}(q, v^{2})\,,
    \\
    {\cal{D}}_{1,2} &\sim - \chi_{1} + {\cal{O}}(q, v)\,.
\end{align}
The non-zero coefficients of $\phi(v)$ and $t(v)$ in Eqs.~\eqref{eq:phi-TF2}-\eqref{eq:t-TF2} 
up to 2PN order read
\allowdisplaybreaks[4]
\begin{align}
    t_{2}^{(-1)} &= -\frac{4}{3} \bar{a}_{2}\,, \qquad t_{3}^{(-1)} = - \frac{8}{5} \bar{a}_{3}\,,
    \\
    t_{4}^{(-1)} &= 2 \left(\bar{a}_{2}^{2} - \bar{a}_{4} \right)\,, \qquad t_{0}^{(0)} = -\bar{a}_{0}^{(1)}\,,
    \\
    t_{2}^{(0)} &= \frac{4}{3} \left(2\bar{a}_{0}^{(1)} \bar{a}_{2} - \bar{a}_{2}^{(1)} \right)\,,
    \\
    t_{3}^{(0)} &= \frac{8}{5} \left(2 \bar{a}_{0}^{(1)} \bar{a}_{3} - \bar{a}_{3}^{(1)} \right)\,,
    \\
    t_{4}^{(0)} &= - 6 \bar{a}_{0}^{(1)} \bar{a}_{2}^{2} + 4 \bar{a}_{2} \bar{a}_{2}^{(1)} + 4 \bar{a}_{0}^{(1)} \bar{a}_{4} - 2 \bar{a}_{4}^{(1)}\,,
    \\
    t_{7}^{(\rm osc)} &= - \frac{608}{9} \chi_{1} \,,
    \\
    t_{8}^{(\rm osc)} &= -\left(\frac{1216}{9} \hat{L}_{z,0} \chi_{1} + \frac{992}{15} \chi_{\rm eff} \right) \chi_{1} \,,
    \\
    \phi_{2}^{(-1)} &= -\frac{5}{3} \bar{a}_{2}\,, \qquad \phi_{3}^{(-1)} = -\frac{5}{2} \bar{a}_{3}\,,
    \\
    \phi_{4}^{(-1)} &= 5 \left(\bar{a}_{2}^{2} - \bar{a}_{4} \right)\,,
    \qquad
    \phi_{0}^{(0)} = -1 - \bar{a}_{0}^{(1)}\,,
    \\
    \phi_{2}^{(0)} &= \frac{5}{3} \left(\bar{a}_{2} + 2 \bar{a}_{0}^{(1)} \bar{a}_{2} - \bar{a}_{2}^{(1)} \right)\,,
    \\
    \phi_{3}^{(0)} &= \frac{5}{2} \left(\bar{a}_{3} + 2 \bar{a}_{0}^{(1)} \bar{a}_{3} - \bar{a}_{3}^{(1)} \right)\,,
    \\
    \phi_{4}^{(0)} &= -5 \left[\left(1 + 3 \bar{a}_{0}^{(1)}\right)\bar{a}_{2}^{2} - 2 \bar{a}_{2} \bar{a}_{2}^{(1)} 
    \right.
    \nn \\
    &\left.
    - \left(1 + 2 \bar{a}_{0}^{(1)} \right) \bar{a}_{4} + \bar{a}_{4}^{(1)} \right]\,,
    \\
    \phi_{7}^{(\rm osc)} &= -\frac{380}{9} \chi_{1}\,,
    \\
    \phi_{8}^{(\rm osc)} &= - \frac{4}{9} \chi_{1} \left(190 \hat{L}_{z,0} \chi_{1} + 93 \chi_{\rm eff} \right)\,.
\end{align}
For the precessional angle in Eq.~\eqref{eq:alpha-F2}, the non-zero PN coefficients up to 2PN order are
\allowdisplaybreaks[4]
\begin{align}
    \alpha_{1}^{(-1)} &= -\frac{3}{2} \chi_{\rm eff}\,, \qquad \alpha_{3}^{(-1)} = 2 \bar{a}_{3}^{(0)} - \frac{3}{2} \chi_{\rm eff} \bar{a}_{2}^{(0)}\,,
    \\
    \alpha_{4}^{(-1)} &= \bar{a}_{4} - \bar{a}_{2}^{2} - \frac{3}{4} \chi_{\rm eff} \bar{a}_{3}\,,
    \qquad
    \alpha_{2}^{l,(-1)} = 2 \bar{a}_{2}\,, 
    \\
    \alpha_{-1}^{(0)} &= \frac{c_{1}}{3 \chi_{1}^{2}}\,,
    \qquad
    \alpha_{0}^{(0)} = -\bar{a}_{0}^{(1)} - \frac{3 c_{1} \chi_{\rm eff}}{8 \chi_{1}^{2}}\,,
    \\
    \alpha_{1}^{(0)} &= \frac{3}{2} \chi_{\rm eff} \bar{a}_{0}^{(1)} - \frac{c_{1}}{\chi_{1}^{2}} \bar{a}_{2}\,,
    \\
    \alpha_{3}^{(0)} &= 2 \bar{a}_{3}^{(1)} - 4 \bar{a}_{3} \bar{a}_{0}^{(1)} + 3 \bar{a}_{2} \bar{a}_{0}^{(1)} \chi_{\rm eff} - \frac{3}{2} \chi_{\rm eff} \bar{a}_{2}^{(1)} 
    \nn \\
    &- \frac{c_{1}}{4 \chi_{1}^{2}} \left(4 \bar{a}_{2}^{2} - 4 \bar{a}_{4} + 3 \chi_{\rm eff} \bar{a}_{3} \right)\,,
    \\
    \alpha_{4}^{(0)} &= \bar{a}_{4}^{(1)} + 3 \bar{a}_{2}^{2} \bar{a}_{0}^{(1)} - 2 \bar{a}_{4} \bar{a}_{0}^{(1)} - 2 \bar{a}_{2} \bar{a}_{2}^{(1)} + \frac{3}{2} \chi_{\rm eff} \bar{a}_{3} \bar{a}_{0}^{(1)} 
    \nn \\
    &- \frac{3}{4} \chi_{\rm eff} \bar{a}_{3}^{(1)} - \frac{c_{1}}{8 \chi_{1}^{2}} \left(8 \bar{a}_{2} \bar{a}_{3} - 3 \bar{a}_{2}^{2} \chi_{\rm eff} + 3 \bar{a}_{4} \chi_{\rm eff} \right)\,,
    \\
    \alpha_{2}^{l,(0)} &= 2 \bar{a}_{2}^{(1)} - 4 \bar{a}_{2} \bar{a}_{0}^{(1)} + \frac{c_{1}}{4 \chi_{1}^{2}} \left(4 \bar{a}_{3} - 3 \bar{a}_{2} \chi_{\rm eff} \right)
    \\
    \alpha_{7}^{(\rm osc)} &= -\frac{152}{9} \chi_{1}\,,
    \\
    \alpha_{8}^{(\rm osc)} &= -\frac{2}{45}\chi_{1} \left(760 \chi_{1} \hat{L}_{z,0} + 87 \chi_{\rm eff} \right)\,.
\end{align}
For the spin angle $\gamma_{P}$ in Eq.~\eqref{eq:gamma-F2}, the non-zero PN coefficients are
\allowdisplaybreaks[4]
\begin{widetext}
    \begin{align}
    \gamma_{1}^{(-1)} &= -\frac{3}{2} \left(\hat{L}_{z,0} \chi_{1} + \chi_{\rm eff} \right)\,,
    \qquad
    \gamma_{2}^{(-1)} = -3\bar{a}_{2} + 3 \chi_{1} \chi_{\rm eff} \hat{L}_{z,0} + \frac{3}{2} \chi_{1}^{2} \left(1 - \hat{L}_{z,0}^{2} \right)\,,
    \\
    \gamma_{4}^{(-1)} &= -3 \left(\bar{a}_{2}^2+\bar{a}_{3} \chi_{\rm eff}-\bar{a}_{4}\right)-3 \hat{L}_{z,0} \chi_{1} (\bar{a}_{3}-\bar{a}_{2} \chi_{\rm eff}) -\frac{3}{2} \bar{a}_{2} \left(\hat{L}_{z,0}^2-1\right) \chi_{1}^2-\frac{3}{2} \hat{L}_{z,0} \left(\hat{L}_{z,0}^2-1\right)\chi_{1}^3 \chi_{\rm eff}
    \nn \\
    &+\frac{3}{8} \left(5 \hat{L}_{z,0}^4-6 \hat{L}_{z,0}^2+1\right) \chi_{1}^4\,,
    \\
    \gamma_{3}^{l,(-1)} &= 3 (\bar{a}_{3}-\bar{a}_{2} \chi_{\rm eff})-3 \bar{a}_{2} \hat{L}_{z,0} \chi_{1}+\frac{3}{2} \hat{L}_{z,0} \left(\hat{L}_{z,0}^2-1\right) \chi_{1}^3-\frac{3}{2} \left(\hat{L}_{z,0}^2-1\right) \chi_{1}^2 \chi_{\rm eff}\,,
    \\
    \bar{\gamma}_{0}^{(0)} &= -\bar{a}_{0}^{(1)} - \frac{2}{3} \frac{c_{1} \hat{L}_{z,0}}{\chi_{1}}\,,
    \qquad
    \bar{\gamma}_{1}^{(0)} = \frac{3}{2} \bar{a}_{0}^{(1)} \left(\chi_{1} \hat{L}_{z,0} + \chi_{\rm eff} \right) + c_{1} \left(1 - \hat{L}_{z,0}^{2} + \frac{3\hat{L}_{z,0} \chi_{\rm efff}}{4 \chi_{1}} \right)\,,
    \\
    \bar{\gamma}_{2}^{(0)} &= 6 \bar{a}_{2} \bar{a}_{0}^{(1)}+c_{1} \left(\frac{2 \bar{a}_{2} \hat{L}_{z,0}}{\chi_{1}} -\frac{3}{2} \left(\hat{L}_{z,0}^2-1\right) (2 \hat{L}_{z,0} \chi_{1}-\chi_{\rm eff})\right)-\frac{3}{2} \bar{a}_{0}^{(1)} \chi_{1} \left(-\hat{L}_{z,0}^2 \chi_{1} +2 \hat{L}_{z,0} \chi_{\rm eff}+\chi_{1}\right)-3 \bar{a}_{2}^{(1)}\,,
    \\
    \gamma_{4}^{(0)} &= \frac{3}{8} \left\{24 \bar{a}_{2}^2 \bar{a}_{0}^{(1)}-8 \bar{a}_{2} \left[\bar{a}_{0}^{(1)} \chi_{1} \left(\hat{L}_{z,0}^2 (-\chi_{1})+2 \hat{L}_{z,0} \chi_{\rm eff}+\chi_{1}\right)+2 \bar{a}_{2}^{(1)}\right]+16 \bar{a}_{3} \bar{a}_{0}^{(1)} \hat{L}_{z,0} \chi_{1}+16\bar{a}_{3} \bar{a}_{0}^{(1)} \chi_{\rm eff}
    \right.
    \nn \\
    &\left.
    -16 \bar{a}_{4} \bar{a}_{0}^{(1)}-5 \bar{a}_{0}^{(1)} \hat{L}_{z,0}^4 \chi_{1}^4+4 \bar{a}_{0}^{(1)} \hat{L}_{z,0}^3 \chi_{1}^3 \chi_{\rm eff}+6 \bar{a}_{0}^{(1)} \hat{L}_{z,0}^2 \chi_{1}^4-4 \bar{a}_{0}^{(1)} \hat{L}_{z,0} \chi_{1}^3 \chi_{\rm eff}-\bar{a}_{0}^{(1)} \chi_{1}^4-4 \bar{a}_{2}^{(1)} \hat{L}_{z,0}^2 \chi_{1}^2
    \right.
    \nn \\
    &\left.
    +8 \bar{a}_{2}^{(1)} \hat{L}_{z,0} \chi_{1} \chi_{\rm eff}+4 \bar{a}_{2}^{(1)} \chi_{1}^2-8 \bar{a}_{3}^{(1)} \hat{L}_{z,0} \chi_{1}-8 \bar{a}_{3}^{(1)} \chi_{\rm eff}+8 \bar{a}_{4}^{(1)}\right\}
    \nn \\
    &+\frac{c_{1}}{4 \chi_{1}} \left[8 \bar{a}_{2}^2 \hat{L}_{z,0}-6 \bar{a}_{2} \left(\hat{L}_{z,0}^2-1\right) \chi_{1}(2 \hat{L}_{z,0} \chi_{1}-\chi_{\rm eff})-8 \bar{a}_{3} \hat{L}_{z,0}^2 \chi_{1}+6 \bar{a}_{3} \hat{L}_{z,0} \chi_{\rm eff}+8 \bar{a}_{3} \chi_{1}-8 \bar{a}_{4} \hat{L}_{z,0}
    \right.
    \nn \\
    &\left.
    +35 \hat{L}_{z,0}^5 \chi_{1}^4-15 \hat{L}_{z,0}^4 \chi_{1}^3 \chi_{\rm eff}-50 \hat{L}_{z,0}^3 \chi_{1}^4+18 \hat{L}_{z,0}^2 \chi_{1}^3 \chi_{\rm eff}+15 \hat{L}_{z,0} \chi_{1}^4-3 \chi_{1}^3 \chi_{\rm eff}\right]\,,
    \\
    \gamma_{3}^{l, (0)} &= \frac{3}{2} \left(4 \bar{a}_{2} \bar{a}_{0}^{(1)} \hat{L}_{z,0} \chi_{1}+4 \bar{a}_{2} \bar{a}_{0}^{(1)} \chi_{\rm eff}-4 \bar{a}_{3} \bar{a}_{0}^{(1)}-\bar{a}_{0}^{(1)} \hat{L}_{z,0}^3 \chi_{1}^3+\bar{a}_{0}^{(1)} \hat{L}_{z,0}^2 \chi_{1}^2 \chi_{\rm eff}+\bar{a}_{0}^{(1)} \hat{L}_{z,0} \chi_{1}^3-\bar{a}_{0}^{(1)} \chi_{1}^2 \chi_{\rm eff}
    \right.
    \nn \\
    &\left.
    -2 \bar{a}_{2}^{(1)} \hat{L}_{z,0} \chi_{1}-2 \bar{a}_{2}^{(1)} \chi_{\rm eff}+2 \bar{a}_{3}^{(1)}\right)+\frac{c_{1}}{4 \chi_{1}} \left[\bar{a}_{2} \left(-8 \hat{L}_{z,0}^2 \chi_{1}+6 \hat{L}_{z,0} \chi_{\rm eff}+8 \chi_{1}\right)-8 \bar{a}_{3} \hat{L}_{z,0}
    \right.
    \nn \\
    &\left.
    +\left(\hat{L}_{z,0}^2-1\right) \chi_{1}^2 \left(20 \hat{L}_{z,0}^2 \chi_{1}-9 \hat{L}_{z,0} \chi_{\rm eff}-4 \chi_{1}\right)\right]\,,
    \\
    \gamma_{7}^{(\rm osc)} &= -\frac{1235}{48} \chi_{1}\,,
    \qquad
    \gamma_{8}^{(\rm osc)} = -\frac{13}{48} \chi_{1} \left(95 \chi_{1} \hat{L}_{z,0} - 2 \chi_{\rm eff} \right)\,.
\end{align}
\end{widetext}
Lastly, for the renormalization angle $\lambda$ in Eq.~\eqref{eq:lambda-F2}, the non-zero PN coefficients 
are given by
\allowdisplaybreaks[4]
\begin{align}
    \lambda_{1}^{(0)} &= -\frac{4}{3} \chi_{\rm eff} + \frac{2}{3} \bar{\chi}_{Q,0}\,,
    \\
    \lambda_{3}^{(0)} &= 2 \bar{a}_{3} - \frac{4}{3} \chi_{\rm eff} \bar{a}_{2} + \frac{2}{3} \bar{\chi}_{Q,0} \bar{a}_{2} 
    \\
    \lambda_{4}^{(0)} &= -\bar{a}_{2}^2 + \bar{a}_{4} - \frac{1}{3} \left(2\bar{a}_{3} \chi_{\rm eff} + \bar{a}_{3} \bar{\chi}_{Q,0} + \bar{a}_{2} \chi_{\rm eff} \bar{\chi}_{Q,0} \right)\,,
    \\
    \lambda_{2}^{l,(0)} &= 2\bar{a}_{2} + \frac{2}{3} \chi_{\rm eff} \bar{\chi}_{Q,0}\,,
    \\
    \lambda_{7}^{(\rm osc)} &= -\frac{152}{9} \chi_{1}\,,
    \\
    \lambda_{8}^{(\rm osc)} &= -\frac{8}{135} \chi_{1} \left(570 \chi_{1} \hat{L}_{z,0} + 583 \chi_{\rm eff} - 152 \bar{\chi}_{Q,0} \right)\,.
\end{align}

\section{Coefficients of the Waveform Precession Phase}
\label{app:prec-coeffs}

The $(b_{i}, d_{i})$ coefficients appearing in the reduced evolution equation for $\delta\Phi$ in Eq.~\eqref{eq:dPhi-red} are
\allowdisplaybreaks[4]
\begin{align}
    b_{0} &= \hat{L}_{z,0} \left(1 - \hat{L}_{z,0}^{2} \right)\cos^{2}\theta_{N}\,,
    \\
    b_{1} &= \left(1 - 2 \hat{L}_{z,0}^{2} \right) \sqrt{1 - \hat{L}_{z,0}^{2}} \cos\theta_{N} \sin\theta_{N}\,,
    \\
    b_{2} &= \hat{L}_{z,0}\left(1 - \hat{L}_{z,0}^{2}\right) \sin^{2}\theta_{N}\,,
    \\
    d_{0} &= 1 - \hat{L}_{z,0}^{2} \cos^{2} \theta_{N}\,,
    \\
    d_{1} &= \hat{L}_{z,0} \sqrt{1 - \hat{L}_{z,0}^{2}} \sin(2\theta_{N})\,,
    \\
    d_{2} &= \left(1 - \hat{L}_{z,0}^{2} \right) \sin^{2}\theta_{N}\,.
\end{align}
From these, and defining $\zeta = 1 + q \nu_{XY}$, the coefficients appearing the solution in Eq.~\eqref{eq:dPhi-expression} are
\allowdisplaybreaks[4]
\begin{align}
    \Delta_{0} &= d_{1}^{2} + 4 d_{0} d_{2}\,,
    \\
    \Delta_{\pm} &= d_{1}^{2} - 2 d_{2} (-d_{0} + d_{2}) \mp d_{1} \sqrt{\Delta_{0}}\,,
    \\
    e_{\pm} &= \sqrt{\Delta}_{0} \mp d_{1} \pm 2d_{2}\,,
    \\
    n_{\pm} &= \frac{e_{\pm}}{\sqrt{2 \Delta_{\pm}}}\,,
    \qquad
    \beta_{\pm} = \frac{n_{\pm} - 1}{n_{\pm} + 1}\,,
    \\
    \label{eq:hpm-eq}
    h_{\pm} &= \frac{b_{1} d_{2} + b_{2} d_{1}}{\zeta \sqrt{\Delta_{\pm}}} \mp \frac{b_{2} d_{1}^{2} + 2 b_{2} d_{0} d_{2} + b_{1} d_{1} d_{2} - 2 b_{0} d_{2}^{2} }{\zeta \sqrt{\Delta_{0} \Delta_{\pm}}}\,,
    \\
    \label{eq:Nphi-eq}
    {\cal{N}}_{\Phi} &= \frac{-\sqrt{2}b_{2} - h_{+} + h_{-}}{\sqrt{2} \zeta d_{2}}\,,
\end{align}
The discriminants $\Delta_{0}$ and $\Delta_{\pm}$ are all positive definite for any value of $(\beta_{0}, \theta_{N})$. 

For the solutions at linear order in $q$, the coefficients appearing in $\delta\Phi_{0}^{({\cal{B}}_{1})}$ are
\begin{align}
    b_{0}^{(1)} &= \frac{\hat{L}_{z,0}}{\nu_{RP}} \left(1 - \hat{L}_{z,0}^{2}\right) \Bigg[
    \nu_{1} \nu_{RP} 
    \nn \\
    &+ \nu_{SL} \bar{\chi}_{Q,0} \left(\nu_{RP} \hat{L}_{z,0} - \nu_{QP} \sqrt{1-\hat{L}_{z,0}^{2}} \right)\Bigg] \cos^{2}\theta_{N}\,,
    \\
    b_{1}^{(1)} &= \frac{(1 - 2 \hat{L}_{z,0}^{2}}{2\nu_{RP}} \Bigg[\nu_{RP} \sqrt{1 - \hat{L}_{z,0}^{2}} \left(\nu_{1} + \hat{L}_{z,0} \nu_{SL} \bar{\chi}_{Q,0}\right) 
    \nn \\
    &- \left(1 - \hat{L}_{z,0}^{2}\right) \nu_{QP} \nu_{SL} \bar{\chi}_{Q,0} \Bigg] \sin(2\theta_{N})\,,
    \\
    b_{2}^{(1)} &= -b_{0}^{(1)} \tan^{2}(\theta_{N})\,,
\end{align}
The quantities $[{\cal{N}}_{\phi}^{({\cal{B}}_{1})}, h_{\pm}^{({\cal{B}}_{1})}]$ are given by Eqs.~\eqref{eq:hpm-eq}-\eqref{eq:Nphi-eq}, with the replacements $h_{\pm}\rightarrow h_{\pm}^{({\cal{B}}_{1})}$ and $b_{i}\rightarrow b_{i}^{(1)}$ for $i=0,1,2$. For $\delta\Phi^{({\cal{B}}_{2})}_{0}$, the coefficients are
\allowdisplaybreaks
\begin{widetext}
\begin{align}
    c_{1}^{(1)} &= -\frac{1}{4} \left(2 - 3 \hat{L}_{z,0}^{2} + \hat{L}_{z,0}^{2} \cos(2\theta_{N}) \right) {\cal{L}}_{+} \sin(2\theta_{N})\,,
    \\
    c_{2}^{(1)} &= 2 \hat{L}_{z,0} \sqrt{1 - \hat{L}_{z,0}^{2}} {\cal{L}}_{+} \sin^{4}(\theta_{N})\,,
    \\
    c_{3}^{(1)} &= - \left(1 - \hat{L}_{z,0}^{2}\right) {\cal{L}}_{+} \cos\theta_{N} \sin^{3}\theta_{N}\,,
    \\
    s_{0}^{(1)} &= - \left(1 - 2 \hat{L}_{z,0}^{2} + \hat{L}_{z,0}^{2} \cos^{2}\theta_{N} \right) {\cal{L}}_{-} \cos\theta_{N} \sin\theta_{N}\,,
    \\
    s_{1}^{(1)} &= 2 \hat{L}_{z,0} \sqrt{1 - \hat{L}_{z,0}^{2}} {\cal{L}}_{-} \sin^{4}\theta_{N}\,,
    \\
    s_{2}^{(1)} &= - \left(1 - \hat{L}_{z,0}^{2} \right) {\cal{L}}_{-} \cos\theta_{N} \sin^{3}\theta_{N}\,,
    \\
    {\cal{L}}_{\pm} &= \sqrt{\ell_{x}^{2} + \ell_{y}^{2}} \cos\phi_{0} \cos\phi_{1} \pm \sqrt{\dot{\ell}_{x}^{2} + \dot{\ell}_{y}^{2}} \cos\psi_{1}\sin\phi_{0}\,,
    \\
    \phi_{0} &= \phi_{N} - \phi_{L}\,,
    \\
    \phi_{1} &= \tan^{-1}(\ell_{y}/\ell_{x}) - \phi_{N}\,,
    \\
    \psi_{1} &= \tan^{-1}(\dot{\ell}_{y}/\dot{\ell}_{x}) - \phi_{N}\,,
    \\
    \Delta_{1} &= -d_{0}^{2} + d_{1}^{2} + 2d_{0} d_{2} - d_{2}^{2}\,,
    \\
    \kappa &= -2 d_{2} s_{0}^{(1)} + d_{1} s_{1}^{(1)} + 2 d_{0} s_{2}^{(1)}\,,
    \\
    g_{+} &= c_{1}^{(1)} d_{2} \left(2 d_{0}^2 \left(d_{1}-\sqrt{\Delta_{0}}\right)+2 d_{0} d_{2} \left(\sqrt{\Delta_{0}}-4 d_{1}\right)+d_{1} \left(-3d_{1}^2+\sqrt{\Delta_{0}} d_{1}+2 d_{2}^2\right)\right)
    \nn \\
    &+c_{2}^{(1)} \left(-d_{0}^2 \left(d_{1}^2-\sqrt{\Delta_{0}} d_{1}+4d_{2}^2\right)+d_{0} d_{2} \left(d_{1}^2-3 \sqrt{\Delta_{0}} d_{1}+4 d_{2}^2\right)+d_{1}^3 \left(d_{1}-\sqrt{\Delta_{0}}\right)\right)
    \nn \\
    &+c_{3}^{(1)} \left(-2 d_{0}^3 \left(d_{1}-\sqrt{\Delta_{0}}\right)+d_{0}^2 \left(4 d_{1} d_{2}-6 \sqrt{\Delta_{0}}d_{2}\right)+2 d_{0} \left(d_{1}^3-\sqrt{\Delta_{0}} d_{1}^2-3 d_{1} d_{2}^2+2 \sqrt{\Delta_{0}} d_{2}^2\right)+d_{1}^2 d_{2}\left(\sqrt{\Delta_{0}}-d_{1}\right)\right)\,,
   \\
   g_{-} &= -c_{1}^{(1)} d_{2} \left(-2 d_{0}^2 \left(\sqrt{\Delta_{0}}+d_{1}\right)+2 d_{0} d_{2} \left(\sqrt{\Delta_{0}}+4 d_{1}\right)+d_{1}\left(3 d_{1}^2+\sqrt{\Delta_{0}} d_{1}-2 d_{2}^2\right)\right)
   \nn \\
   &-c_{2}^{(1)} \left(d_{0}^2 \left(d_{1}^2+\sqrt{\Delta_{0}} d_{1}+4d_{2}^2\right)-d_{0} d_{2} \left(d_{1}^2+3 \sqrt{\Delta_{0}} d_{1}+4 d_{2}^2\right)-\left(d_{1}^3 \left(\sqrt{\Delta_{0}}+d_{1}\right)\right)\right)
   \nn \\
   &-c_{3}^{(1)} \left(2 d_{0}^3 \left(\sqrt{\Delta_{0}}+d_{1}\right)-2 d_{0}^2 d_{2} \left(3 \sqrt{\Delta_{0}}+2d_{1}\right)-2 d_{0} \left(d_{1}^3+\sqrt{\Delta_{0}} d_{1}^2-3 d_{1} d_{2}^2-2 \sqrt{\Delta_{0}} d_{2}^2\right)+d_{1}^2 d_{2}\left(\sqrt{\Delta_{0}}+d_{1}\right)\right)\,,
    \\
    \sigma_{0} &= 2 d_{0}^3 d_{1} s_{2}^{(1)}-4 d_{0}^3 d_{2} s_{1}^{(1)}-2 d_{0}^2 d_{1} d_{2} s_{0}^{(1)}-4 d_{0}^2 d_{1} d_{2} s_{2}^{(1)}+8 d_{0}^2d_{2}^2 s_{1}^{(1)}-2 d_{0} d_{1}^3 s_{2}^{(1)}+4 d_{0} d_{1}^2 d_{2} s_{1}^{(1)}+4 d_{0} d_{1} d_{2}^2 s_{0}^{(1)}
    \nn \\
    &+2 d_{0} d_{1}d_{2}^2 s_{2}^{(1)}-4 d_{0} d_{2}^3 s_{1}^{(1)}+2 d_{1}^3 d_{2} s_{0}^{(1)}-2 d_{1} d_{2}^3 s_{0}^{(1)}\,,
    \\
    \sigma_{C} &= 2 \left(-d_{0}^2+2 d_{0} d_{2}+d_{1}^2-d_{2}^2\right) \left(2 d_{0} d_{2} s_{2}^{(1)}+d_{1}^2 s_{2}^{(1)}-d_{1} d_{2} s_{1}^{(1)}+2d_{2}^2 s_{0}^{(1)}\right)\,,
   \\
   \sigma_{S}^{(1)} &= 2 d_{0} d_{2} \left(c_{1}^{(1)} \left(2 d_{2} (d_{0}-d_{2})+d_{1}^2\right)+c_{2}^{(1)} d_{1} (d_{0}+d_{2})+c_{3}^{(1)} \left(2 d_{0}^2-2d_{0} d_{2}-d_{1}^2\right)\right)\,,
   \\
   \sigma_{S}^{(2)} &= c_{1}^{(1)} d_{0} d_{1} d_{2}^2+c_{1}^{(1)} d_{1} d_{2}^3+2 c_{2}^{(1)} d_{0}^2 d_{2}^2-2 c_{2}^{(1)} d_{0} d_{2}^3-c_{2}^{(1)} d_{1}^2d_{2}^2-c_{3}^{(1)} d_{0}^2 d_{1} d_{2}+3 c_{3}^{(1)} d_{0} d_{1} d_{2}^2+c_{3}^{(1)} d_{1}^3 d_{2}\,,
   \\
   {\cal{N}}_{\Phi}^{({\cal{B}}_{2})} &= \frac{1}{\sqrt{2} \Delta_{0}^{3/2}\sqrt{\Delta_{-}} \sqrt{\Delta_{+}} \Delta_{1}} \Bigg\{\sqrt{\Delta_{+}} \Bigg[c_{1}^{(1)} d_{2} \left(2 d_{1} \left(d_{0}^2-4 d_{0} d_{2}+d_{2}^2\right)-\sqrt{\Delta_{0}} \left(2 d_{0}(d_{2}-d_{0})+d_{1}^2\right)-3 d_{1}^3\right)
   \nn \\
   &+c_{2}^{(1)} \left(-d_{0}^2 \left(d_{1} \left(\sqrt{\Delta_{0}}+d_{1}\right)+4d_{2}^2\right)+d_{0} d_{2} \left(d_{1}^2+3 \sqrt{\Delta_{0}} d_{1}+4 d_{2}^2\right)+d_{1}^3 \left(\sqrt{\Delta_{0}}+d_{1}\right)\right)\Bigg]
   \nn \\
   &+\sqrt{\Delta_{-}} \Bigg[c_{1}^{(1)} d_{2} \left(2 d_{1} \left(d_{0}^2-4 d_{0}d_{2}+d_{2}^2\right)+\sqrt{\Delta_{0}} \left(2 d_{0} (d_{2}-d_{0})+d_{1}^2\right)-3 d_{1}^3\right)
   \nn \\
   &+c_{2}^{(1)} \left(\sqrt{\Delta_{0}}d_{1} \left(d_{0}^2-3 d_{0} d_{2}-d_{1}^2\right)-d_{0}^2 d_{1}^2-4 d_{0}^2 d_{2}^2+d_{0} d_{1}^2 d_{2}+4 d_{0}d_{2}^3+d_{1}^4\right)\Bigg]
   \nn \\
   &+c_{3}^{(1)} \sqrt{\Delta_{-}} \left(\sqrt{\Delta_{0}} \left(2 d_{0}^3-6 d_{0}^2 d_{2}-2 d_{0} d_{1}^2+4d_{0} d_{2}^2+d_{1}^2 d_{2}\right)+d_{1} \left(-2 d_{0}^3+4 d_{0}^2 d_{2}+2 d_{0} \left(d_{1}^2-3 d_{2}^2\right)-d_{1}^2d_{2}\right)\right)
   \nn \\
   &-c_{3}^{(1)} \sqrt{\Delta_{+}} \left(\sqrt{\Delta_{0}} \left(2 d_{0}^3-6 d_{0}^2 d_{2}-2 d_{0} d_{1}^2+4 d_{0}d_{2}^2+d_{1}^2 d_{2}\right)+d_{1} \left(2 d_{0}^3-4 d_{0}^2 d_{2}-2 d_{0} d_{1}^2+6 d_{0} d_{2}^2+d_{1}^2d_{2}\right)\right)\Bigg\}\,.
\end{align}    
\end{widetext}

For the oscillatory correction in Eq.~\eqref{eq:dPhi-1-osc}, the functions ${\cal{B}}_{1,2}^{C,S}(\bar{\alpha})$ can be decomposed as
\begin{align}
    {\cal{B}}_{1}^{C,S} &= \nu_{SL} \sum_{k=0}^{2} \left[{\cal{H}}^{C,S}_{(k)} \cos(k \bar{\alpha}) + {\cal{K}}^{C,S}_{(k)} \sin(k \bar{\alpha})\right]\,,
    \\
    {\cal{B}}_{2}^{C,S} &= \frac{v \omega_{SL} \omega_{L}}{\omega_{P}} {\cal{I}}(\bar{\alpha}) {\cal{J}}^{C,S}(\alpha)\,,
\end{align}
with
\begin{align}
    {\cal{H}}_{(0)}^{C} &= \frac{1}{2} \hat{L}_{z,0}^{2} \nu_{QP} \bar{\chi}_{P,0} \left(\sin^{2}\theta_{N} - 2 \cos^{2}\theta_{N} \right)\,,
    \\
    {\cal{H}}_{(1)}^{C} &= \frac{\hat{L}_{z,0} (2 \hat{L}_{z,0}^{2} - 1)}{\sqrt{1 - \hat{L}_{z,0}^{2}}} \nu_{QP} \bar{\chi}_{P,0} \cos\theta_{N} \sin\theta_{N} \,,
    \\
    {\cal{H}}_{(2)}^{C} &= \frac{1}{2} \hat{L}_{z,0}^{2} \nu_{QP} \bar{\chi}_{P,0} \sin^{2}\theta_{N}\,,
    \\
    {\cal{K}}_{(1)}^{C} &= \frac{\hat{L}_{z,0}}{\sqrt{1 - \hat{L}_{z,0}^{2}}} \dot{\bar{\chi}}_{P,0} \cos\theta_{N} \sin\theta_{N}\,,
    \\
    {\cal{K}}_{(2)}^{C} &= \frac{1}{2} \dot{\bar{\chi}}_{P,0} \sin^{2}\theta_{N}\,,
    \\
    {\cal{H}}_{(0)}^{S} &= \frac{1}{2} \hat{L}_{z,0}^{2} \nu_{QP} \dot{\bar{\chi}}_{P,0} \left(\sin^{2}\theta_{N} - 2 \cos^{2}\theta_{N} \right)
    \\
    {\cal{H}}_{(1)}^{S} &= \frac{\hat{L}_{z,0} (2 \hat{L}_{z,0}^{2} - 1)}{\sqrt{1 - \hat{L}_{z,0}^{2}}} \nu_{QP} \dot{\bar{\chi}}_{P,0} \cos\theta_{N} \sin\theta_{N}\,,
    \\
    {\cal{H}}_{(2)}^{S} &= \frac{1}{2} \hat{L}_{z,0}^{2} \nu_{QP} \dot{\bar{\chi}}_{P,0} \sin^{2}\theta_{N}\,,
    \\
    {\cal{K}}_{(1)}^{S} &= - \frac{\hat{L}_{z,0}}{\sqrt{1 - \hat{L}_{z,0}^{2}}} \bar{\chi}_{P,0} \cos\theta_{N} \sin\theta_{N}\,,
    \\
    {\cal{K}}_{(2)}^{S} &= - \frac{1}{2} \bar{\chi}_{P,0} \sin^{2}\theta_{N}
    \\
    {\cal{I}}(\bar{\alpha}) &= \left(1 - 2 \hat{L}_{z,0}^{2} + \hat{L}_{z,0}^{2} \cos^{2}\theta_{N}\right)\cos\theta_{N} 
    \nn \\
    &- 2 \hat{L}_{z,0} \sqrt{1 - \hat{L}_{z,0}^{2}} \sin^{3}\theta_{N} \cos\bar{\alpha} 
    \nn \\
    &+ \left(1 + \hat{L}_{z,0}^{2} \right) \cos\theta_{N} \sin\theta_{N} \cos^{2}\theta_{N}\,,
    \\
    {\cal{J}}^{S}(\alpha) &= \dot{\bar{\chi}}_{P,0} \cos\theta_{N} 
    \nn \\
    &+ \left[S_{x}^{(a)} \cos\phi_{N} + S_{y}^{(a)} \sin\phi_{N} \right]\sin\theta_{N} \sin\alpha 
    \nn \\
    &+ \left[C_{x}^{(a)} \cos\phi_{N} + C_{y}^{(a)} \sin\phi_{N} \right] \sin\theta_{N} \cos\alpha\,,
    \\
    {\cal{J}}^{C}(\alpha) &= \bar{\chi}_{P,0} \cos\theta_{N} 
    \nn \\
    &+ \left[C_{x}^{(s)} \cos\phi_{N} + C_{y}^{(s)} \sin\phi_{N} \right]\sin\theta_{N} \sin\alpha 
    \nn \\
    &+ \left[S_{x}^{(s)} \cos\phi_{N} + S_{y}^{(s)} \sin\phi_{N} \right] \sin\theta_{N} \cos\alpha\,,
    \\
    C_{x,y}^{(a)} &= C_{x,y}^{(+)} - C_{x,y}^{(-)}\,,
    \qquad
    C_{x,y}^{(s)} = C_{x,y}^{(+)} + C_{x,y}^{(-)}\,,
    \\
    S_{x,y}^{(a)} &= S_{x,y}^{(+)} - S_{x,y}^{(-)}\,,
    \qquad
    S_{x,y}^{(s)} = S_{x,y}^{(+)} + S_{x,y}^{(-)}\,,
\end{align}
where $S_{x,y}^{(\pm)}$ and $C_{x,y}^{(\pm)}$ are given in Eqs.~\eqref{eq:CS-coeffs-1}-\eqref{eq:CS-coeffs-2}.

\section{Obtaining $\hat{L}(q)$ through the Renormalization Group method}
\label{app:RG}
In addition to the MSA technique used to integrate $\hat{L}(q)$ as discussed in Sec.~\ref{sec:L1}, we now employ \textit{Renormalization group} (RG) method to obtain $\hat{L}(q)$. The RG method is a perturbative approach, and particularly useful to study nonlinear equations~\cite{PhysRevLett.73.1311}. For example, in order to solve an equation of the following form
\begin{equation}\label{eq:nonlinear_Oscilator}
    \ddot{x}+\omega x=\epsilon x^3,
\end{equation}
the RG method can be useful. Note that in this example, the RHS contains a nonlinear perturbation (proportional to $x^3$), and $\epsilon$ is the perturbation parameter. Within a perturbative scheme, the zeroth order solution acts as a source of perturbation for the first order, and it introduces a diverging feature. In the present context, it turns out that $\hat{L}$ can also be written as Eq.~\eqref{eq:nonlinear_Oscilator}, and the RG method can be implemented. Our aim is to compare MSA and RG methods, and understand how well these analytical techniques match with the numerical estimation. 

In order to employ RG method in the present paper, we are interested to write $\hat{L}$ as in Eq.~\eqref{eq:nonlinear_Oscilator}. By using Eqs.~\eqref{eq:prec-1-emri})-\eqref{eq:OmegaL1}, and ignoring terms $\sim$ $\mathcal{O}(q^2)$, we arrive at the following expression:
\allowdisplaybreaks[4]
\begin{widetext}
\begin{eqnarray}
\dfrac{d^2 \hat{L}_{ x,y}}{d\tau^2} +\omega^2_L \hat{L}_{x,y} &=&q \Big\{-2 \Big[-\omega^{(1)}_L \omega_L +v \omega_{ SL} \omega_L \chi_{2z}\Big]\hat{L}_{x,y}+v \omega_{ SL} \omega_L\hat{L}_{z}\chi_{2x,y} \mp v\omega_{SL} \omega_{SJ}\hat{L}_{y,x}\Big\}\label{eq:d2Lx1}.
\end{eqnarray}
\end{widetext}
Here we have ignored the z-component of $\hat{L}$ as our interest is to address the artificial resonance or divergence, which appears from the x and y components of $\hat{L}$. We can now seek a perturbative solution of $\hat{L}_x$ $\hat{L}_{x,y}=\hat{L}^{(0)}_{x,y}+q L^{(1)}_{x,y}$, and arrive at the following expression
%
\begin{widetext}
\begin{eqnarray}\label{eq:d2LX}
\dfrac{d^2 \hat{L}^{(1)}_{x,y}}{d\tau^2} +\omega^2_L \hat{L}^{(1)}_{x,y} &=& \Big\{-2 \Big[-\omega^{(1)}_L \omega_L +v \omega_{ SL} \omega_L \chi^{(0)}_{2z}\Big]\hat{L}^{(0)}_{x,y}+v \omega_{ SL}\omega_L \hat{L}^{(0)}_{z}\chi^{(0)}_{2x,y} \mp v\omega_{SL} \omega_{SJ}\hat{L}^{(0)}_{y,x}\Big\}.
\end{eqnarray}
\end{widetext}
From this point, we will only focus on the x-component, whereas the y-component can be derived similarly. To proceed further, we will adopt the notion used in Sec.~\ref{sec:prec}, and obtain $\chi^{(0)}_{2x}$, $L^{(0)}_{x,y}$ accordingly. Let us first introduce the following expressions:
\begin{eqnarray}
X=\sin\beta_0\cos\Psi_0,\quad Y=-\sin\beta_0\cos\Psi_0, \quad Z=\cos\beta_0,\nonumber\\
\end{eqnarray}
where we have dropped the $\tilde{\tau}$ term in bracket. However, as can be understood from the discussion in Sec.~\ref{sec:prec}, the above quantities only change over the radiation reaction timescale. With this substitution,  
we obtain $\hat{P}$ and $\hat{Q}$ as follows:
%
\begin{eqnarray}    \hat{P}&=&\Big(\sin(\alpha+\Psi_0),-\cos(\alpha+\Psi_0),0\Big), \nonumber \\ \hat{Q}&=&\Big(\cos(\alpha+\Psi_0),\sin(\alpha+\Psi_0),0\Big).
\end{eqnarray}
Therefore, the expression for $\hat{L}^{(0)}_{x,y}$ has become:
\begin{equation}
    \hat{L}^{(0)}_{x}(\tau)=\sin\beta_0 \cos(\alpha+\Psi_0), \hat{L}^{(0)}_{y}(\tau)=\sin\beta_0 \sin(\alpha+\Psi_0),
    \label{eq:Lx0_Ly0_RG}
\end{equation}
where the values of $\Psi_0$ and $\beta_0$ can be fixed from the initial conditions. With these implementation, we now write down the expression of $\chi^{(0)}_{2x}$ and $\chi^{(0)}_{2z}$ as follows:
%
%
\begin{equation}
\chi^{(0)}_{2x} =\chi_{2,P}(\hat{P}\cdot \hat{x})+\chi_{2,Q}(\hat{Q}\cdot \hat{x})=\chi_{Q,0}\cos(\alpha+\Psi_0)+\mathcal{F}_x
\end{equation}
where 
    \begin{eqnarray}
\mathcal{F}_x &=& \mathcal{F}_{1x}\cos[\delta^{+}+\Psi_0]+ \mathcal{F}_{2x}\cos[\delta^{-}+\Psi_0]+ \nonumber\\
&&
\mathcal{F}_{3x}\sin[\delta^{+}+\Psi_0]+ \mathcal{F}_{4x}\sin[\delta^{-}+\Psi_0], 
\label{eq:Fx}
\end{eqnarray}
with $\delta^{+}=(\omega_L+\omega_P)\tau$, and $\delta^{-}=(\omega_L-\omega_P)\tau$. The expressions for $\mathcal{F}_{1x}$, $\mathcal{F}_{2x}$, $\mathcal{F}_{3x}$ and $\mathcal{F}_{4x}$ are given by
\begin{eqnarray}
\mathcal{F}_{1x}&=&-(1/2)(\nu_{Q}+1)\dot{\chi}_{P,0}, \nonumber \\ \mathcal{F}_{2x} &=& -(1/2)(\nu_{Q}-1)\dot{\chi}_{P,0}, \nonumber \\
\mathcal{F}_{3x}&=&(1/2)(\nu_Q+1)\chi_{P,0}, \nonumber \\ \mathcal{F}_{4x} &=& -(1/2)(\nu_Q-1)\chi_{P,0}.
\label{eq:F1_F4}
\end{eqnarray}
To expand Eq.~\eqref{eq:d2LX}, the other quantity of particular interest is $\chi^{(0)}_{2z}$. By using Eq.~\eqref{eq:chi2-co-p} and Eq.~\eqref{eq:chiJ-sol}, we arrive at
\begin{widetext}
\begin{eqnarray}
\chi^{(0)}_{2z} \hat{L}^{(0)}_{ x} &=& \chi_{J,0}\hat{L}^{(0)}_{x}+\nu_{R}\Big[-\chi_{P,0}\sin(\gamma_P)+\dot{\chi}_{P,0}\cos(\gamma_P)\Big]\hat{L}^{(0)}_{ x}, \nonumber \\
 &=& \chi_{J,0}\hat{L}^{(0)}_{x}+\nu_{R} \sin\beta_0\Big[-\chi_{P,0}\sin(\gamma_P )+\dot{\chi}_{P,0}\cos(\omega_P \tau)\Big] \cos(\alpha+\Psi_0), \nonumber \\
 &=& \chi_{J,0} \hat{L}^{(0)}_{x}+\mathcal{G}_x,
\end{eqnarray}
\end{widetext}
where 
\begin{eqnarray}
\mathcal{G}_x &=&\mathcal{G}_{1x} \cos[\delta^{+}+\Psi_0]+\mathcal{G}_{2x} \cos[\delta^{-}+\Psi_0]\nonumber \\
&&+\mathcal{G}_{3x} \sin[\delta^{+}+\Psi_0]+\mathcal{G}_{4x} \sin[\delta^{-}+\Psi_0],
\end{eqnarray}
and
\begin{eqnarray}
\mathcal{G}_{1x}&=&\mathcal{G}_{2x}=\dfrac{\nu_{R} \sin\beta_0\dot{\chi}_{P,0}}{2},   \nonumber \\ \mathcal{G}_{3x}&=&-\mathcal{G}_{4x}= -\dfrac{\nu_{R} \sin\beta_0\chi_{P,0}}{2}\,.
\label{eq:G1_G2}
\end{eqnarray}
Therefore, the final expression reads 
\begin{widetext}
\begin{eqnarray}
      \dfrac{d^2 \hat{L}^{(1)}_{x}}{d\tau^2} +\omega^2_L \hat{L}^{(1)}_{x} &=& -2 \Big[-\omega^{(1)}_L \omega_L +v \omega_{ SL} \omega_L \chi^{(0)}_{2z}\Big]\hat{L}^{(0)}_{x}+v \omega_{ SL} \omega_L\hat{L}^{(0)}_{z}\chi^{(0)}_{2x} - v\omega_{SL} \omega_{SJ}\hat{L}^{(0)}_{y}, \nonumber \\
     & =& -2 \Big[-\omega^{(1)}_L \omega_L +v \omega_{ SL} \omega_L \chi_{J,0}\Big]\hat{L}^{(0)}_{x}+v \omega_{ SL} \omega_L\hat{L}^{(0)}_{z}\chi_{Q,0}\cos (\omega_L \tau+\Psi_0) - v\omega_{SL} \omega_{SJ}\hat{L}^{(0)}_{y} \nonumber \\
     &&\hspace{9cm} + v\omega_L \omega_{SL}\hat{L}^{(0)}_{z}\mathcal{F}_x-2v \omega_L \omega_{SL} \mathcal{G}_x, \nonumber \\
     & =& -2 \Big[-\omega^{(1)}_L \omega_L +v \omega_{ SL} \omega_L \chi_{J,0}\Big]\hat{L}^{(0)}_{x}+v \omega_{ SL} \omega_L\cos\beta_0\chi_{Q,0}\cos (\omega_L \tau+\Psi_0) - v\omega_{SL} \omega_{SJ}\sin\beta_0 \sin(\omega_L \tau+\Psi_0)\nonumber \\
      && \hspace{9cm}+ v\omega_L \omega_{SL}\cos\beta_0\mathcal{F}_x-2v \omega_L \omega_{SL} \mathcal{G}_x.
\end{eqnarray}
\end{widetext}
%
In the last line of the above expression, we have used $\hat{L}^{(0)}_{z}=\cos\beta_0$, and expressed $\hat{L}^{(0)}_{y}$ from Eq.~\eqref{eq:Lx0_Ly0_RG}. Before solving the above equation using RG method, we can clean it a bit by introducing the following notations, along with setting $\chi_{Q,0}=\nu_R \chi_{J,0}$ (as discussed at the end of Sec.~\ref{sec:chi20}): 
\begin{eqnarray}
\label{eq:ABCD}
\mathcal{A}&=&-2(-\omega^{(1)}_L \omega_L +v \omega_{ SL} \omega_L \chi_{J,0}), \nonumber \\
\mathcal{B}&=&v \omega_{ SL} \omega_L\cos\beta_0\nu_R \chi_{J,0} \nonumber \\
\mathcal{C}&=&-v\omega_{SL} \omega_{SJ} \sin\beta_0,\nonumber \\
\mathcal{D}_1&=&v\omega_L \omega_{SL}\cos \beta_0, \nonumber \\
\mathcal{D}_2&=&2v \omega_L \omega_{SL} \mathcal{G}_x.
\end{eqnarray}
With these expressions, we arrive at
\begin{eqnarray}
      \dfrac{d^2 \hat{L}^{(1)}_{x}}{d\tau^2} +\omega^2_L \hat{L}^{(1)}_{x}= \mathcal{A} \hat{L}^{(0)}_{x}+\mathcal{B}\cos(\omega_L \tau+\Psi_0)+\nonumber \\
      \mathcal{C}\sin(\omega_L \tau+\Psi_0)+(\mathcal{D}_1 \mathcal{F}_x-\mathcal{D}_2 \mathcal{G}_x).  
      \label{eq:D2LXRG}
      \end{eqnarray}
%
The perturbations that lead to diverging solutions are given by the first three terms, whereas the last term ($\mathcal{D}_1 \mathcal{F}_x-\mathcal{D}_2 \mathcal{G}_x$) contains $\delta^{\pm}$ which provide regular solution. Therefore, our goal is to renormalize the first three terms of the above expression.

With this equation at hand, we are now in a position to employ RG method. By following the standard convention to employ RG techniques in literature~\cite{RG.2001}, we require the leading order solution to be $A_{q}(\tau)\cos(\omega_L \tau+\Psi_{q}(\tau))$. With this, the final solution, truncated at the first order, can be written as~\cite{RG.2001}:
\begin{equation}\label{eq:LXRG}
   \hat{L}_{x}(\tau) = A_{q}(\tau)\cos(\omega_L \tau+\Psi_{q}(\tau))+q \tilde{y}_1(\tau,A_{q},\Psi_{q})+q \mathcal{R}_{x},
\end{equation}
in which $A_{q}(\tau)$ and $\Psi_{q}(\tau)$ are to be expanded in the perturbation parameter, i.e., mass ratio $q$. 
%
%
In the above equation, $\tilde{y}_1$ is obtained by obtaining the Fourier transform of the perturbation term, i.e., the entire RHS of Eq.~\eqref{eq:d2LX}. To be precise, $\tilde{y}_1$ is the non-secular part of this Fourier transformation and therefore, gives a regular solution. In order to achieve this, we re-introduce the long timescale $\tilde{\tau}=q\tau$, and use a near-identity transformation on $A_{q}$ and $\Psi_{q}$:
\begin{eqnarray}
A_{q}(\tau)=\tilde{A_q}(\tilde{\tau})+q \alpha(\tau,\tilde{A_q}), \Psi_{q}(\tau)=\tilde{\Psi_q}(\tilde{\tau})+q \beta(\tau,\tilde{A_q}).  \nonumber \\ 
\end{eqnarray}
In a naive sense, the values of $\tilde{A_q}$, $\tilde{\Psi_q}$, $\alpha$ and $\beta$ will be fixed by ensuring that all the secular terms are eliminated. 
The other quantity in Eq.~\eqref{eq:LXRG} that is yet to be defined is given by $\mathcal{R}_{x}$. It appears due to ($\mathcal{D}_1 \mathcal{F}_x-\mathcal{D}_2 \mathcal{G}_x$) in Eq.~\eqref{eq:D2LXRG}, and is given by:
\begin{widetext}
\begin{eqnarray}
\mathcal{R}_{x}&=&\mathcal{R}_0 \cos(\omega_L \tau +\mathcal{R}_1)+v \omega_{L}\omega_{SL} \cos\beta_0\Big[\mathcal{F}_{1x}\dfrac{\cos(\delta^{+}+\Psi_0)}{(\omega_L+\omega_P)^2-\omega^2_L}+\mathcal{F}_{2x}\dfrac{\cos(\delta^{-}+\Psi_0)}{(\omega_L-\omega_P)^2-\omega^2_L}+\mathcal{F}_{3x}\dfrac{\sin(\delta^{+}+\Psi_0)}{(\omega_L+\omega_P)^2-\omega^2_L}\nonumber \\
&+&\mathcal{F}_{4x}\dfrac{\sin(\delta^{-}+\Psi_0)}{(\omega_L-\omega_P)^2-\omega^2_L}\Big]-2 v \omega_L \omega_{SL}\Big[\mathcal{G}_{1x}\dfrac{\cos(\delta^{+}+\Psi_0)}{(\omega_L+\omega_P)^2-\omega^2_L}+\mathcal{G}_{2x}\dfrac{\cos(\delta^{-}+\Psi_0)}{(\omega_L-\omega_P)^2-\omega^2_L}+\mathcal{G}_{3x}\dfrac{\sin(\delta^{+}+\Psi_0)}{(\omega_L+\omega_P)^2-\omega^2_L} \nonumber \\
&& \hspace{11cm} +  \mathcal{G}_{4x}\dfrac{\sin(\delta^{-}+\Psi_0)}{(\omega_L-\omega_P)^2-\omega^2_L}\Big].
\end{eqnarray}
\end{widetext}
In the above expression, the first term on the RHS is homogeneous part of the solution. Here $\mathcal{R}_0$ and $\mathcal{R}_1$ are free parameters which can be determined from the initial conditions. Finally, we arrive at 
\begin{eqnarray}
 A_{q}(\tau)&=&A_{q}(0)-\dfrac{q\tau}{2\omega_L}\mathcal{C}, \tilde{y_1}=0, \nonumber \\
 \Psi_{q}(\tau)&=&\Psi_{q}(0)-\dfrac{q\tau}{2\omega_L A_{q}(0)}\Big(\mathcal{B}+A_{q}(0)\mathcal{A}\Big). 
\end{eqnarray}
It is easy to note that $A_{q}(0)=\sin\beta_0$, and $\Psi_{q}(0)=\Psi_0$. Therefore, we write down the expression for $\hat{L}_{x}(\tau)$ as
\begin{widetext}

\begin{eqnarray}
    \label{eq:Lx-RG}
    \hat{L}_{x}(\tau)=\Big(\sin\beta_0-\dfrac{q\tau}{2\omega_L}\mathcal{C}\Big)\cos\Big(\omega_L \tau+\Psi_0-\dfrac{q\tau}{2\omega_L A_{q}(0)}(\mathcal{B}+\sin\beta_0\mathcal{A})\Big)+q \mathcal{R}_x. \nonumber\\
\end{eqnarray}
\end{widetext}
By setting $q=0$, we end up with $\hat{L}^{(0)}_x(\tau)=\sin\beta_0 \cos(\omega_L \tau+\Psi_0)$, which is consistent with the previously obtained result in Eq.~\eqref{eq:Lx0_Ly0_RG}. 

\begin{figure}[hbt!]
    \centering
\includegraphics[width=\columnwidth]{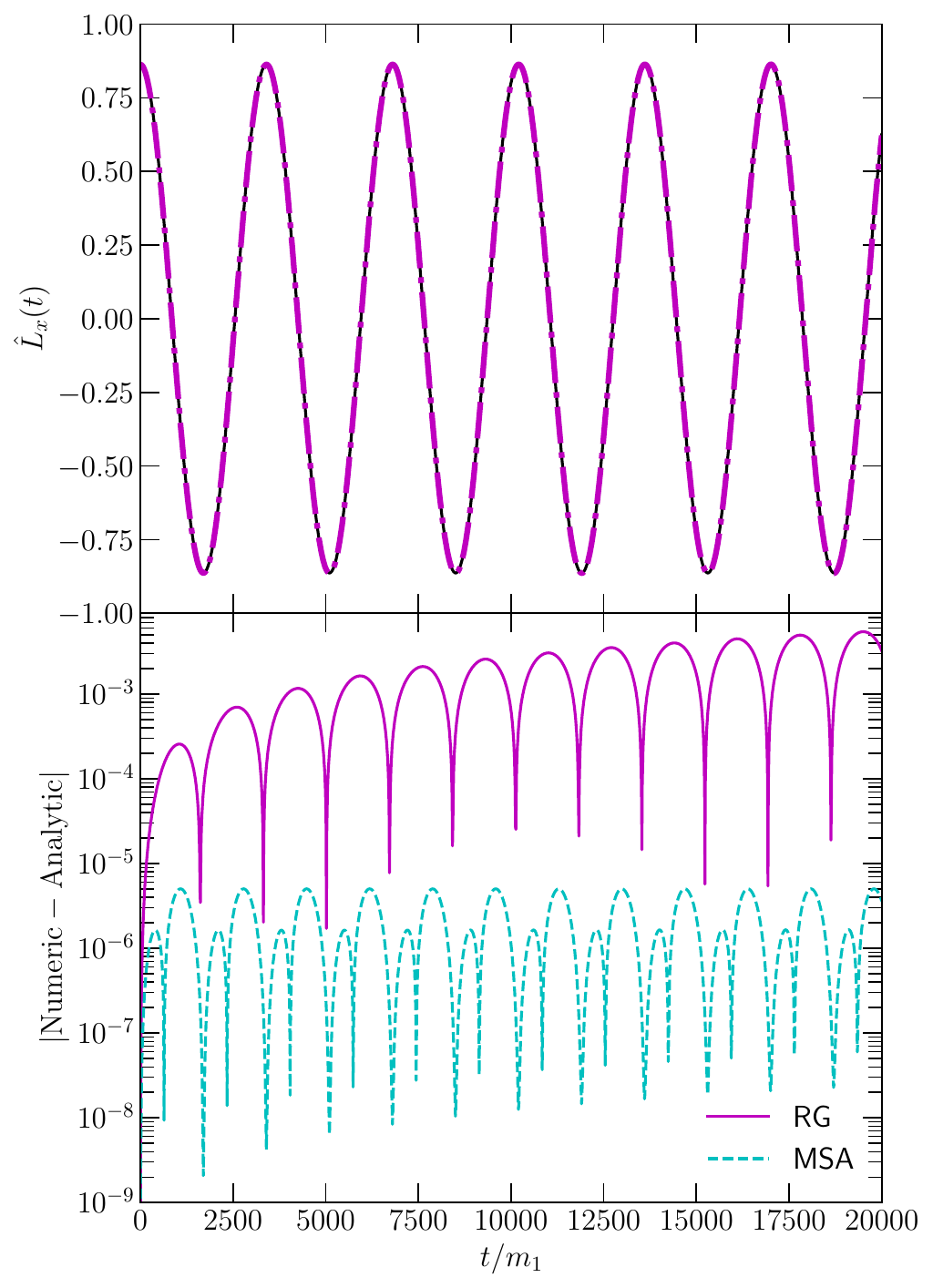}
    \caption{Top: Comparison of the RG solution (purple dot-dashed line) in Eq.~\eqref{eq:Lx-RG} for $\hat{L}_{x}(t)$ to the numerical solution (black solid line) of the PN precession equations for the same EMRI considered in Figs.~\ref{fig:chi2-comp}-\ref{fig:L-comp} Bottom: Difference between the numerical and analytic RG solutions (purple solid line). The same quantity, but with the MSA solution (cyan dashed line), is provided for comparison between the two techniques. Generally, the RG performs worse due to the linearly growing component of the amplitude in Eq.~\eqref{eq:Lx-RG}, which can be corrected by proceeded to higher order in the RG method.}     \label{Fig:Compare_RG_MSA}
\end{figure}

We can now compare the results from MSA and RG methods for a given set of parameters. This is shown in Fig.~\ref{Fig:Compare_RG_MSA} for a mass ratio of $q=10^{-4}$ and representative binary parameters. 
%

\bibliography{refs}

\begin{thebibliography}{104}%
\makeatletter
\providecommand \@ifxundefined [1]{%
 \@ifx{#1\undefined}
}%
\providecommand \@ifnum [1]{%
 \ifnum #1\expandafter \@firstoftwo
 \else \expandafter \@secondoftwo
 \fi
}%
\providecommand \@ifx [1]{%
 \ifx #1\expandafter \@firstoftwo
 \else \expandafter \@secondoftwo
 \fi
}%
\providecommand \natexlab [1]{#1}%
\providecommand \enquote  [1]{``#1''}%
\providecommand \bibnamefont  [1]{#1}%
\providecommand \bibfnamefont [1]{#1}%
\providecommand \citenamefont [1]{#1}%
\providecommand \href@noop [0]{\@secondoftwo}%
\providecommand \href [0]{\begingroup \@sanitize@url \@href}%
\providecommand \@href[1]{\@@startlink{#1}\@@href}%
\providecommand \@@href[1]{\endgroup#1\@@endlink}%
\providecommand \@sanitize@url [0]{\catcode `\\12\catcode `\$12\catcode
  `\&12\catcode `\#12\catcode `\^12\catcode `\_12\catcode `\%12\relax}%
\providecommand \@@startlink[1]{}%
\providecommand \@@endlink[0]{}%
\providecommand \url  [0]{\begingroup\@sanitize@url \@url }%
\providecommand \@url [1]{\endgroup\@href {#1}{\urlprefix }}%
\providecommand \urlprefix  [0]{URL }%
\providecommand \Eprint [0]{\href }%
\providecommand \doibase [0]{http://dx.doi.org/}%
\providecommand \selectlanguage [0]{\@gobble}%
\providecommand \bibinfo  [0]{\@secondoftwo}%
\providecommand \bibfield  [0]{\@secondoftwo}%
\providecommand \translation [1]{[#1]}%
\providecommand \BibitemOpen [0]{}%
\providecommand \bibitemStop [0]{}%
\providecommand \bibitemNoStop [0]{.\EOS\space}%
\providecommand \EOS [0]{\spacefactor3000\relax}%
\providecommand \BibitemShut  [1]{\csname bibitem#1\endcsname}%
\let\auto@bib@innerbib\@empty
\bibitem [{\citenamefont {Amaro-Seoane}\ \emph {et~al.}(2017)\citenamefont
  {Amaro-Seoane} \emph {et~al.}}]{LISA:2017pwj}%
  \BibitemOpen
  \bibfield  {author} {\bibinfo {author} {\bibfnamefont {P.}~\bibnamefont
  {Amaro-Seoane}} \emph {et~al.} (\bibinfo {collaboration} {LISA}),\
  }\href@noop {} {\  (\bibinfo {year} {2017})},\ \Eprint
  {http://arxiv.org/abs/1702.00786} {arXiv:1702.00786 [astro-ph.IM]}
  \BibitemShut {NoStop}%
\bibitem [{\citenamefont {Harms}\ \emph {et~al.}(2021)\citenamefont {Harms}
  \emph {et~al.}}]{LGWA:2020mma}%
  \BibitemOpen
  \bibfield  {author} {\bibinfo {author} {\bibfnamefont {J.}~\bibnamefont
  {Harms}} \emph {et~al.} (\bibinfo {collaboration} {LGWA}),\ }\href {\doibase
  10.3847/1538-4357/abe5a7} {\bibfield  {journal} {\bibinfo  {journal}
  {Astrophys. J.}\ }\textbf {\bibinfo {volume} {910}},\ \bibinfo {pages} {1}
  (\bibinfo {year} {2021})},\ \Eprint {http://arxiv.org/abs/2010.13726}
  {arXiv:2010.13726 [gr-qc]} \BibitemShut {NoStop}%
\bibitem [{\citenamefont {Miller}\ \emph {et~al.}(2021)\citenamefont {Miller},
  \citenamefont {Clesse}, \citenamefont {De~Lillo}, \citenamefont {Bruno},
  \citenamefont {Depasse},\ and\ \citenamefont {Tanasijczuk}}]{Miller:2020kmv}%
  \BibitemOpen
  \bibfield  {author} {\bibinfo {author} {\bibfnamefont {A.~L.}\ \bibnamefont
  {Miller}}, \bibinfo {author} {\bibfnamefont {S.}~\bibnamefont {Clesse}},
  \bibinfo {author} {\bibfnamefont {F.}~\bibnamefont {De~Lillo}}, \bibinfo
  {author} {\bibfnamefont {G.}~\bibnamefont {Bruno}}, \bibinfo {author}
  {\bibfnamefont {A.}~\bibnamefont {Depasse}}, \ and\ \bibinfo {author}
  {\bibfnamefont {A.}~\bibnamefont {Tanasijczuk}},\ }\href {\doibase
  10.1016/j.dark.2021.100836} {\bibfield  {journal} {\bibinfo  {journal} {Phys.
  Dark Univ.}\ }\textbf {\bibinfo {volume} {32}},\ \bibinfo {pages} {100836}
  (\bibinfo {year} {2021})},\ \Eprint {http://arxiv.org/abs/2012.12983}
  {arXiv:2012.12983 [astro-ph.HE]} \BibitemShut {NoStop}%
\bibitem [{\citenamefont {Barsanti}\ \emph
  {et~al.}(2022{\natexlab{a}})\citenamefont {Barsanti}, \citenamefont
  {De~Luca}, \citenamefont {Maselli},\ and\ \citenamefont
  {Pani}}]{Barsanti:2021ydd}%
  \BibitemOpen
  \bibfield  {author} {\bibinfo {author} {\bibfnamefont {S.}~\bibnamefont
  {Barsanti}}, \bibinfo {author} {\bibfnamefont {V.}~\bibnamefont {De~Luca}},
  \bibinfo {author} {\bibfnamefont {A.}~\bibnamefont {Maselli}}, \ and\
  \bibinfo {author} {\bibfnamefont {P.}~\bibnamefont {Pani}},\ }\href {\doibase
  10.1103/PhysRevLett.128.111104} {\bibfield  {journal} {\bibinfo  {journal}
  {Phys. Rev. Lett.}\ }\textbf {\bibinfo {volume} {128}},\ \bibinfo {pages}
  {111104} (\bibinfo {year} {2022}{\natexlab{a}})},\ \Eprint
  {http://arxiv.org/abs/2109.02170} {arXiv:2109.02170 [gr-qc]} \BibitemShut
  {NoStop}%
\bibitem [{\citenamefont {Babak}\ \emph {et~al.}(2017)\citenamefont {Babak},
  \citenamefont {Gair}, \citenamefont {Sesana}, \citenamefont {Barausse},
  \citenamefont {Sopuerta}, \citenamefont {Berry}, \citenamefont {Berti},
  \citenamefont {Amaro-Seoane}, \citenamefont {Petiteau},\ and\ \citenamefont
  {Klein}}]{Babak:2017tow}%
  \BibitemOpen
  \bibfield  {author} {\bibinfo {author} {\bibfnamefont {S.}~\bibnamefont
  {Babak}}, \bibinfo {author} {\bibfnamefont {J.}~\bibnamefont {Gair}},
  \bibinfo {author} {\bibfnamefont {A.}~\bibnamefont {Sesana}}, \bibinfo
  {author} {\bibfnamefont {E.}~\bibnamefont {Barausse}}, \bibinfo {author}
  {\bibfnamefont {C.~F.}\ \bibnamefont {Sopuerta}}, \bibinfo {author}
  {\bibfnamefont {C.~P.~L.}\ \bibnamefont {Berry}}, \bibinfo {author}
  {\bibfnamefont {E.}~\bibnamefont {Berti}}, \bibinfo {author} {\bibfnamefont
  {P.}~\bibnamefont {Amaro-Seoane}}, \bibinfo {author} {\bibfnamefont
  {A.}~\bibnamefont {Petiteau}}, \ and\ \bibinfo {author} {\bibfnamefont
  {A.}~\bibnamefont {Klein}},\ }\href {\doibase 10.1103/PhysRevD.95.103012}
  {\bibfield  {journal} {\bibinfo  {journal} {Phys. Rev. D}\ }\textbf {\bibinfo
  {volume} {95}},\ \bibinfo {pages} {103012} (\bibinfo {year} {2017})},\
  \Eprint {http://arxiv.org/abs/1703.09722} {arXiv:1703.09722 [gr-qc]}
  \BibitemShut {NoStop}%
\bibitem [{\citenamefont {Barausse}\ \emph {et~al.}(2020)\citenamefont
  {Barausse} \emph {et~al.}}]{Barausse:2020rsu}%
  \BibitemOpen
  \bibfield  {author} {\bibinfo {author} {\bibfnamefont {E.}~\bibnamefont
  {Barausse}} \emph {et~al.},\ }\href {\doibase 10.1007/s10714-020-02691-1}
  {\bibfield  {journal} {\bibinfo  {journal} {Gen. Rel. Grav.}\ }\textbf
  {\bibinfo {volume} {52}},\ \bibinfo {pages} {81} (\bibinfo {year} {2020})},\
  \Eprint {http://arxiv.org/abs/2001.09793} {arXiv:2001.09793 [gr-qc]}
  \BibitemShut {NoStop}%
\bibitem [{\citenamefont {Arun}\ \emph {et~al.}(2022)\citenamefont {Arun} \emph
  {et~al.}}]{LISA:2022kgy}%
  \BibitemOpen
  \bibfield  {author} {\bibinfo {author} {\bibfnamefont {K.~G.}\ \bibnamefont
  {Arun}} \emph {et~al.} (\bibinfo {collaboration} {LISA}),\ }\href {\doibase
  10.1007/s41114-022-00036-9} {\bibfield  {journal} {\bibinfo  {journal}
  {Living Rev. Rel.}\ }\textbf {\bibinfo {volume} {25}},\ \bibinfo {pages} {4}
  (\bibinfo {year} {2022})},\ \Eprint {http://arxiv.org/abs/2205.01597}
  {arXiv:2205.01597 [gr-qc]} \BibitemShut {NoStop}%
\bibitem [{\citenamefont {Baibhav}\ \emph {et~al.}(2021)\citenamefont {Baibhav}
  \emph {et~al.}}]{Baibhav:2019rsa}%
  \BibitemOpen
  \bibfield  {author} {\bibinfo {author} {\bibfnamefont {V.}~\bibnamefont
  {Baibhav}} \emph {et~al.},\ }\href {\doibase 10.1007/s10686-021-09741-9}
  {\bibfield  {journal} {\bibinfo  {journal} {Exper. Astron.}\ }\textbf
  {\bibinfo {volume} {51}},\ \bibinfo {pages} {1385} (\bibinfo {year}
  {2021})},\ \Eprint {http://arxiv.org/abs/1908.11390} {arXiv:1908.11390
  [astro-ph.HE]} \BibitemShut {NoStop}%
\bibitem [{\citenamefont {Cardoso}\ \emph {et~al.}(2011)\citenamefont
  {Cardoso}, \citenamefont {Chakrabarti}, \citenamefont {Pani}, \citenamefont
  {Berti},\ and\ \citenamefont {Gualtieri}}]{Cardoso:2011xi}%
  \BibitemOpen
  \bibfield  {author} {\bibinfo {author} {\bibfnamefont {V.}~\bibnamefont
  {Cardoso}}, \bibinfo {author} {\bibfnamefont {S.}~\bibnamefont
  {Chakrabarti}}, \bibinfo {author} {\bibfnamefont {P.}~\bibnamefont {Pani}},
  \bibinfo {author} {\bibfnamefont {E.}~\bibnamefont {Berti}}, \ and\ \bibinfo
  {author} {\bibfnamefont {L.}~\bibnamefont {Gualtieri}},\ }\href {\doibase
  10.1103/PhysRevLett.107.241101} {\bibfield  {journal} {\bibinfo  {journal}
  {Phys. Rev. Lett.}\ }\textbf {\bibinfo {volume} {107}},\ \bibinfo {pages}
  {241101} (\bibinfo {year} {2011})},\ \Eprint {http://arxiv.org/abs/1109.6021}
  {arXiv:1109.6021 [gr-qc]} \BibitemShut {NoStop}%
\bibitem [{\citenamefont {Yunes}\ \emph {et~al.}(2012)\citenamefont {Yunes},
  \citenamefont {Pani},\ and\ \citenamefont {Cardoso}}]{Yunes:2011aa}%
  \BibitemOpen
  \bibfield  {author} {\bibinfo {author} {\bibfnamefont {N.}~\bibnamefont
  {Yunes}}, \bibinfo {author} {\bibfnamefont {P.}~\bibnamefont {Pani}}, \ and\
  \bibinfo {author} {\bibfnamefont {V.}~\bibnamefont {Cardoso}},\ }\href
  {\doibase 10.1103/PhysRevD.85.102003} {\bibfield  {journal} {\bibinfo
  {journal} {Phys. Rev. D}\ }\textbf {\bibinfo {volume} {85}},\ \bibinfo
  {pages} {102003} (\bibinfo {year} {2012})},\ \Eprint
  {http://arxiv.org/abs/1112.3351} {arXiv:1112.3351 [gr-qc]} \BibitemShut
  {NoStop}%
\bibitem [{\citenamefont {Pani}\ \emph {et~al.}(2011)\citenamefont {Pani},
  \citenamefont {Cardoso},\ and\ \citenamefont {Gualtieri}}]{Pani:2011xj}%
  \BibitemOpen
  \bibfield  {author} {\bibinfo {author} {\bibfnamefont {P.}~\bibnamefont
  {Pani}}, \bibinfo {author} {\bibfnamefont {V.}~\bibnamefont {Cardoso}}, \
  and\ \bibinfo {author} {\bibfnamefont {L.}~\bibnamefont {Gualtieri}},\ }\href
  {\doibase 10.1103/PhysRevD.83.104048} {\bibfield  {journal} {\bibinfo
  {journal} {Phys. Rev. D}\ }\textbf {\bibinfo {volume} {83}},\ \bibinfo
  {pages} {104048} (\bibinfo {year} {2011})},\ \Eprint
  {http://arxiv.org/abs/1104.1183} {arXiv:1104.1183 [gr-qc]} \BibitemShut
  {NoStop}%
\bibitem [{\citenamefont {Maselli}\ \emph {et~al.}(2020)\citenamefont
  {Maselli}, \citenamefont {Franchini}, \citenamefont {Gualtieri},\ and\
  \citenamefont {Sotiriou}}]{Maselli:2020zgv}%
  \BibitemOpen
  \bibfield  {author} {\bibinfo {author} {\bibfnamefont {A.}~\bibnamefont
  {Maselli}}, \bibinfo {author} {\bibfnamefont {N.}~\bibnamefont {Franchini}},
  \bibinfo {author} {\bibfnamefont {L.}~\bibnamefont {Gualtieri}}, \ and\
  \bibinfo {author} {\bibfnamefont {T.~P.}\ \bibnamefont {Sotiriou}},\ }\href
  {\doibase 10.1103/PhysRevLett.125.141101} {\bibfield  {journal} {\bibinfo
  {journal} {Phys. Rev. Lett.}\ }\textbf {\bibinfo {volume} {125}},\ \bibinfo
  {pages} {141101} (\bibinfo {year} {2020})},\ \Eprint
  {http://arxiv.org/abs/2004.11895} {arXiv:2004.11895 [gr-qc]} \BibitemShut
  {NoStop}%
\bibitem [{\citenamefont {Maselli}\ \emph {et~al.}(2022)\citenamefont
  {Maselli}, \citenamefont {Franchini}, \citenamefont {Gualtieri},
  \citenamefont {Sotiriou}, \citenamefont {Barsanti},\ and\ \citenamefont
  {Pani}}]{Maselli:2021men}%
  \BibitemOpen
  \bibfield  {author} {\bibinfo {author} {\bibfnamefont {A.}~\bibnamefont
  {Maselli}}, \bibinfo {author} {\bibfnamefont {N.}~\bibnamefont {Franchini}},
  \bibinfo {author} {\bibfnamefont {L.}~\bibnamefont {Gualtieri}}, \bibinfo
  {author} {\bibfnamefont {T.~P.}\ \bibnamefont {Sotiriou}}, \bibinfo {author}
  {\bibfnamefont {S.}~\bibnamefont {Barsanti}}, \ and\ \bibinfo {author}
  {\bibfnamefont {P.}~\bibnamefont {Pani}},\ }\href {\doibase
  10.1038/s41550-021-01589-5} {\bibfield  {journal} {\bibinfo  {journal}
  {Nature Astron.}\ }\textbf {\bibinfo {volume} {6}},\ \bibinfo {pages} {464}
  (\bibinfo {year} {2022})},\ \Eprint {http://arxiv.org/abs/2106.11325}
  {arXiv:2106.11325 [gr-qc]} \BibitemShut {NoStop}%
\bibitem [{\citenamefont {Barsanti}\ \emph
  {et~al.}(2022{\natexlab{b}})\citenamefont {Barsanti}, \citenamefont
  {Franchini}, \citenamefont {Gualtieri}, \citenamefont {Maselli},\ and\
  \citenamefont {Sotiriou}}]{Barsanti:2022ana}%
  \BibitemOpen
  \bibfield  {author} {\bibinfo {author} {\bibfnamefont {S.}~\bibnamefont
  {Barsanti}}, \bibinfo {author} {\bibfnamefont {N.}~\bibnamefont {Franchini}},
  \bibinfo {author} {\bibfnamefont {L.}~\bibnamefont {Gualtieri}}, \bibinfo
  {author} {\bibfnamefont {A.}~\bibnamefont {Maselli}}, \ and\ \bibinfo
  {author} {\bibfnamefont {T.~P.}\ \bibnamefont {Sotiriou}},\ }\href {\doibase
  10.1103/PhysRevD.106.044029} {\bibfield  {journal} {\bibinfo  {journal}
  {Phys. Rev. D}\ }\textbf {\bibinfo {volume} {106}},\ \bibinfo {pages}
  {044029} (\bibinfo {year} {2022}{\natexlab{b}})},\ \Eprint
  {http://arxiv.org/abs/2203.05003} {arXiv:2203.05003 [gr-qc]} \BibitemShut
  {NoStop}%
\bibitem [{\citenamefont {Barsanti}\ \emph {et~al.}(2023)\citenamefont
  {Barsanti}, \citenamefont {Maselli}, \citenamefont {Sotiriou},\ and\
  \citenamefont {Gualtieri}}]{Barsanti:2022vvl}%
  \BibitemOpen
  \bibfield  {author} {\bibinfo {author} {\bibfnamefont {S.}~\bibnamefont
  {Barsanti}}, \bibinfo {author} {\bibfnamefont {A.}~\bibnamefont {Maselli}},
  \bibinfo {author} {\bibfnamefont {T.~P.}\ \bibnamefont {Sotiriou}}, \ and\
  \bibinfo {author} {\bibfnamefont {L.}~\bibnamefont {Gualtieri}},\ }\href
  {\doibase 10.1103/PhysRevLett.131.051401} {\bibfield  {journal} {\bibinfo
  {journal} {Phys. Rev. Lett.}\ }\textbf {\bibinfo {volume} {131}},\ \bibinfo
  {pages} {051401} (\bibinfo {year} {2023})},\ \Eprint
  {http://arxiv.org/abs/2212.03888} {arXiv:2212.03888 [gr-qc]} \BibitemShut
  {NoStop}%
\bibitem [{\citenamefont {Liang}\ \emph {et~al.}(2023)\citenamefont {Liang},
  \citenamefont {Xu}, \citenamefont {Mai},\ and\ \citenamefont
  {Shao}}]{Liang:2022gdk}%
  \BibitemOpen
  \bibfield  {author} {\bibinfo {author} {\bibfnamefont {D.}~\bibnamefont
  {Liang}}, \bibinfo {author} {\bibfnamefont {R.}~\bibnamefont {Xu}}, \bibinfo
  {author} {\bibfnamefont {Z.-F.}\ \bibnamefont {Mai}}, \ and\ \bibinfo
  {author} {\bibfnamefont {L.}~\bibnamefont {Shao}},\ }\href {\doibase
  10.1103/PhysRevD.107.044053} {\bibfield  {journal} {\bibinfo  {journal}
  {Phys. Rev. D}\ }\textbf {\bibinfo {volume} {107}},\ \bibinfo {pages}
  {044053} (\bibinfo {year} {2023})},\ \Eprint
  {http://arxiv.org/abs/2212.09346} {arXiv:2212.09346 [gr-qc]} \BibitemShut
  {NoStop}%
\bibitem [{\citenamefont {Zhang}\ \emph {et~al.}(2023)\citenamefont {Zhang},
  \citenamefont {Guo}, \citenamefont {Gong},\ and\ \citenamefont
  {Wang}}]{Zhang:2023vok}%
  \BibitemOpen
  \bibfield  {author} {\bibinfo {author} {\bibfnamefont {C.}~\bibnamefont
  {Zhang}}, \bibinfo {author} {\bibfnamefont {H.}~\bibnamefont {Guo}}, \bibinfo
  {author} {\bibfnamefont {Y.}~\bibnamefont {Gong}}, \ and\ \bibinfo {author}
  {\bibfnamefont {B.}~\bibnamefont {Wang}},\ }\href {\doibase
  10.1088/1475-7516/2023/06/020} {\bibfield  {journal} {\bibinfo  {journal}
  {JCAP}\ }\textbf {\bibinfo {volume} {06}},\ \bibinfo {pages} {020} (\bibinfo
  {year} {2023})},\ \Eprint {http://arxiv.org/abs/2301.05915} {arXiv:2301.05915
  [gr-qc]} \BibitemShut {NoStop}%
\bibitem [{\citenamefont {Zi}\ \emph {et~al.}(2023)\citenamefont {Zi},
  \citenamefont {Zhou}, \citenamefont {Wang}, \citenamefont {Li}, \citenamefont
  {Zhang},\ and\ \citenamefont {Chen}}]{Zi:2022hcc}%
  \BibitemOpen
  \bibfield  {author} {\bibinfo {author} {\bibfnamefont {T.}~\bibnamefont
  {Zi}}, \bibinfo {author} {\bibfnamefont {Z.}~\bibnamefont {Zhou}}, \bibinfo
  {author} {\bibfnamefont {H.-T.}\ \bibnamefont {Wang}}, \bibinfo {author}
  {\bibfnamefont {P.-C.}\ \bibnamefont {Li}}, \bibinfo {author} {\bibfnamefont
  {J.-d.}\ \bibnamefont {Zhang}}, \ and\ \bibinfo {author} {\bibfnamefont
  {B.}~\bibnamefont {Chen}},\ }\href {\doibase 10.1103/PhysRevD.107.023005}
  {\bibfield  {journal} {\bibinfo  {journal} {Phys. Rev. D}\ }\textbf {\bibinfo
  {volume} {107}},\ \bibinfo {pages} {023005} (\bibinfo {year} {2023})},\
  \Eprint {http://arxiv.org/abs/2205.00425} {arXiv:2205.00425 [gr-qc]}
  \BibitemShut {NoStop}%
\bibitem [{\citenamefont {Lestingi}\ \emph {et~al.}(2023)\citenamefont
  {Lestingi}, \citenamefont {Cannizzaro},\ and\ \citenamefont
  {Pani}}]{Lestingi:2023ovn}%
  \BibitemOpen
  \bibfield  {author} {\bibinfo {author} {\bibfnamefont {J.}~\bibnamefont
  {Lestingi}}, \bibinfo {author} {\bibfnamefont {E.}~\bibnamefont
  {Cannizzaro}}, \ and\ \bibinfo {author} {\bibfnamefont {P.}~\bibnamefont
  {Pani}},\ }\href@noop {} {\  (\bibinfo {year} {2023})},\ \Eprint
  {http://arxiv.org/abs/2310.07772} {arXiv:2310.07772 [gr-qc]} \BibitemShut
  {NoStop}%
\bibitem [{\citenamefont {Barack}\ and\ \citenamefont
  {Cutler}(2007)}]{Barack:2006pq}%
  \BibitemOpen
  \bibfield  {author} {\bibinfo {author} {\bibfnamefont {L.}~\bibnamefont
  {Barack}}\ and\ \bibinfo {author} {\bibfnamefont {C.}~\bibnamefont
  {Cutler}},\ }\href {\doibase 10.1103/PhysRevD.75.042003} {\bibfield
  {journal} {\bibinfo  {journal} {Phys. Rev. D}\ }\textbf {\bibinfo {volume}
  {75}},\ \bibinfo {pages} {042003} (\bibinfo {year} {2007})},\ \Eprint
  {http://arxiv.org/abs/gr-qc/0612029} {arXiv:gr-qc/0612029} \BibitemShut
  {NoStop}%
\bibitem [{\citenamefont {Fransen}\ and\ \citenamefont
  {Mayerson}(2022)}]{Fransen:2022jtw}%
  \BibitemOpen
  \bibfield  {author} {\bibinfo {author} {\bibfnamefont {K.}~\bibnamefont
  {Fransen}}\ and\ \bibinfo {author} {\bibfnamefont {D.~R.}\ \bibnamefont
  {Mayerson}},\ }\href {\doibase 10.1103/PhysRevD.106.064035} {\bibfield
  {journal} {\bibinfo  {journal} {Phys. Rev. D}\ }\textbf {\bibinfo {volume}
  {106}},\ \bibinfo {pages} {064035} (\bibinfo {year} {2022})},\ \Eprint
  {http://arxiv.org/abs/2201.03569} {arXiv:2201.03569 [gr-qc]} \BibitemShut
  {NoStop}%
\bibitem [{\citenamefont {Raposo}\ \emph {et~al.}(2019)\citenamefont {Raposo},
  \citenamefont {Pani},\ and\ \citenamefont {Emparan}}]{Raposo:2018xkf}%
  \BibitemOpen
  \bibfield  {author} {\bibinfo {author} {\bibfnamefont {G.}~\bibnamefont
  {Raposo}}, \bibinfo {author} {\bibfnamefont {P.}~\bibnamefont {Pani}}, \ and\
  \bibinfo {author} {\bibfnamefont {R.}~\bibnamefont {Emparan}},\ }\href
  {\doibase 10.1103/PhysRevD.99.104050} {\bibfield  {journal} {\bibinfo
  {journal} {Phys. Rev. D}\ }\textbf {\bibinfo {volume} {99}},\ \bibinfo
  {pages} {104050} (\bibinfo {year} {2019})},\ \Eprint
  {http://arxiv.org/abs/1812.07615} {arXiv:1812.07615 [gr-qc]} \BibitemShut
  {NoStop}%
\bibitem [{\citenamefont {Bena}\ and\ \citenamefont
  {Mayerson}(2020)}]{Bena:2020see}%
  \BibitemOpen
  \bibfield  {author} {\bibinfo {author} {\bibfnamefont {I.}~\bibnamefont
  {Bena}}\ and\ \bibinfo {author} {\bibfnamefont {D.~R.}\ \bibnamefont
  {Mayerson}},\ }\href {\doibase 10.1103/PhysRevLett.125.221602} {\bibfield
  {journal} {\bibinfo  {journal} {Phys. Rev. Lett.}\ }\textbf {\bibinfo
  {volume} {125}},\ \bibinfo {pages} {221602} (\bibinfo {year} {2020})},\
  \Eprint {http://arxiv.org/abs/2006.10750} {arXiv:2006.10750 [hep-th]}
  \BibitemShut {NoStop}%
\bibitem [{\citenamefont {Bianchi}\ \emph {et~al.}(2020)\citenamefont
  {Bianchi}, \citenamefont {Consoli}, \citenamefont {Grillo}, \citenamefont
  {Morales}, \citenamefont {Pani},\ and\ \citenamefont
  {Raposo}}]{Bianchi:2020bxa}%
  \BibitemOpen
  \bibfield  {author} {\bibinfo {author} {\bibfnamefont {M.}~\bibnamefont
  {Bianchi}}, \bibinfo {author} {\bibfnamefont {D.}~\bibnamefont {Consoli}},
  \bibinfo {author} {\bibfnamefont {A.}~\bibnamefont {Grillo}}, \bibinfo
  {author} {\bibfnamefont {J.~F.}\ \bibnamefont {Morales}}, \bibinfo {author}
  {\bibfnamefont {P.}~\bibnamefont {Pani}}, \ and\ \bibinfo {author}
  {\bibfnamefont {G.}~\bibnamefont {Raposo}},\ }\href {\doibase
  10.1103/PhysRevLett.125.221601} {\bibfield  {journal} {\bibinfo  {journal}
  {Phys. Rev. Lett.}\ }\textbf {\bibinfo {volume} {125}},\ \bibinfo {pages}
  {221601} (\bibinfo {year} {2020})},\ \Eprint
  {http://arxiv.org/abs/2007.01743} {arXiv:2007.01743 [hep-th]} \BibitemShut
  {NoStop}%
\bibitem [{\citenamefont {Loutrel}\ \emph {et~al.}(2022)\citenamefont
  {Loutrel}, \citenamefont {Brito}, \citenamefont {Maselli},\ and\
  \citenamefont {Pani}}]{Loutrel:2022ant}%
  \BibitemOpen
  \bibfield  {author} {\bibinfo {author} {\bibfnamefont {N.}~\bibnamefont
  {Loutrel}}, \bibinfo {author} {\bibfnamefont {R.}~\bibnamefont {Brito}},
  \bibinfo {author} {\bibfnamefont {A.}~\bibnamefont {Maselli}}, \ and\
  \bibinfo {author} {\bibfnamefont {P.}~\bibnamefont {Pani}},\ }\href {\doibase
  10.1103/PhysRevD.105.124050} {\bibfield  {journal} {\bibinfo  {journal}
  {Phys. Rev. D}\ }\textbf {\bibinfo {volume} {105}},\ \bibinfo {pages}
  {124050} (\bibinfo {year} {2022})},\ \Eprint
  {http://arxiv.org/abs/2203.01725} {arXiv:2203.01725 [gr-qc]} \BibitemShut
  {NoStop}%
\bibitem [{\citenamefont {Kumar}\ \emph {et~al.}(2023)\citenamefont {Kumar},
  \citenamefont {Chowdhuri},\ and\ \citenamefont
  {Bhattacharyya}}]{Kumar:2023bdf}%
  \BibitemOpen
  \bibfield  {author} {\bibinfo {author} {\bibfnamefont {S.}~\bibnamefont
  {Kumar}}, \bibinfo {author} {\bibfnamefont {A.}~\bibnamefont {Chowdhuri}}, \
  and\ \bibinfo {author} {\bibfnamefont {A.}~\bibnamefont {Bhattacharyya}},\
  }\href@noop {} {\  (\bibinfo {year} {2023})},\ \Eprint
  {http://arxiv.org/abs/2311.05983} {arXiv:2311.05983 [gr-qc]} \BibitemShut
  {NoStop}%
\bibitem [{\citenamefont {Piovano}\ \emph
  {et~al.}(2020{\natexlab{a}})\citenamefont {Piovano}, \citenamefont
  {Maselli},\ and\ \citenamefont {Pani}}]{Piovano:2020ooe}%
  \BibitemOpen
  \bibfield  {author} {\bibinfo {author} {\bibfnamefont {G.~A.}\ \bibnamefont
  {Piovano}}, \bibinfo {author} {\bibfnamefont {A.}~\bibnamefont {Maselli}}, \
  and\ \bibinfo {author} {\bibfnamefont {P.}~\bibnamefont {Pani}},\ }\href
  {\doibase 10.1016/j.physletb.2020.135860} {\bibfield  {journal} {\bibinfo
  {journal} {Phys. Lett. B}\ }\textbf {\bibinfo {volume} {811}},\ \bibinfo
  {pages} {135860} (\bibinfo {year} {2020}{\natexlab{a}})},\ \Eprint
  {http://arxiv.org/abs/2003.08448} {arXiv:2003.08448 [gr-qc]} \BibitemShut
  {NoStop}%
\bibitem [{\citenamefont {Pani}\ and\ \citenamefont
  {Maselli}(2019)}]{Pani:2019cyc}%
  \BibitemOpen
  \bibfield  {author} {\bibinfo {author} {\bibfnamefont {P.}~\bibnamefont
  {Pani}}\ and\ \bibinfo {author} {\bibfnamefont {A.}~\bibnamefont {Maselli}},\
  }\href {\doibase 10.1142/S0218271819440012} {\bibfield  {journal} {\bibinfo
  {journal} {Int. J. Mod. Phys. D}\ }\textbf {\bibinfo {volume} {28}},\
  \bibinfo {pages} {1944001} (\bibinfo {year} {2019})},\ \Eprint
  {http://arxiv.org/abs/1905.03947} {arXiv:1905.03947 [gr-qc]} \BibitemShut
  {NoStop}%
\bibitem [{\citenamefont {Piovano}\ \emph {et~al.}(2023)\citenamefont
  {Piovano}, \citenamefont {Maselli},\ and\ \citenamefont
  {Pani}}]{Piovano:2022ojl}%
  \BibitemOpen
  \bibfield  {author} {\bibinfo {author} {\bibfnamefont {G.~A.}\ \bibnamefont
  {Piovano}}, \bibinfo {author} {\bibfnamefont {A.}~\bibnamefont {Maselli}}, \
  and\ \bibinfo {author} {\bibfnamefont {P.}~\bibnamefont {Pani}},\ }\href
  {\doibase 10.1103/PhysRevD.107.024021} {\bibfield  {journal} {\bibinfo
  {journal} {Phys. Rev. D}\ }\textbf {\bibinfo {volume} {107}},\ \bibinfo
  {pages} {024021} (\bibinfo {year} {2023})},\ \Eprint
  {http://arxiv.org/abs/2207.07452} {arXiv:2207.07452 [gr-qc]} \BibitemShut
  {NoStop}%
\bibitem [{\citenamefont {Datta}(2022)}]{Datta:2021hvm}%
  \BibitemOpen
  \bibfield  {author} {\bibinfo {author} {\bibfnamefont {S.}~\bibnamefont
  {Datta}},\ }\href {\doibase 10.1088/1361-6382/ac9ae4} {\bibfield  {journal}
  {\bibinfo  {journal} {Class. Quant. Grav.}\ }\textbf {\bibinfo {volume}
  {39}},\ \bibinfo {pages} {225016} (\bibinfo {year} {2022})},\ \Eprint
  {http://arxiv.org/abs/2107.07258} {arXiv:2107.07258 [gr-qc]} \BibitemShut
  {NoStop}%
\bibitem [{\citenamefont {Datta}\ \emph {et~al.}(2020)\citenamefont {Datta},
  \citenamefont {Brito}, \citenamefont {Bose}, \citenamefont {Pani},\ and\
  \citenamefont {Hughes}}]{Datta:2019epe}%
  \BibitemOpen
  \bibfield  {author} {\bibinfo {author} {\bibfnamefont {S.}~\bibnamefont
  {Datta}}, \bibinfo {author} {\bibfnamefont {R.}~\bibnamefont {Brito}},
  \bibinfo {author} {\bibfnamefont {S.}~\bibnamefont {Bose}}, \bibinfo {author}
  {\bibfnamefont {P.}~\bibnamefont {Pani}}, \ and\ \bibinfo {author}
  {\bibfnamefont {S.~A.}\ \bibnamefont {Hughes}},\ }\href {\doibase
  10.1103/PhysRevD.101.044004} {\bibfield  {journal} {\bibinfo  {journal}
  {Phys. Rev. D}\ }\textbf {\bibinfo {volume} {101}},\ \bibinfo {pages}
  {044004} (\bibinfo {year} {2020})},\ \Eprint
  {http://arxiv.org/abs/1910.07841} {arXiv:1910.07841 [gr-qc]} \BibitemShut
  {NoStop}%
\bibitem [{\citenamefont {Datta}\ and\ \citenamefont
  {Bose}(2019)}]{Datta:2019euh}%
  \BibitemOpen
  \bibfield  {author} {\bibinfo {author} {\bibfnamefont {S.}~\bibnamefont
  {Datta}}\ and\ \bibinfo {author} {\bibfnamefont {S.}~\bibnamefont {Bose}},\
  }\href {\doibase 10.1103/PhysRevD.99.084001} {\bibfield  {journal} {\bibinfo
  {journal} {Phys. Rev. D}\ }\textbf {\bibinfo {volume} {99}},\ \bibinfo
  {pages} {084001} (\bibinfo {year} {2019})},\ \Eprint
  {http://arxiv.org/abs/1902.01723} {arXiv:1902.01723 [gr-qc]} \BibitemShut
  {NoStop}%
\bibitem [{\citenamefont {Maggio}\ \emph {et~al.}(2021)\citenamefont {Maggio},
  \citenamefont {van~de Meent},\ and\ \citenamefont {Pani}}]{Maggio:2021uge}%
  \BibitemOpen
  \bibfield  {author} {\bibinfo {author} {\bibfnamefont {E.}~\bibnamefont
  {Maggio}}, \bibinfo {author} {\bibfnamefont {M.}~\bibnamefont {van~de
  Meent}}, \ and\ \bibinfo {author} {\bibfnamefont {P.}~\bibnamefont {Pani}},\
  }\href {\doibase 10.1103/PhysRevD.104.104026} {\bibfield  {journal} {\bibinfo
   {journal} {Phys. Rev. D}\ }\textbf {\bibinfo {volume} {104}},\ \bibinfo
  {pages} {104026} (\bibinfo {year} {2021})},\ \Eprint
  {http://arxiv.org/abs/2106.07195} {arXiv:2106.07195 [gr-qc]} \BibitemShut
  {NoStop}%
\bibitem [{\citenamefont {Pani}\ \emph {et~al.}(2010)\citenamefont {Pani},
  \citenamefont {Berti}, \citenamefont {Cardoso}, \citenamefont {Chen},\ and\
  \citenamefont {Norte}}]{Pani:2010em}%
  \BibitemOpen
  \bibfield  {author} {\bibinfo {author} {\bibfnamefont {P.}~\bibnamefont
  {Pani}}, \bibinfo {author} {\bibfnamefont {E.}~\bibnamefont {Berti}},
  \bibinfo {author} {\bibfnamefont {V.}~\bibnamefont {Cardoso}}, \bibinfo
  {author} {\bibfnamefont {Y.}~\bibnamefont {Chen}}, \ and\ \bibinfo {author}
  {\bibfnamefont {R.}~\bibnamefont {Norte}},\ }\href {\doibase
  10.1103/PhysRevD.81.084011} {\bibfield  {journal} {\bibinfo  {journal} {Phys.
  Rev. D}\ }\textbf {\bibinfo {volume} {81}},\ \bibinfo {pages} {084011}
  (\bibinfo {year} {2010})},\ \Eprint {http://arxiv.org/abs/1001.3031}
  {arXiv:1001.3031 [gr-qc]} \BibitemShut {NoStop}%
\bibitem [{\citenamefont {Macedo}\ \emph {et~al.}(2013)\citenamefont {Macedo},
  \citenamefont {Pani}, \citenamefont {Cardoso},\ and\ \citenamefont
  {Crispino}}]{Macedo:2013qea}%
  \BibitemOpen
  \bibfield  {author} {\bibinfo {author} {\bibfnamefont {C.~F.~B.}\
  \bibnamefont {Macedo}}, \bibinfo {author} {\bibfnamefont {P.}~\bibnamefont
  {Pani}}, \bibinfo {author} {\bibfnamefont {V.}~\bibnamefont {Cardoso}}, \
  and\ \bibinfo {author} {\bibfnamefont {L.~C.~B.}\ \bibnamefont {Crispino}},\
  }\href {\doibase 10.1088/0004-637X/774/1/48} {\bibfield  {journal} {\bibinfo
  {journal} {Astrophys. J.}\ }\textbf {\bibinfo {volume} {774}},\ \bibinfo
  {pages} {48} (\bibinfo {year} {2013})},\ \Eprint
  {http://arxiv.org/abs/1302.2646} {arXiv:1302.2646 [gr-qc]} \BibitemShut
  {NoStop}%
\bibitem [{\citenamefont {Destounis}\ \emph {et~al.}(2023)\citenamefont
  {Destounis}, \citenamefont {Angeloni}, \citenamefont {Vaglio},\ and\
  \citenamefont {Pani}}]{Destounis:2023khj}%
  \BibitemOpen
  \bibfield  {author} {\bibinfo {author} {\bibfnamefont {K.}~\bibnamefont
  {Destounis}}, \bibinfo {author} {\bibfnamefont {F.}~\bibnamefont {Angeloni}},
  \bibinfo {author} {\bibfnamefont {M.}~\bibnamefont {Vaglio}}, \ and\ \bibinfo
  {author} {\bibfnamefont {P.}~\bibnamefont {Pani}},\ }\href {\doibase
  10.1103/PhysRevD.108.084062} {\bibfield  {journal} {\bibinfo  {journal}
  {Phys. Rev. D}\ }\textbf {\bibinfo {volume} {108}},\ \bibinfo {pages}
  {084062} (\bibinfo {year} {2023})},\ \Eprint
  {http://arxiv.org/abs/2305.05691} {arXiv:2305.05691 [gr-qc]} \BibitemShut
  {NoStop}%
\bibitem [{\citenamefont {Hannuksela}\ \emph {et~al.}(2019)\citenamefont
  {Hannuksela}, \citenamefont {Wong}, \citenamefont {Brito}, \citenamefont
  {Berti},\ and\ \citenamefont {Li}}]{Hannuksela:2018izj}%
  \BibitemOpen
  \bibfield  {author} {\bibinfo {author} {\bibfnamefont {O.~A.}\ \bibnamefont
  {Hannuksela}}, \bibinfo {author} {\bibfnamefont {K.~W.~K.}\ \bibnamefont
  {Wong}}, \bibinfo {author} {\bibfnamefont {R.}~\bibnamefont {Brito}},
  \bibinfo {author} {\bibfnamefont {E.}~\bibnamefont {Berti}}, \ and\ \bibinfo
  {author} {\bibfnamefont {T.~G.~F.}\ \bibnamefont {Li}},\ }\href {\doibase
  10.1038/s41550-019-0712-4} {\bibfield  {journal} {\bibinfo  {journal} {Nature
  Astron.}\ }\textbf {\bibinfo {volume} {3}},\ \bibinfo {pages} {447} (\bibinfo
  {year} {2019})},\ \Eprint {http://arxiv.org/abs/1804.09659} {arXiv:1804.09659
  [astro-ph.HE]} \BibitemShut {NoStop}%
\bibitem [{\citenamefont {Hannuksela}\ \emph {et~al.}(2020)\citenamefont
  {Hannuksela}, \citenamefont {Ng},\ and\ \citenamefont
  {Li}}]{Hannuksela:2019vip}%
  \BibitemOpen
  \bibfield  {author} {\bibinfo {author} {\bibfnamefont {O.~A.}\ \bibnamefont
  {Hannuksela}}, \bibinfo {author} {\bibfnamefont {K.~C.~Y.}\ \bibnamefont
  {Ng}}, \ and\ \bibinfo {author} {\bibfnamefont {T.~G.~F.}\ \bibnamefont
  {Li}},\ }\href {\doibase 10.1103/PhysRevD.102.103022} {\bibfield  {journal}
  {\bibinfo  {journal} {Phys. Rev. D}\ }\textbf {\bibinfo {volume} {102}},\
  \bibinfo {pages} {103022} (\bibinfo {year} {2020})},\ \Eprint
  {http://arxiv.org/abs/1906.11845} {arXiv:1906.11845 [astro-ph.CO]}
  \BibitemShut {NoStop}%
\bibitem [{\citenamefont {Brito}\ and\ \citenamefont
  {Shah}(2023)}]{Brito:2023pyl}%
  \BibitemOpen
  \bibfield  {author} {\bibinfo {author} {\bibfnamefont {R.}~\bibnamefont
  {Brito}}\ and\ \bibinfo {author} {\bibfnamefont {S.}~\bibnamefont {Shah}},\
  }\href {\doibase 10.1103/PhysRevD.108.084019} {\bibfield  {journal} {\bibinfo
   {journal} {Phys. Rev. D}\ }\textbf {\bibinfo {volume} {108}},\ \bibinfo
  {pages} {084019} (\bibinfo {year} {2023})},\ \Eprint
  {http://arxiv.org/abs/2307.16093} {arXiv:2307.16093 [gr-qc]} \BibitemShut
  {NoStop}%
\bibitem [{\citenamefont {Duque}\ \emph {et~al.}(2023)\citenamefont {Duque},
  \citenamefont {Macedo}, \citenamefont {Vicente},\ and\ \citenamefont
  {Cardoso}}]{Duque:2023cac}%
  \BibitemOpen
  \bibfield  {author} {\bibinfo {author} {\bibfnamefont {F.}~\bibnamefont
  {Duque}}, \bibinfo {author} {\bibfnamefont {C.~F.~B.}\ \bibnamefont
  {Macedo}}, \bibinfo {author} {\bibfnamefont {R.}~\bibnamefont {Vicente}}, \
  and\ \bibinfo {author} {\bibfnamefont {V.}~\bibnamefont {Cardoso}},\
  }\href@noop {} {\  (\bibinfo {year} {2023})},\ \Eprint
  {http://arxiv.org/abs/2312.06767} {arXiv:2312.06767 [gr-qc]} \BibitemShut
  {NoStop}%
\bibitem [{\citenamefont {Afshordi}\ \emph {et~al.}(2023)\citenamefont
  {Afshordi} \emph {et~al.}}]{LISAConsortiumWaveformWorkingGroup:2023arg}%
  \BibitemOpen
  \bibfield  {author} {\bibinfo {author} {\bibfnamefont {N.}~\bibnamefont
  {Afshordi}} \emph {et~al.} (\bibinfo {collaboration} {LISA Consortium
  Waveform Working Group}),\ }\href@noop {} {\  (\bibinfo {year} {2023})},\
  \Eprint {http://arxiv.org/abs/2311.01300} {arXiv:2311.01300 [gr-qc]}
  \BibitemShut {NoStop}%
\bibitem [{\citenamefont {Abbott}\ \emph {et~al.}(2023)\citenamefont {Abbott}
  \emph {et~al.}}]{GWTC-3}%
  \BibitemOpen
  \bibfield  {author} {\bibinfo {author} {\bibfnamefont {R.}~\bibnamefont
  {Abbott}} \emph {et~al.} (\bibinfo {collaboration} {KAGRA, VIRGO, LIGO
  Scientific}),\ }\href {\doibase 10.1103/PhysRevX.13.041039} {\bibfield
  {journal} {\bibinfo  {journal} {Phys. Rev. X}\ }\textbf {\bibinfo {volume}
  {13}},\ \bibinfo {pages} {041039} (\bibinfo {year} {2023})},\ \Eprint
  {http://arxiv.org/abs/2111.03606} {arXiv:2111.03606 [gr-qc]} \BibitemShut
  {NoStop}%
\bibitem [{\citenamefont {Pound}\ \emph {et~al.}(2020)\citenamefont {Pound},
  \citenamefont {Wardell}, \citenamefont {Warburton},\ and\ \citenamefont
  {Miller}}]{Pound:2019lzj}%
  \BibitemOpen
  \bibfield  {author} {\bibinfo {author} {\bibfnamefont {A.}~\bibnamefont
  {Pound}}, \bibinfo {author} {\bibfnamefont {B.}~\bibnamefont {Wardell}},
  \bibinfo {author} {\bibfnamefont {N.}~\bibnamefont {Warburton}}, \ and\
  \bibinfo {author} {\bibfnamefont {J.}~\bibnamefont {Miller}},\ }\href
  {\doibase 10.1103/PhysRevLett.124.021101} {\bibfield  {journal} {\bibinfo
  {journal} {Phys. Rev. Lett.}\ }\textbf {\bibinfo {volume} {124}},\ \bibinfo
  {pages} {021101} (\bibinfo {year} {2020})},\ \Eprint
  {http://arxiv.org/abs/1908.07419} {arXiv:1908.07419 [gr-qc]} \BibitemShut
  {NoStop}%
\bibitem [{\citenamefont {Warburton}\ \emph {et~al.}(2021)\citenamefont
  {Warburton}, \citenamefont {Pound}, \citenamefont {Wardell}, \citenamefont
  {Miller},\ and\ \citenamefont {Durkan}}]{Warburton:2021kwk}%
  \BibitemOpen
  \bibfield  {author} {\bibinfo {author} {\bibfnamefont {N.}~\bibnamefont
  {Warburton}}, \bibinfo {author} {\bibfnamefont {A.}~\bibnamefont {Pound}},
  \bibinfo {author} {\bibfnamefont {B.}~\bibnamefont {Wardell}}, \bibinfo
  {author} {\bibfnamefont {J.}~\bibnamefont {Miller}}, \ and\ \bibinfo {author}
  {\bibfnamefont {L.}~\bibnamefont {Durkan}},\ }\href {\doibase
  10.1103/PhysRevLett.127.151102} {\bibfield  {journal} {\bibinfo  {journal}
  {Phys. Rev. Lett.}\ }\textbf {\bibinfo {volume} {127}},\ \bibinfo {pages}
  {151102} (\bibinfo {year} {2021})},\ \Eprint
  {http://arxiv.org/abs/2107.01298} {arXiv:2107.01298 [gr-qc]} \BibitemShut
  {NoStop}%
\bibitem [{\citenamefont {Wardell}\ \emph {et~al.}(2023)\citenamefont
  {Wardell}, \citenamefont {Pound}, \citenamefont {Warburton}, \citenamefont
  {Miller}, \citenamefont {Durkan},\ and\ \citenamefont
  {Le~Tiec}}]{Wardell:2021fyy}%
  \BibitemOpen
  \bibfield  {author} {\bibinfo {author} {\bibfnamefont {B.}~\bibnamefont
  {Wardell}}, \bibinfo {author} {\bibfnamefont {A.}~\bibnamefont {Pound}},
  \bibinfo {author} {\bibfnamefont {N.}~\bibnamefont {Warburton}}, \bibinfo
  {author} {\bibfnamefont {J.}~\bibnamefont {Miller}}, \bibinfo {author}
  {\bibfnamefont {L.}~\bibnamefont {Durkan}}, \ and\ \bibinfo {author}
  {\bibfnamefont {A.}~\bibnamefont {Le~Tiec}},\ }\href {\doibase
  10.1103/PhysRevLett.130.241402} {\bibfield  {journal} {\bibinfo  {journal}
  {Phys. Rev. Lett.}\ }\textbf {\bibinfo {volume} {130}},\ \bibinfo {pages}
  {241402} (\bibinfo {year} {2023})},\ \Eprint
  {http://arxiv.org/abs/2112.12265} {arXiv:2112.12265 [gr-qc]} \BibitemShut
  {NoStop}%
\bibitem [{\citenamefont {Durkan}\ and\ \citenamefont
  {Warburton}(2022)}]{Durkan:2022fvm}%
  \BibitemOpen
  \bibfield  {author} {\bibinfo {author} {\bibfnamefont {L.}~\bibnamefont
  {Durkan}}\ and\ \bibinfo {author} {\bibfnamefont {N.}~\bibnamefont
  {Warburton}},\ }\href {\doibase 10.1103/PhysRevD.106.084023} {\bibfield
  {journal} {\bibinfo  {journal} {Phys. Rev. D}\ }\textbf {\bibinfo {volume}
  {106}},\ \bibinfo {pages} {084023} (\bibinfo {year} {2022})},\ \Eprint
  {http://arxiv.org/abs/2206.08179} {arXiv:2206.08179 [gr-qc]} \BibitemShut
  {NoStop}%
\bibitem [{\citenamefont {Miller}\ and\ \citenamefont
  {Pound}(2021)}]{Miller:2020bft}%
  \BibitemOpen
  \bibfield  {author} {\bibinfo {author} {\bibfnamefont {J.}~\bibnamefont
  {Miller}}\ and\ \bibinfo {author} {\bibfnamefont {A.}~\bibnamefont {Pound}},\
  }\href {\doibase 10.1103/PhysRevD.103.064048} {\bibfield  {journal} {\bibinfo
   {journal} {Phys. Rev. D}\ }\textbf {\bibinfo {volume} {103}},\ \bibinfo
  {pages} {064048} (\bibinfo {year} {2021})},\ \Eprint
  {http://arxiv.org/abs/2006.11263} {arXiv:2006.11263 [gr-qc]} \BibitemShut
  {NoStop}%
\bibitem [{\citenamefont {Green}\ \emph {et~al.}(2020)\citenamefont {Green},
  \citenamefont {Hollands},\ and\ \citenamefont {Zimmerman}}]{Green:2019nam}%
  \BibitemOpen
  \bibfield  {author} {\bibinfo {author} {\bibfnamefont {S.~R.}\ \bibnamefont
  {Green}}, \bibinfo {author} {\bibfnamefont {S.}~\bibnamefont {Hollands}}, \
  and\ \bibinfo {author} {\bibfnamefont {P.}~\bibnamefont {Zimmerman}},\ }\href
  {\doibase 10.1088/1361-6382/ab7075} {\bibfield  {journal} {\bibinfo
  {journal} {Class. Quant. Grav.}\ }\textbf {\bibinfo {volume} {37}},\ \bibinfo
  {pages} {075001} (\bibinfo {year} {2020})},\ \Eprint
  {http://arxiv.org/abs/1908.09095} {arXiv:1908.09095 [gr-qc]} \BibitemShut
  {NoStop}%
\bibitem [{\citenamefont {Dolan}\ \emph {et~al.}(2022)\citenamefont {Dolan},
  \citenamefont {Kavanagh},\ and\ \citenamefont {Wardell}}]{Dolan:2021ijg}%
  \BibitemOpen
  \bibfield  {author} {\bibinfo {author} {\bibfnamefont {S.~R.}\ \bibnamefont
  {Dolan}}, \bibinfo {author} {\bibfnamefont {C.}~\bibnamefont {Kavanagh}}, \
  and\ \bibinfo {author} {\bibfnamefont {B.}~\bibnamefont {Wardell}},\ }\href
  {\doibase 10.1103/PhysRevLett.128.151101} {\bibfield  {journal} {\bibinfo
  {journal} {Phys. Rev. Lett.}\ }\textbf {\bibinfo {volume} {128}},\ \bibinfo
  {pages} {151101} (\bibinfo {year} {2022})},\ \Eprint
  {http://arxiv.org/abs/2108.06344} {arXiv:2108.06344 [gr-qc]} \BibitemShut
  {NoStop}%
\bibitem [{\citenamefont {Upton}\ and\ \citenamefont
  {Pound}(2021)}]{Upton:2021oxf}%
  \BibitemOpen
  \bibfield  {author} {\bibinfo {author} {\bibfnamefont {S.~D.}\ \bibnamefont
  {Upton}}\ and\ \bibinfo {author} {\bibfnamefont {A.}~\bibnamefont {Pound}},\
  }\href {\doibase 10.1103/PhysRevD.103.124016} {\bibfield  {journal} {\bibinfo
   {journal} {Phys. Rev. D}\ }\textbf {\bibinfo {volume} {103}},\ \bibinfo
  {pages} {124016} (\bibinfo {year} {2021})},\ \Eprint
  {http://arxiv.org/abs/2101.11409} {arXiv:2101.11409 [gr-qc]} \BibitemShut
  {NoStop}%
\bibitem [{\citenamefont {Toomani}\ \emph {et~al.}(2022)\citenamefont
  {Toomani}, \citenamefont {Zimmerman}, \citenamefont {Spiers}, \citenamefont
  {Hollands}, \citenamefont {Pound},\ and\ \citenamefont
  {Green}}]{Toomani:2021jlo}%
  \BibitemOpen
  \bibfield  {author} {\bibinfo {author} {\bibfnamefont {V.}~\bibnamefont
  {Toomani}}, \bibinfo {author} {\bibfnamefont {P.}~\bibnamefont {Zimmerman}},
  \bibinfo {author} {\bibfnamefont {A.}~\bibnamefont {Spiers}}, \bibinfo
  {author} {\bibfnamefont {S.}~\bibnamefont {Hollands}}, \bibinfo {author}
  {\bibfnamefont {A.}~\bibnamefont {Pound}}, \ and\ \bibinfo {author}
  {\bibfnamefont {S.~R.}\ \bibnamefont {Green}},\ }\href {\doibase
  10.1088/1361-6382/ac37a5} {\bibfield  {journal} {\bibinfo  {journal} {Class.
  Quant. Grav.}\ }\textbf {\bibinfo {volume} {39}},\ \bibinfo {pages} {015019}
  (\bibinfo {year} {2022})},\ \Eprint {http://arxiv.org/abs/2108.04273}
  {arXiv:2108.04273 [gr-qc]} \BibitemShut {NoStop}%
\bibitem [{\citenamefont {Osburn}\ and\ \citenamefont
  {Nishimura}(2022)}]{Osburn:2022bby}%
  \BibitemOpen
  \bibfield  {author} {\bibinfo {author} {\bibfnamefont {T.}~\bibnamefont
  {Osburn}}\ and\ \bibinfo {author} {\bibfnamefont {N.}~\bibnamefont
  {Nishimura}},\ }\href {\doibase 10.1103/PhysRevD.106.044056} {\bibfield
  {journal} {\bibinfo  {journal} {Phys. Rev. D}\ }\textbf {\bibinfo {volume}
  {106}},\ \bibinfo {pages} {044056} (\bibinfo {year} {2022})},\ \Eprint
  {http://arxiv.org/abs/2206.07031} {arXiv:2206.07031 [gr-qc]} \BibitemShut
  {NoStop}%
\bibitem [{\citenamefont {Spiers}\ \emph
  {et~al.}(2023{\natexlab{a}})\citenamefont {Spiers}, \citenamefont {Pound},\
  and\ \citenamefont {Moxon}}]{Spiers:2023cip}%
  \BibitemOpen
  \bibfield  {author} {\bibinfo {author} {\bibfnamefont {A.}~\bibnamefont
  {Spiers}}, \bibinfo {author} {\bibfnamefont {A.}~\bibnamefont {Pound}}, \
  and\ \bibinfo {author} {\bibfnamefont {J.}~\bibnamefont {Moxon}},\ }\href
  {\doibase 10.1103/PhysRevD.108.064002} {\bibfield  {journal} {\bibinfo
  {journal} {Phys. Rev. D}\ }\textbf {\bibinfo {volume} {108}},\ \bibinfo
  {pages} {064002} (\bibinfo {year} {2023}{\natexlab{a}})},\ \Eprint
  {http://arxiv.org/abs/2305.19332} {arXiv:2305.19332 [gr-qc]} \BibitemShut
  {NoStop}%
\bibitem [{\citenamefont {Spiers}\ \emph
  {et~al.}(2023{\natexlab{b}})\citenamefont {Spiers}, \citenamefont {Pound},\
  and\ \citenamefont {Wardell}}]{Spiers:2023mor}%
  \BibitemOpen
  \bibfield  {author} {\bibinfo {author} {\bibfnamefont {A.}~\bibnamefont
  {Spiers}}, \bibinfo {author} {\bibfnamefont {A.}~\bibnamefont {Pound}}, \
  and\ \bibinfo {author} {\bibfnamefont {B.}~\bibnamefont {Wardell}},\
  }\href@noop {} {\  (\bibinfo {year} {2023}{\natexlab{b}})},\ \Eprint
  {http://arxiv.org/abs/2306.17847} {arXiv:2306.17847 [gr-qc]} \BibitemShut
  {NoStop}%
\bibitem [{\citenamefont {Miller}\ \emph {et~al.}(2023)\citenamefont {Miller},
  \citenamefont {Leather}, \citenamefont {Pound},\ and\ \citenamefont
  {Warburton}}]{Miller:2023ers}%
  \BibitemOpen
  \bibfield  {author} {\bibinfo {author} {\bibfnamefont {J.}~\bibnamefont
  {Miller}}, \bibinfo {author} {\bibfnamefont {B.}~\bibnamefont {Leather}},
  \bibinfo {author} {\bibfnamefont {A.}~\bibnamefont {Pound}}, \ and\ \bibinfo
  {author} {\bibfnamefont {N.}~\bibnamefont {Warburton}},\ }\href@noop {} {\
  (\bibinfo {year} {2023})},\ \Eprint {http://arxiv.org/abs/2401.00455}
  {arXiv:2401.00455 [gr-qc]} \BibitemShut {NoStop}%
\bibitem [{\citenamefont {Nasipak}\ and\ \citenamefont
  {Evans}(2021)}]{Nasipak:2021qfu}%
  \BibitemOpen
  \bibfield  {author} {\bibinfo {author} {\bibfnamefont {Z.}~\bibnamefont
  {Nasipak}}\ and\ \bibinfo {author} {\bibfnamefont {C.~R.}\ \bibnamefont
  {Evans}},\ }\href {\doibase 10.1103/PhysRevD.104.084011} {\bibfield
  {journal} {\bibinfo  {journal} {Phys. Rev. D}\ }\textbf {\bibinfo {volume}
  {104}},\ \bibinfo {pages} {084011} (\bibinfo {year} {2021})},\ \Eprint
  {http://arxiv.org/abs/2105.15188} {arXiv:2105.15188 [gr-qc]} \BibitemShut
  {NoStop}%
\bibitem [{\citenamefont {Piovano}\ \emph
  {et~al.}(2020{\natexlab{b}})\citenamefont {Piovano}, \citenamefont
  {Maselli},\ and\ \citenamefont {Pani}}]{Piovano:2020zin}%
  \BibitemOpen
  \bibfield  {author} {\bibinfo {author} {\bibfnamefont {G.~A.}\ \bibnamefont
  {Piovano}}, \bibinfo {author} {\bibfnamefont {A.}~\bibnamefont {Maselli}}, \
  and\ \bibinfo {author} {\bibfnamefont {P.}~\bibnamefont {Pani}},\ }\href
  {\doibase 10.1103/PhysRevD.102.024041} {\bibfield  {journal} {\bibinfo
  {journal} {Phys. Rev. D}\ }\textbf {\bibinfo {volume} {102}},\ \bibinfo
  {pages} {024041} (\bibinfo {year} {2020}{\natexlab{b}})},\ \Eprint
  {http://arxiv.org/abs/2004.02654} {arXiv:2004.02654 [gr-qc]} \BibitemShut
  {NoStop}%
\bibitem [{\citenamefont {Mathews}\ \emph {et~al.}(2022)\citenamefont
  {Mathews}, \citenamefont {Pound},\ and\ \citenamefont
  {Wardell}}]{Mathews:2021rod}%
  \BibitemOpen
  \bibfield  {author} {\bibinfo {author} {\bibfnamefont {J.}~\bibnamefont
  {Mathews}}, \bibinfo {author} {\bibfnamefont {A.}~\bibnamefont {Pound}}, \
  and\ \bibinfo {author} {\bibfnamefont {B.}~\bibnamefont {Wardell}},\ }\href
  {\doibase 10.1103/PhysRevD.105.084031} {\bibfield  {journal} {\bibinfo
  {journal} {Phys. Rev. D}\ }\textbf {\bibinfo {volume} {105}},\ \bibinfo
  {pages} {084031} (\bibinfo {year} {2022})},\ \Eprint
  {http://arxiv.org/abs/2112.13069} {arXiv:2112.13069 [gr-qc]} \BibitemShut
  {NoStop}%
\bibitem [{\citenamefont {Drummond}\ and\ \citenamefont
  {Hughes}(2022)}]{Drummond:2022xej}%
  \BibitemOpen
  \bibfield  {author} {\bibinfo {author} {\bibfnamefont {L.~V.}\ \bibnamefont
  {Drummond}}\ and\ \bibinfo {author} {\bibfnamefont {S.~A.}\ \bibnamefont
  {Hughes}},\ }\href {\doibase 10.1103/PhysRevD.105.124040} {\bibfield
  {journal} {\bibinfo  {journal} {Phys. Rev. D}\ }\textbf {\bibinfo {volume}
  {105}},\ \bibinfo {pages} {124040} (\bibinfo {year} {2022})},\ \Eprint
  {http://arxiv.org/abs/2201.13334} {arXiv:2201.13334 [gr-qc]} \BibitemShut
  {NoStop}%
\bibitem [{\citenamefont {Upton}(2023)}]{Upton:2023tcv}%
  \BibitemOpen
  \bibfield  {author} {\bibinfo {author} {\bibfnamefont {S.~D.}\ \bibnamefont
  {Upton}},\ }\href@noop {} {\  (\bibinfo {year} {2023})},\ \Eprint
  {http://arxiv.org/abs/2309.03778} {arXiv:2309.03778 [gr-qc]} \BibitemShut
  {NoStop}%
\bibitem [{\citenamefont {Drummond}\ \emph {et~al.}(2023)\citenamefont
  {Drummond}, \citenamefont {Hanselman}, \citenamefont {Becker},\ and\
  \citenamefont {Hughes}}]{Drummond:2023loz}%
  \BibitemOpen
  \bibfield  {author} {\bibinfo {author} {\bibfnamefont {L.~V.}\ \bibnamefont
  {Drummond}}, \bibinfo {author} {\bibfnamefont {A.~G.}\ \bibnamefont
  {Hanselman}}, \bibinfo {author} {\bibfnamefont {D.~R.}\ \bibnamefont
  {Becker}}, \ and\ \bibinfo {author} {\bibfnamefont {S.~A.}\ \bibnamefont
  {Hughes}},\ }\href@noop {} {\  (\bibinfo {year} {2023})},\ \Eprint
  {http://arxiv.org/abs/2305.08919} {arXiv:2305.08919 [gr-qc]} \BibitemShut
  {NoStop}%
\bibitem [{\citenamefont {Buonanno}\ and\ \citenamefont
  {Damour}(1999)}]{Buonanno:1998gg}%
  \BibitemOpen
  \bibfield  {author} {\bibinfo {author} {\bibfnamefont {A.}~\bibnamefont
  {Buonanno}}\ and\ \bibinfo {author} {\bibfnamefont {T.}~\bibnamefont
  {Damour}},\ }\href {\doibase 10.1103/PhysRevD.59.084006} {\bibfield
  {journal} {\bibinfo  {journal} {Phys. Rev. D}\ }\textbf {\bibinfo {volume}
  {59}},\ \bibinfo {pages} {084006} (\bibinfo {year} {1999})},\ \Eprint
  {http://arxiv.org/abs/gr-qc/9811091} {arXiv:gr-qc/9811091} \BibitemShut
  {NoStop}%
\bibitem [{\citenamefont {Buonanno}\ \emph {et~al.}(2006)\citenamefont
  {Buonanno}, \citenamefont {Chen},\ and\ \citenamefont
  {Damour}}]{Buonanno:2005xu}%
  \BibitemOpen
  \bibfield  {author} {\bibinfo {author} {\bibfnamefont {A.}~\bibnamefont
  {Buonanno}}, \bibinfo {author} {\bibfnamefont {Y.}~\bibnamefont {Chen}}, \
  and\ \bibinfo {author} {\bibfnamefont {T.}~\bibnamefont {Damour}},\ }\href
  {\doibase 10.1103/PhysRevD.74.104005} {\bibfield  {journal} {\bibinfo
  {journal} {Phys. Rev. D}\ }\textbf {\bibinfo {volume} {74}},\ \bibinfo
  {pages} {104005} (\bibinfo {year} {2006})},\ \Eprint
  {http://arxiv.org/abs/gr-qc/0508067} {arXiv:gr-qc/0508067} \BibitemShut
  {NoStop}%
\bibitem [{\citenamefont {Damour}\ and\ \citenamefont
  {Nagar}(2007)}]{Damour:2007xr}%
  \BibitemOpen
  \bibfield  {author} {\bibinfo {author} {\bibfnamefont {T.}~\bibnamefont
  {Damour}}\ and\ \bibinfo {author} {\bibfnamefont {A.}~\bibnamefont {Nagar}},\
  }\href {\doibase 10.1103/PhysRevD.76.064028} {\bibfield  {journal} {\bibinfo
  {journal} {Phys. Rev. D}\ }\textbf {\bibinfo {volume} {76}},\ \bibinfo
  {pages} {064028} (\bibinfo {year} {2007})},\ \Eprint
  {http://arxiv.org/abs/0705.2519} {arXiv:0705.2519 [gr-qc]} \BibitemShut
  {NoStop}%
\bibitem [{\citenamefont {Ramos-Buades}\ \emph {et~al.}(2023)\citenamefont
  {Ramos-Buades}, \citenamefont {Buonanno}, \citenamefont {Estell\'es},
  \citenamefont {Khalil}, \citenamefont {Mihaylov}, \citenamefont {Ossokine},
  \citenamefont {Pompili},\ and\ \citenamefont
  {Shiferaw}}]{Ramos-Buades:2023ehm}%
  \BibitemOpen
  \bibfield  {author} {\bibinfo {author} {\bibfnamefont {A.}~\bibnamefont
  {Ramos-Buades}}, \bibinfo {author} {\bibfnamefont {A.}~\bibnamefont
  {Buonanno}}, \bibinfo {author} {\bibfnamefont {H.}~\bibnamefont
  {Estell\'es}}, \bibinfo {author} {\bibfnamefont {M.}~\bibnamefont {Khalil}},
  \bibinfo {author} {\bibfnamefont {D.~P.}\ \bibnamefont {Mihaylov}}, \bibinfo
  {author} {\bibfnamefont {S.}~\bibnamefont {Ossokine}}, \bibinfo {author}
  {\bibfnamefont {L.}~\bibnamefont {Pompili}}, \ and\ \bibinfo {author}
  {\bibfnamefont {M.}~\bibnamefont {Shiferaw}},\ }\href {\doibase
  10.1103/PhysRevD.108.124037} {\bibfield  {journal} {\bibinfo  {journal}
  {Phys. Rev. D}\ }\textbf {\bibinfo {volume} {108}},\ \bibinfo {pages}
  {124037} (\bibinfo {year} {2023})},\ \Eprint
  {http://arxiv.org/abs/2303.18046} {arXiv:2303.18046 [gr-qc]} \BibitemShut
  {NoStop}%
\bibitem [{\citenamefont {Gamba}\ \emph {et~al.}(2022)\citenamefont {Gamba},
  \citenamefont {Ak\c{c}ay}, \citenamefont {Bernuzzi},\ and\ \citenamefont
  {Williams}}]{Gamba:2021ydi}%
  \BibitemOpen
  \bibfield  {author} {\bibinfo {author} {\bibfnamefont {R.}~\bibnamefont
  {Gamba}}, \bibinfo {author} {\bibfnamefont {S.}~\bibnamefont {Ak\c{c}ay}},
  \bibinfo {author} {\bibfnamefont {S.}~\bibnamefont {Bernuzzi}}, \ and\
  \bibinfo {author} {\bibfnamefont {J.}~\bibnamefont {Williams}},\ }\href
  {\doibase 10.1103/PhysRevD.106.024020} {\bibfield  {journal} {\bibinfo
  {journal} {Phys. Rev. D}\ }\textbf {\bibinfo {volume} {106}},\ \bibinfo
  {pages} {024020} (\bibinfo {year} {2022})},\ \Eprint
  {http://arxiv.org/abs/2111.03675} {arXiv:2111.03675 [gr-qc]} \BibitemShut
  {NoStop}%
\bibitem [{\citenamefont {Albertini}\ \emph {et~al.}(2024)\citenamefont
  {Albertini}, \citenamefont {Gamba}, \citenamefont {Nagar},\ and\
  \citenamefont {Bernuzzi}}]{Albertini:2023aol}%
  \BibitemOpen
  \bibfield  {author} {\bibinfo {author} {\bibfnamefont {A.}~\bibnamefont
  {Albertini}}, \bibinfo {author} {\bibfnamefont {R.}~\bibnamefont {Gamba}},
  \bibinfo {author} {\bibfnamefont {A.}~\bibnamefont {Nagar}}, \ and\ \bibinfo
  {author} {\bibfnamefont {S.}~\bibnamefont {Bernuzzi}},\ }\href {\doibase
  10.1103/PhysRevD.109.044022} {\bibfield  {journal} {\bibinfo  {journal}
  {Phys. Rev. D}\ }\textbf {\bibinfo {volume} {109}},\ \bibinfo {pages}
  {044022} (\bibinfo {year} {2024})},\ \Eprint
  {http://arxiv.org/abs/2310.13578} {arXiv:2310.13578 [gr-qc]} \BibitemShut
  {NoStop}%
\bibitem [{\citenamefont {Spiers}\ \emph
  {et~al.}(2023{\natexlab{c}})\citenamefont {Spiers}, \citenamefont {Maselli},\
  and\ \citenamefont {Sotiriou}}]{Spiers:2023cva}%
  \BibitemOpen
  \bibfield  {author} {\bibinfo {author} {\bibfnamefont {A.}~\bibnamefont
  {Spiers}}, \bibinfo {author} {\bibfnamefont {A.}~\bibnamefont {Maselli}}, \
  and\ \bibinfo {author} {\bibfnamefont {T.~P.}\ \bibnamefont {Sotiriou}},\
  }\href@noop {} {\  (\bibinfo {year} {2023}{\natexlab{c}})},\ \Eprint
  {http://arxiv.org/abs/2310.02315} {arXiv:2310.02315 [gr-qc]} \BibitemShut
  {NoStop}%
\bibitem [{\citenamefont {Huerta}\ and\ \citenamefont
  {Gair}(2011)}]{Huerta:2011kt}%
  \BibitemOpen
  \bibfield  {author} {\bibinfo {author} {\bibfnamefont {E.~A.}\ \bibnamefont
  {Huerta}}\ and\ \bibinfo {author} {\bibfnamefont {J.~R.}\ \bibnamefont
  {Gair}},\ }\href {\doibase 10.1103/PhysRevD.84.064023} {\bibfield  {journal}
  {\bibinfo  {journal} {Phys. Rev. D}\ }\textbf {\bibinfo {volume} {84}},\
  \bibinfo {pages} {064023} (\bibinfo {year} {2011})},\ \Eprint
  {http://arxiv.org/abs/1105.3567} {arXiv:1105.3567 [gr-qc]} \BibitemShut
  {NoStop}%
\bibitem [{\citenamefont {Huerta}\ \emph {et~al.}(2012)\citenamefont {Huerta},
  \citenamefont {Gair},\ and\ \citenamefont {Brown}}]{Huerta:2011zi}%
  \BibitemOpen
  \bibfield  {author} {\bibinfo {author} {\bibfnamefont {E.~A.}\ \bibnamefont
  {Huerta}}, \bibinfo {author} {\bibfnamefont {J.~R.}\ \bibnamefont {Gair}}, \
  and\ \bibinfo {author} {\bibfnamefont {D.~A.}\ \bibnamefont {Brown}},\ }\href
  {\doibase 10.1103/PhysRevD.85.064023} {\bibfield  {journal} {\bibinfo
  {journal} {Phys. Rev. D}\ }\textbf {\bibinfo {volume} {85}},\ \bibinfo
  {pages} {064023} (\bibinfo {year} {2012})},\ \Eprint
  {http://arxiv.org/abs/1111.3243} {arXiv:1111.3243 [gr-qc]} \BibitemShut
  {NoStop}%
\bibitem [{\citenamefont {Piovano}\ \emph {et~al.}(2021)\citenamefont
  {Piovano}, \citenamefont {Brito}, \citenamefont {Maselli},\ and\
  \citenamefont {Pani}}]{Piovano:2021iwv}%
  \BibitemOpen
  \bibfield  {author} {\bibinfo {author} {\bibfnamefont {G.~A.}\ \bibnamefont
  {Piovano}}, \bibinfo {author} {\bibfnamefont {R.}~\bibnamefont {Brito}},
  \bibinfo {author} {\bibfnamefont {A.}~\bibnamefont {Maselli}}, \ and\
  \bibinfo {author} {\bibfnamefont {P.}~\bibnamefont {Pani}},\ }\href {\doibase
  10.1103/PhysRevD.104.124019} {\bibfield  {journal} {\bibinfo  {journal}
  {Phys. Rev. D}\ }\textbf {\bibinfo {volume} {104}},\ \bibinfo {pages}
  {124019} (\bibinfo {year} {2021})},\ \Eprint
  {http://arxiv.org/abs/2105.07083} {arXiv:2105.07083 [gr-qc]} \BibitemShut
  {NoStop}%
\bibitem [{\citenamefont {Lynch}\ \emph {et~al.}(2023)\citenamefont {Lynch},
  \citenamefont {van~de Meent},\ and\ \citenamefont
  {Warburton}}]{Lynch:2023gpu}%
  \BibitemOpen
  \bibfield  {author} {\bibinfo {author} {\bibfnamefont {P.}~\bibnamefont
  {Lynch}}, \bibinfo {author} {\bibfnamefont {M.}~\bibnamefont {van~de Meent}},
  \ and\ \bibinfo {author} {\bibfnamefont {N.}~\bibnamefont {Warburton}},\
  }\href@noop {} {\  (\bibinfo {year} {2023})},\ \Eprint
  {http://arxiv.org/abs/2305.10533} {arXiv:2305.10533 [gr-qc]} \BibitemShut
  {NoStop}%
\bibitem [{\citenamefont {Drummond}\ \emph {et~al.}(2024)\citenamefont
  {Drummond}, \citenamefont {Lynch}, \citenamefont {Hanselman}, \citenamefont
  {Becker},\ and\ \citenamefont {Hughes}}]{Drummond:2023wqc}%
  \BibitemOpen
  \bibfield  {author} {\bibinfo {author} {\bibfnamefont {L.~V.}\ \bibnamefont
  {Drummond}}, \bibinfo {author} {\bibfnamefont {P.}~\bibnamefont {Lynch}},
  \bibinfo {author} {\bibfnamefont {A.~G.}\ \bibnamefont {Hanselman}}, \bibinfo
  {author} {\bibfnamefont {D.~R.}\ \bibnamefont {Becker}}, \ and\ \bibinfo
  {author} {\bibfnamefont {S.~A.}\ \bibnamefont {Hughes}},\ }\href {\doibase
  10.1103/PhysRevD.109.064030} {\bibfield  {journal} {\bibinfo  {journal}
  {Phys. Rev. D}\ }\textbf {\bibinfo {volume} {109}},\ \bibinfo {pages}
  {064030} (\bibinfo {year} {2024})},\ \Eprint
  {http://arxiv.org/abs/2310.08438} {arXiv:2310.08438 [gr-qc]} \BibitemShut
  {NoStop}%
\bibitem [{\citenamefont {Loutrel}\ \emph {et~al.}(2023)\citenamefont
  {Loutrel}, \citenamefont {Brito}, \citenamefont {Maselli},\ and\
  \citenamefont {Pani}}]{Loutrel:2023boq}%
  \BibitemOpen
  \bibfield  {author} {\bibinfo {author} {\bibfnamefont {N.}~\bibnamefont
  {Loutrel}}, \bibinfo {author} {\bibfnamefont {R.}~\bibnamefont {Brito}},
  \bibinfo {author} {\bibfnamefont {A.}~\bibnamefont {Maselli}}, \ and\
  \bibinfo {author} {\bibfnamefont {P.}~\bibnamefont {Pani}},\ }\href@noop {}
  {\  (\bibinfo {year} {2023})},\ \Eprint {http://arxiv.org/abs/2309.17404}
  {arXiv:2309.17404 [gr-qc]} \BibitemShut {NoStop}%
\bibitem [{\citenamefont {Fujita}(2012)}]{Fujita:2012cm}%
  \BibitemOpen
  \bibfield  {author} {\bibinfo {author} {\bibfnamefont {R.}~\bibnamefont
  {Fujita}},\ }\href {\doibase 10.1143/PTP.128.971} {\bibfield  {journal}
  {\bibinfo  {journal} {Prog. Theor. Phys.}\ }\textbf {\bibinfo {volume}
  {128}},\ \bibinfo {pages} {971} (\bibinfo {year} {2012})},\ \Eprint
  {http://arxiv.org/abs/1211.5535} {arXiv:1211.5535 [gr-qc]} \BibitemShut
  {NoStop}%
\bibitem [{\citenamefont {Yunes}\ and\ \citenamefont
  {Berti}(2008)}]{Yunes:2008tw}%
  \BibitemOpen
  \bibfield  {author} {\bibinfo {author} {\bibfnamefont {N.}~\bibnamefont
  {Yunes}}\ and\ \bibinfo {author} {\bibfnamefont {E.}~\bibnamefont {Berti}},\
  }\href {\doibase 10.1103/PhysRevD.77.124006} {\bibfield  {journal} {\bibinfo
  {journal} {Phys. Rev. D}\ }\textbf {\bibinfo {volume} {77}},\ \bibinfo
  {pages} {124006} (\bibinfo {year} {2008})},\ \bibinfo {note} {[Erratum:
  Phys.Rev.D 83, 109901 (2011)]},\ \Eprint {http://arxiv.org/abs/0803.1853}
  {arXiv:0803.1853 [gr-qc]} \BibitemShut {NoStop}%
\bibitem [{\citenamefont {Futamase}\ and\ \citenamefont
  {Schutz}(1983)}]{Futamase:1983vsr}%
  \BibitemOpen
  \bibfield  {author} {\bibinfo {author} {\bibfnamefont {T.}~\bibnamefont
  {Futamase}}\ and\ \bibinfo {author} {\bibfnamefont {B.~F.}\ \bibnamefont
  {Schutz}},\ }\href {\doibase 10.1103/PhysRevD.28.2363} {\bibfield  {journal}
  {\bibinfo  {journal} {Phys. Rev. D}\ }\textbf {\bibinfo {volume} {28}},\
  \bibinfo {pages} {2363} (\bibinfo {year} {1983})}\BibitemShut {NoStop}%
\bibitem [{\citenamefont {Chua}\ \emph {et~al.}(2021)\citenamefont {Chua},
  \citenamefont {Katz}, \citenamefont {Warburton},\ and\ \citenamefont
  {Hughes}}]{Chua:2020stf}%
  \BibitemOpen
  \bibfield  {author} {\bibinfo {author} {\bibfnamefont {A.~J.~K.}\
  \bibnamefont {Chua}}, \bibinfo {author} {\bibfnamefont {M.~L.}\ \bibnamefont
  {Katz}}, \bibinfo {author} {\bibfnamefont {N.}~\bibnamefont {Warburton}}, \
  and\ \bibinfo {author} {\bibfnamefont {S.~A.}\ \bibnamefont {Hughes}},\
  }\href {\doibase 10.1103/PhysRevLett.126.051102} {\bibfield  {journal}
  {\bibinfo  {journal} {Phys. Rev. Lett.}\ }\textbf {\bibinfo {volume} {126}},\
  \bibinfo {pages} {051102} (\bibinfo {year} {2021})},\ \Eprint
  {http://arxiv.org/abs/2008.06071} {arXiv:2008.06071 [gr-qc]} \BibitemShut
  {NoStop}%
\bibitem [{\citenamefont {Hughes}\ \emph {et~al.}(2021)\citenamefont {Hughes},
  \citenamefont {Warburton}, \citenamefont {Khanna}, \citenamefont {Chua},\
  and\ \citenamefont {Katz}}]{Hughes:2021exa}%
  \BibitemOpen
  \bibfield  {author} {\bibinfo {author} {\bibfnamefont {S.~A.}\ \bibnamefont
  {Hughes}}, \bibinfo {author} {\bibfnamefont {N.}~\bibnamefont {Warburton}},
  \bibinfo {author} {\bibfnamefont {G.}~\bibnamefont {Khanna}}, \bibinfo
  {author} {\bibfnamefont {A.~J.~K.}\ \bibnamefont {Chua}}, \ and\ \bibinfo
  {author} {\bibfnamefont {M.~L.}\ \bibnamefont {Katz}},\ }\href {\doibase
  10.1103/PhysRevD.103.104014} {\bibfield  {journal} {\bibinfo  {journal}
  {Phys. Rev. D}\ }\textbf {\bibinfo {volume} {103}},\ \bibinfo {pages}
  {104014} (\bibinfo {year} {2021})},\ \bibinfo {note} {[Erratum: Phys.Rev.D
  107, 089901 (2023)]},\ \Eprint {http://arxiv.org/abs/2102.02713}
  {arXiv:2102.02713 [gr-qc]} \BibitemShut {NoStop}%
\bibitem [{\citenamefont {Katz}\ \emph {et~al.}(2021)\citenamefont {Katz},
  \citenamefont {Chua}, \citenamefont {Speri}, \citenamefont {Warburton},\ and\
  \citenamefont {Hughes}}]{Katz:2021yft}%
  \BibitemOpen
  \bibfield  {author} {\bibinfo {author} {\bibfnamefont {M.~L.}\ \bibnamefont
  {Katz}}, \bibinfo {author} {\bibfnamefont {A.~J.~K.}\ \bibnamefont {Chua}},
  \bibinfo {author} {\bibfnamefont {L.}~\bibnamefont {Speri}}, \bibinfo
  {author} {\bibfnamefont {N.}~\bibnamefont {Warburton}}, \ and\ \bibinfo
  {author} {\bibfnamefont {S.~A.}\ \bibnamefont {Hughes}},\ }\href {\doibase
  10.1103/PhysRevD.104.064047} {\bibfield  {journal} {\bibinfo  {journal}
  {Phys. Rev. D}\ }\textbf {\bibinfo {volume} {104}},\ \bibinfo {pages}
  {064047} (\bibinfo {year} {2021})},\ \Eprint
  {http://arxiv.org/abs/2104.04582} {arXiv:2104.04582 [gr-qc]} \BibitemShut
  {NoStop}%
\bibitem [{\citenamefont {Speri}\ \emph {et~al.}(2023)\citenamefont {Speri},
  \citenamefont {Katz}, \citenamefont {Chua}, \citenamefont {Hughes},
  \citenamefont {Warburton}, \citenamefont {Thompson}, \citenamefont
  {Chapman-Bird},\ and\ \citenamefont {Gair}}]{Speri:2023jte}%
  \BibitemOpen
  \bibfield  {author} {\bibinfo {author} {\bibfnamefont {L.}~\bibnamefont
  {Speri}}, \bibinfo {author} {\bibfnamefont {M.~L.}\ \bibnamefont {Katz}},
  \bibinfo {author} {\bibfnamefont {A.~J.~K.}\ \bibnamefont {Chua}}, \bibinfo
  {author} {\bibfnamefont {S.~A.}\ \bibnamefont {Hughes}}, \bibinfo {author}
  {\bibfnamefont {N.}~\bibnamefont {Warburton}}, \bibinfo {author}
  {\bibfnamefont {J.~E.}\ \bibnamefont {Thompson}}, \bibinfo {author}
  {\bibfnamefont {C.~E.~A.}\ \bibnamefont {Chapman-Bird}}, \ and\ \bibinfo
  {author} {\bibfnamefont {J.~R.}\ \bibnamefont {Gair}},\ }\href@noop {} {\
  (\bibinfo {year} {2023})},\ \Eprint {http://arxiv.org/abs/2307.12585}
  {arXiv:2307.12585 [gr-qc]} \BibitemShut {NoStop}%
\bibitem [{\citenamefont {Albertini}\ \emph {et~al.}(2022)\citenamefont
  {Albertini}, \citenamefont {Nagar}, \citenamefont {Pound}, \citenamefont
  {Warburton}, \citenamefont {Wardell}, \citenamefont {Durkan},\ and\
  \citenamefont {Miller}}]{Albertini:2022dmc}%
  \BibitemOpen
  \bibfield  {author} {\bibinfo {author} {\bibfnamefont {A.}~\bibnamefont
  {Albertini}}, \bibinfo {author} {\bibfnamefont {A.}~\bibnamefont {Nagar}},
  \bibinfo {author} {\bibfnamefont {A.}~\bibnamefont {Pound}}, \bibinfo
  {author} {\bibfnamefont {N.}~\bibnamefont {Warburton}}, \bibinfo {author}
  {\bibfnamefont {B.}~\bibnamefont {Wardell}}, \bibinfo {author} {\bibfnamefont
  {L.}~\bibnamefont {Durkan}}, \ and\ \bibinfo {author} {\bibfnamefont
  {J.}~\bibnamefont {Miller}},\ }\href {\doibase 10.1103/PhysRevD.106.084062}
  {\bibfield  {journal} {\bibinfo  {journal} {Phys. Rev. D}\ }\textbf {\bibinfo
  {volume} {106}},\ \bibinfo {pages} {084062} (\bibinfo {year} {2022})},\
  \Eprint {http://arxiv.org/abs/2208.02055} {arXiv:2208.02055 [gr-qc]}
  \BibitemShut {NoStop}%
\bibitem [{\citenamefont {van~de Meent}\ \emph {et~al.}(2023)\citenamefont
  {van~de Meent}, \citenamefont {Buonanno}, \citenamefont {Mihaylov},
  \citenamefont {Ossokine}, \citenamefont {Pompili}, \citenamefont {Warburton},
  \citenamefont {Pound}, \citenamefont {Wardell}, \citenamefont {Durkan},\ and\
  \citenamefont {Miller}}]{vandeMeent:2023ols}%
  \BibitemOpen
  \bibfield  {author} {\bibinfo {author} {\bibfnamefont {M.}~\bibnamefont
  {van~de Meent}}, \bibinfo {author} {\bibfnamefont {A.}~\bibnamefont
  {Buonanno}}, \bibinfo {author} {\bibfnamefont {D.~P.}\ \bibnamefont
  {Mihaylov}}, \bibinfo {author} {\bibfnamefont {S.}~\bibnamefont {Ossokine}},
  \bibinfo {author} {\bibfnamefont {L.}~\bibnamefont {Pompili}}, \bibinfo
  {author} {\bibfnamefont {N.}~\bibnamefont {Warburton}}, \bibinfo {author}
  {\bibfnamefont {A.}~\bibnamefont {Pound}}, \bibinfo {author} {\bibfnamefont
  {B.}~\bibnamefont {Wardell}}, \bibinfo {author} {\bibfnamefont
  {L.}~\bibnamefont {Durkan}}, \ and\ \bibinfo {author} {\bibfnamefont
  {J.}~\bibnamefont {Miller}},\ }\href {\doibase 10.1103/PhysRevD.108.124038}
  {\bibfield  {journal} {\bibinfo  {journal} {Phys. Rev. D}\ }\textbf {\bibinfo
  {volume} {108}},\ \bibinfo {pages} {124038} (\bibinfo {year} {2023})},\
  \Eprint {http://arxiv.org/abs/2303.18026} {arXiv:2303.18026 [gr-qc]}
  \BibitemShut {NoStop}%
\bibitem [{\citenamefont {Loutrel}\ \emph {et~al.}()\citenamefont {Loutrel},
  \citenamefont {Mukherjee}, \citenamefont {Maselli},\ and\ \citenamefont
  {Pani}}]{followup}%
  \BibitemOpen
  \bibfield  {author} {\bibinfo {author} {\bibfnamefont {N.}~\bibnamefont
  {Loutrel}}, \bibinfo {author} {\bibfnamefont {S.}~\bibnamefont {Mukherjee}},
  \bibinfo {author} {\bibfnamefont {A.}~\bibnamefont {Maselli}}, \ and\
  \bibinfo {author} {\bibfnamefont {P.}~\bibnamefont {Pani}},\ }\href@noop {}
  {\ }\bibinfo {note} {In preparation (2024)}\BibitemShut {NoStop}%
\bibitem [{\citenamefont {Khan}\ \emph {et~al.}(2019)\citenamefont {Khan},
  \citenamefont {Chatziioannou}, \citenamefont {Hannam},\ and\ \citenamefont
  {Ohme}}]{Khan:2018fmp}%
  \BibitemOpen
  \bibfield  {author} {\bibinfo {author} {\bibfnamefont {S.}~\bibnamefont
  {Khan}}, \bibinfo {author} {\bibfnamefont {K.}~\bibnamefont {Chatziioannou}},
  \bibinfo {author} {\bibfnamefont {M.}~\bibnamefont {Hannam}}, \ and\ \bibinfo
  {author} {\bibfnamefont {F.}~\bibnamefont {Ohme}},\ }\href {\doibase
  10.1103/PhysRevD.100.024059} {\bibfield  {journal} {\bibinfo  {journal}
  {Phys. Rev. D}\ }\textbf {\bibinfo {volume} {100}},\ \bibinfo {pages}
  {024059} (\bibinfo {year} {2019})},\ \Eprint
  {http://arxiv.org/abs/1809.10113} {arXiv:1809.10113 [gr-qc]} \BibitemShut
  {NoStop}%
\bibitem [{\citenamefont {Kesden}\ \emph {et~al.}(2015)\citenamefont {Kesden},
  \citenamefont {Gerosa}, \citenamefont {O'Shaughnessy}, \citenamefont
  {Berti},\ and\ \citenamefont {Sperhake}}]{Kesden:2014sla}%
  \BibitemOpen
  \bibfield  {author} {\bibinfo {author} {\bibfnamefont {M.}~\bibnamefont
  {Kesden}}, \bibinfo {author} {\bibfnamefont {D.}~\bibnamefont {Gerosa}},
  \bibinfo {author} {\bibfnamefont {R.}~\bibnamefont {O'Shaughnessy}}, \bibinfo
  {author} {\bibfnamefont {E.}~\bibnamefont {Berti}}, \ and\ \bibinfo {author}
  {\bibfnamefont {U.}~\bibnamefont {Sperhake}},\ }\href {\doibase
  10.1103/PhysRevLett.114.081103} {\bibfield  {journal} {\bibinfo  {journal}
  {Phys. Rev. Lett.}\ }\textbf {\bibinfo {volume} {114}},\ \bibinfo {pages}
  {081103} (\bibinfo {year} {2015})},\ \Eprint {http://arxiv.org/abs/1411.0674}
  {arXiv:1411.0674 [gr-qc]} \BibitemShut {NoStop}%
\bibitem [{\citenamefont {Gerosa}\ \emph {et~al.}(2015)\citenamefont {Gerosa},
  \citenamefont {Kesden}, \citenamefont {Sperhake}, \citenamefont {Berti},\
  and\ \citenamefont {O'Shaughnessy}}]{Gerosa:2015tea}%
  \BibitemOpen
  \bibfield  {author} {\bibinfo {author} {\bibfnamefont {D.}~\bibnamefont
  {Gerosa}}, \bibinfo {author} {\bibfnamefont {M.}~\bibnamefont {Kesden}},
  \bibinfo {author} {\bibfnamefont {U.}~\bibnamefont {Sperhake}}, \bibinfo
  {author} {\bibfnamefont {E.}~\bibnamefont {Berti}}, \ and\ \bibinfo {author}
  {\bibfnamefont {R.}~\bibnamefont {O'Shaughnessy}},\ }\href {\doibase
  10.1103/PhysRevD.92.064016} {\bibfield  {journal} {\bibinfo  {journal} {Phys.
  Rev. D}\ }\textbf {\bibinfo {volume} {92}},\ \bibinfo {pages} {064016}
  (\bibinfo {year} {2015})},\ \Eprint {http://arxiv.org/abs/1506.03492}
  {arXiv:1506.03492 [gr-qc]} \BibitemShut {NoStop}%
\bibitem [{\citenamefont {Chatziioannou}\ \emph {et~al.}(2017)\citenamefont
  {Chatziioannou}, \citenamefont {Klein}, \citenamefont {Yunes},\ and\
  \citenamefont {Cornish}}]{Chatziioannou:2017tdw}%
  \BibitemOpen
  \bibfield  {author} {\bibinfo {author} {\bibfnamefont {K.}~\bibnamefont
  {Chatziioannou}}, \bibinfo {author} {\bibfnamefont {A.}~\bibnamefont
  {Klein}}, \bibinfo {author} {\bibfnamefont {N.}~\bibnamefont {Yunes}}, \ and\
  \bibinfo {author} {\bibfnamefont {N.}~\bibnamefont {Cornish}},\ }\href
  {\doibase 10.1103/PhysRevD.95.104004} {\bibfield  {journal} {\bibinfo
  {journal} {Phys. Rev. D}\ }\textbf {\bibinfo {volume} {95}},\ \bibinfo
  {pages} {104004} (\bibinfo {year} {2017})},\ \Eprint
  {http://arxiv.org/abs/1703.03967} {arXiv:1703.03967 [gr-qc]} \BibitemShut
  {NoStop}%
\bibitem [{\citenamefont {Apostolatos}\ \emph {et~al.}(1994)\citenamefont
  {Apostolatos}, \citenamefont {Cutler}, \citenamefont {Sussman},\ and\
  \citenamefont {Thorne}}]{ApostolatosCutler}%
  \BibitemOpen
  \bibfield  {author} {\bibinfo {author} {\bibfnamefont {T.~A.}\ \bibnamefont
  {Apostolatos}}, \bibinfo {author} {\bibfnamefont {C.}~\bibnamefont {Cutler}},
  \bibinfo {author} {\bibfnamefont {G.~J.}\ \bibnamefont {Sussman}}, \ and\
  \bibinfo {author} {\bibfnamefont {K.~S.}\ \bibnamefont {Thorne}},\ }\href
  {\doibase 10.1103/PhysRevD.49.6274} {\bibfield  {journal} {\bibinfo
  {journal} {Phys. Rev. D}\ }\textbf {\bibinfo {volume} {49}},\ \bibinfo
  {pages} {6274} (\bibinfo {year} {1994})}\BibitemShut {NoStop}%
\bibitem [{\citenamefont {Lang}\ and\ \citenamefont
  {Hughes}(2006)}]{Lang:2006bsg}%
  \BibitemOpen
  \bibfield  {author} {\bibinfo {author} {\bibfnamefont {R.~N.}\ \bibnamefont
  {Lang}}\ and\ \bibinfo {author} {\bibfnamefont {S.~A.}\ \bibnamefont
  {Hughes}},\ }\href {\doibase 10.1103/PhysRevD.75.089902} {\bibfield
  {journal} {\bibinfo  {journal} {Phys. Rev. D}\ }\textbf {\bibinfo {volume}
  {74}},\ \bibinfo {pages} {122001} (\bibinfo {year} {2006})},\ \bibinfo {note}
  {[Erratum: Phys.Rev.D 75, 089902 (2007), Erratum: Phys.Rev.D 77, 109901
  (2008)]},\ \Eprint {http://arxiv.org/abs/gr-qc/0608062} {arXiv:gr-qc/0608062}
  \BibitemShut {NoStop}%
\bibitem [{\citenamefont {Bender}\ and\ \citenamefont {Orszag}(1999)}]{Bender}%
  \BibitemOpen
  \bibfield  {author} {\bibinfo {author} {\bibfnamefont {C.~M.}\ \bibnamefont
  {Bender}}\ and\ \bibinfo {author} {\bibfnamefont {S.}~\bibnamefont
  {Orszag}},\ }\href {\doibase 10.1007/978-1-4757-3069-2} {\emph {\bibinfo
  {title} {{Advanced Mathematical Methods for Scientists and Engineers I:
  Asymptotic Methods and Perturbation Theory}}}}\ (\bibinfo  {publisher}
  {Springer},\ \bibinfo {address} {New York, NY},\ \bibinfo {year}
  {1999})\BibitemShut {NoStop}%
\bibitem [{\citenamefont {Klein}\ \emph {et~al.}(2014)\citenamefont {Klein},
  \citenamefont {Cornish},\ and\ \citenamefont {Yunes}}]{Klein:2014bua}%
  \BibitemOpen
  \bibfield  {author} {\bibinfo {author} {\bibfnamefont {A.}~\bibnamefont
  {Klein}}, \bibinfo {author} {\bibfnamefont {N.}~\bibnamefont {Cornish}}, \
  and\ \bibinfo {author} {\bibfnamefont {N.}~\bibnamefont {Yunes}},\ }\href
  {\doibase 10.1103/PhysRevD.90.124029} {\bibfield  {journal} {\bibinfo
  {journal} {Phys. Rev. D}\ }\textbf {\bibinfo {volume} {90}},\ \bibinfo
  {pages} {124029} (\bibinfo {year} {2014})},\ \Eprint
  {http://arxiv.org/abs/1408.5158} {arXiv:1408.5158 [gr-qc]} \BibitemShut
  {NoStop}%
\bibitem [{\citenamefont {Lang}\ \emph {et~al.}(2011)\citenamefont {Lang},
  \citenamefont {Hughes},\ and\ \citenamefont {Cornish}}]{Lang:2011je}%
  \BibitemOpen
  \bibfield  {author} {\bibinfo {author} {\bibfnamefont {R.~N.}\ \bibnamefont
  {Lang}}, \bibinfo {author} {\bibfnamefont {S.~A.}\ \bibnamefont {Hughes}}, \
  and\ \bibinfo {author} {\bibfnamefont {N.~J.}\ \bibnamefont {Cornish}},\
  }\href {\doibase 10.1103/PhysRevD.84.022002} {\bibfield  {journal} {\bibinfo
  {journal} {Phys. Rev. D}\ }\textbf {\bibinfo {volume} {84}},\ \bibinfo
  {pages} {022002} (\bibinfo {year} {2011})},\ \Eprint
  {http://arxiv.org/abs/1101.3591} {arXiv:1101.3591 [gr-qc]} \BibitemShut
  {NoStop}%
\bibitem [{\citenamefont {Yagi}\ and\ \citenamefont
  {Tanaka}(2010)}]{Yagi:2009zm}%
  \BibitemOpen
  \bibfield  {author} {\bibinfo {author} {\bibfnamefont {K.}~\bibnamefont
  {Yagi}}\ and\ \bibinfo {author} {\bibfnamefont {T.}~\bibnamefont {Tanaka}},\
  }\href {\doibase 10.1103/PhysRevD.81.109902} {\bibfield  {journal} {\bibinfo
  {journal} {Phys. Rev. D}\ }\textbf {\bibinfo {volume} {81}},\ \bibinfo
  {pages} {064008} (\bibinfo {year} {2010})},\ \bibinfo {note} {[Erratum:
  Phys.Rev.D 81, 109902 (2010)]},\ \Eprint {http://arxiv.org/abs/0906.4269}
  {arXiv:0906.4269 [gr-qc]} \BibitemShut {NoStop}%
\bibitem [{\citenamefont {Damour}\ \emph {et~al.}(2001)\citenamefont {Damour},
  \citenamefont {Iyer},\ and\ \citenamefont {Sathyaprakash}}]{Damour:2000zb}%
  \BibitemOpen
  \bibfield  {author} {\bibinfo {author} {\bibfnamefont {T.}~\bibnamefont
  {Damour}}, \bibinfo {author} {\bibfnamefont {B.~R.}\ \bibnamefont {Iyer}}, \
  and\ \bibinfo {author} {\bibfnamefont {B.~S.}\ \bibnamefont
  {Sathyaprakash}},\ }\href {\doibase 10.1103/PhysRevD.63.044023} {\bibfield
  {journal} {\bibinfo  {journal} {Phys. Rev. D}\ }\textbf {\bibinfo {volume}
  {63}},\ \bibinfo {pages} {044023} (\bibinfo {year} {2001})},\ \bibinfo {note}
  {[Erratum: Phys.Rev.D 72, 029902 (2005)]},\ \Eprint
  {http://arxiv.org/abs/gr-qc/0010009} {arXiv:gr-qc/0010009} \BibitemShut
  {NoStop}%
\bibitem [{\citenamefont {Mac~Uilliam}\ \emph {et~al.}(2024)\citenamefont
  {Mac~Uilliam}, \citenamefont {Akcay},\ and\ \citenamefont
  {Thompson}}]{MacUilliam:2024oif}%
  \BibitemOpen
  \bibfield  {author} {\bibinfo {author} {\bibfnamefont {J.}~\bibnamefont
  {Mac~Uilliam}}, \bibinfo {author} {\bibfnamefont {S.}~\bibnamefont {Akcay}},
  \ and\ \bibinfo {author} {\bibfnamefont {J.~E.}\ \bibnamefont {Thompson}},\
  }\href@noop {} {\  (\bibinfo {year} {2024})},\ \Eprint
  {http://arxiv.org/abs/2402.06781} {arXiv:2402.06781 [gr-qc]} \BibitemShut
  {NoStop}%
\bibitem [{\citenamefont {Arredondo}\ \emph {et~al.}(2024)\citenamefont
  {Arredondo}, \citenamefont {Klein},\ and\ \citenamefont
  {Yunes}}]{Arredondo:2024nsl}%
  \BibitemOpen
  \bibfield  {author} {\bibinfo {author} {\bibfnamefont {J.~N.}\ \bibnamefont
  {Arredondo}}, \bibinfo {author} {\bibfnamefont {A.}~\bibnamefont {Klein}}, \
  and\ \bibinfo {author} {\bibfnamefont {N.}~\bibnamefont {Yunes}},\
  }\href@noop {} {\  (\bibinfo {year} {2024})},\ \Eprint
  {http://arxiv.org/abs/2402.06804} {arXiv:2402.06804 [gr-qc]} \BibitemShut
  {NoStop}%
\bibitem [{\citenamefont {Lindblom}\ \emph {et~al.}(2008)\citenamefont
  {Lindblom}, \citenamefont {Owen},\ and\ \citenamefont
  {Brown}}]{Lindblom:2008cm}%
  \BibitemOpen
  \bibfield  {author} {\bibinfo {author} {\bibfnamefont {L.}~\bibnamefont
  {Lindblom}}, \bibinfo {author} {\bibfnamefont {B.~J.}\ \bibnamefont {Owen}},
  \ and\ \bibinfo {author} {\bibfnamefont {D.~A.}\ \bibnamefont {Brown}},\
  }\href {\doibase 10.1103/PhysRevD.78.124020} {\bibfield  {journal} {\bibinfo
  {journal} {Phys. Rev. D}\ }\textbf {\bibinfo {volume} {78}},\ \bibinfo
  {pages} {124020} (\bibinfo {year} {2008})},\ \Eprint
  {http://arxiv.org/abs/0809.3844} {arXiv:0809.3844 [gr-qc]} \BibitemShut
  {NoStop}%
\bibitem [{\citenamefont {Witzany}(2019)}]{Witzany:2019nml}%
  \BibitemOpen
  \bibfield  {author} {\bibinfo {author} {\bibfnamefont {V.}~\bibnamefont
  {Witzany}},\ }\href {\doibase 10.1103/PhysRevD.100.104030} {\bibfield
  {journal} {\bibinfo  {journal} {Phys. Rev. D}\ }\textbf {\bibinfo {volume}
  {100}},\ \bibinfo {pages} {104030} (\bibinfo {year} {2019})},\ \Eprint
  {http://arxiv.org/abs/1903.03651} {arXiv:1903.03651 [gr-qc]} \BibitemShut
  {NoStop}%
\bibitem [{\citenamefont {Witzany}\ and\ \citenamefont
  {Piovano}(2023)}]{Witzany:2023bmq}%
  \BibitemOpen
  \bibfield  {author} {\bibinfo {author} {\bibfnamefont {V.}~\bibnamefont
  {Witzany}}\ and\ \bibinfo {author} {\bibfnamefont {G.~A.}\ \bibnamefont
  {Piovano}},\ }\href@noop {} {\  (\bibinfo {year} {2023})},\ \Eprint
  {http://arxiv.org/abs/2308.00021} {arXiv:2308.00021 [gr-qc]} \BibitemShut
  {NoStop}%
\bibitem [{\citenamefont {Skoupy}\ \emph {et~al.}(2023)\citenamefont {Skoupy},
  \citenamefont {Lukes-Gerakopoulos}, \citenamefont {Drummond},\ and\
  \citenamefont {Hughes}}]{Skoupy:2023lih}%
  \BibitemOpen
  \bibfield  {author} {\bibinfo {author} {\bibfnamefont {V.}~\bibnamefont
  {Skoupy}}, \bibinfo {author} {\bibfnamefont {G.}~\bibnamefont
  {Lukes-Gerakopoulos}}, \bibinfo {author} {\bibfnamefont {L.~V.}\ \bibnamefont
  {Drummond}}, \ and\ \bibinfo {author} {\bibfnamefont {S.~A.}\ \bibnamefont
  {Hughes}},\ }\href {\doibase 10.1103/PhysRevD.108.044041} {\bibfield
  {journal} {\bibinfo  {journal} {Phys. Rev. D}\ }\textbf {\bibinfo {volume}
  {108}},\ \bibinfo {pages} {044041} (\bibinfo {year} {2023})},\ \Eprint
  {http://arxiv.org/abs/2303.16798} {arXiv:2303.16798 [gr-qc]} \BibitemShut
  {NoStop}%
\bibitem [{\citenamefont {Burke}\ \emph {et~al.}(2023)\citenamefont {Burke},
  \citenamefont {Piovano}, \citenamefont {Warburton}, \citenamefont {Lynch},
  \citenamefont {Speri}, \citenamefont {Kavanagh}, \citenamefont {Wardell},
  \citenamefont {Pound}, \citenamefont {Durkan},\ and\ \citenamefont
  {Miller}}]{Burke:2023lno}%
  \BibitemOpen
  \bibfield  {author} {\bibinfo {author} {\bibfnamefont {O.}~\bibnamefont
  {Burke}}, \bibinfo {author} {\bibfnamefont {G.~A.}\ \bibnamefont {Piovano}},
  \bibinfo {author} {\bibfnamefont {N.}~\bibnamefont {Warburton}}, \bibinfo
  {author} {\bibfnamefont {P.}~\bibnamefont {Lynch}}, \bibinfo {author}
  {\bibfnamefont {L.}~\bibnamefont {Speri}}, \bibinfo {author} {\bibfnamefont
  {C.}~\bibnamefont {Kavanagh}}, \bibinfo {author} {\bibfnamefont
  {B.}~\bibnamefont {Wardell}}, \bibinfo {author} {\bibfnamefont
  {A.}~\bibnamefont {Pound}}, \bibinfo {author} {\bibfnamefont
  {L.}~\bibnamefont {Durkan}}, \ and\ \bibinfo {author} {\bibfnamefont
  {J.}~\bibnamefont {Miller}},\ }\href@noop {} {\  (\bibinfo {year} {2023})},\
  \Eprint {http://arxiv.org/abs/2310.08927} {arXiv:2310.08927 [gr-qc]}
  \BibitemShut {NoStop}%
\bibitem [{\citenamefont {Chen}\ \emph {et~al.}(1994)\citenamefont {Chen},
  \citenamefont {Goldenfeld},\ and\ \citenamefont
  {Oono}}]{PhysRevLett.73.1311}%
  \BibitemOpen
  \bibfield  {author} {\bibinfo {author} {\bibfnamefont {L.~Y.}\ \bibnamefont
  {Chen}}, \bibinfo {author} {\bibfnamefont {N.}~\bibnamefont {Goldenfeld}}, \
  and\ \bibinfo {author} {\bibfnamefont {Y.}~\bibnamefont {Oono}},\ }\href
  {\doibase 10.1103/PhysRevLett.73.1311} {\bibfield  {journal} {\bibinfo
  {journal} {Phys. Rev. Lett.}\ }\textbf {\bibinfo {volume} {73}},\ \bibinfo
  {pages} {1311} (\bibinfo {year} {1994})}\BibitemShut {NoStop}%
\bibitem [{\citenamefont {Mudavanhu}\ and\ \citenamefont
  {O'Malley}(2001)}]{RG.2001}%
  \BibitemOpen
  \bibfield  {author} {\bibinfo {author} {\bibfnamefont {B.}~\bibnamefont
  {Mudavanhu}}\ and\ \bibinfo {author} {\bibfnamefont {R.~E.}\ \bibnamefont
  {O'Malley}, \bibfnamefont {Jr.}},\ }\href {\doibase
  https://doi.org/10.1111/1467-9590.1071178} {\bibfield  {journal} {\bibinfo
  {journal} {Studies in Applied Mathematics}\ }\textbf {\bibinfo {volume}
  {107}},\ \bibinfo {pages} {63} (\bibinfo {year} {2001})},\ \Eprint
  {http://arxiv.org/abs/https://onlinelibrary.wiley.com/doi/pdf/10.1111/1467-9590.1071178}
  {https://onlinelibrary.wiley.com/doi/pdf/10.1111/1467-9590.1071178}
  \BibitemShut {NoStop}%
\end{thebibliography}%
\end{document}